\documentclass[sn-basic]{sn-jnl}

\usepackage[normalem]{ulem}
\usepackage{comment}
\usepackage{graphicx}%
\usepackage{multirow}%
\usepackage{amsmath,amssymb,amsfonts}%
\usepackage{amsthm}%
\usepackage{mathrsfs}%
\usepackage[title]{appendix}%
\usepackage{xcolor}%
\usepackage{textcomp}%
\usepackage{manyfoot}%
\usepackage{booktabs}%
\usepackage{algorithm}%
\usepackage{algorithmicx}%
\usepackage{algpseudocode}%
\usepackage{listings}%

\jyear{2023}%

\theoremstyle{thmstyleone}%
\theoremstyle{thmstyletwo}%

\theoremstyle{thmstylethree}%

\begin{document}

\title[Systems and Complexity in Helio]{Complexity Heliophysics: A lived and living history of systems and complexity science in Heliophysics}


\author*[1]{\fnm{Ryan M.} \sur{McGranaghan}}



\affil*[1]{\orgname{NASA Jet Propulsion Laboratory, California Institute of Technology}, \orgaddress{\city{Pasadena}, \postcode{91109}, \state{CA}, \country{USA}}}




\abstract{

This review examines \textit{complexity science} in the context of Heliophysics, describing it not as a discipline, but as a paradigm. In the context of Heliophysics, complexity science is the study of a star, interplanetary environment, magnetosphere, upper and terrestrial atmospheres, and planetary surface as interacting subsystems. Complexity science studies entities in a system (e.g., electrons in an atom, planets in a solar system, individuals in a society) and their interactions, and is the nature of what emerges from these interactions. It is a paradigm that employs systems approaches and is inherently multi- and cross-scale. Heliophysics processes span at least 15 orders of magnitude in space and another 15 in time, and its reaches go well beyond our own solar system and Earth’s space environment to touch planetary, exoplanetary, and astrophysical domains. It is an uncommon domain within which to explore complexity science.

After first outlining the dimensions of complexity science, the review proceeds in three epochal parts: 1) A pivotal year in the Complexity Heliophysics paradigm: 1996; 2) The transitional years that established foundations of the paradigm (1996-2010); and 3) The emergent literature largely beyond 2010. 

This review article excavates the lived and living history of complexity science in Heliophysics. It identifies five dimensions of complexity science, some enjoying much scholarship in Heliophysics, others that represent relative gaps in the existing research. The history reveals a grand challenge that confronts Heliophysics, as with most physical sciences, to understand the research intersection between fundamental science (e.g., complexity science) and applied science (e.g., artificial intelligence and
machine learning (AI/ML)). A risk science framework is suggested as a way of formulating the grand scientific and societal challenges in a way that AI/ML and complexity science converge.

The intention is to provide inspiration, help researchers think more coherently about ideas of complexity science in Heliophysics, and guide future research. It will be instructive to Heliophysics researchers, but also to any reader interested in or hoping to advance the frontier of systems and complexity science. 

}




\maketitle

\section{Introduction}\label{introduction section}



\begin{quote}
    The [21st] century will be the century of complexity. 
    -Stephen Hawking 
\end{quote}

Heliophysics is `a fundamental science discipline that is the study of the very nature of plasmas throughout space, originating with our own Sun and heliosphere and extending to planetary atmospheres and magnetospheres, stellar atmospheres and astrospheres, and interstellar space' \cite{Cohen2023ReimaginingHA}. Heliophysics processes span at least 15 orders of magnitude in space (from, for instance, the gyroradii of particles in the solar or Earth atmosphere important to wave-particle and particle-particle interactions at sub-micrometer lengths to the $10\times^{10}$ sizes solar phenomena can reach or the interplanetary distances over which these processes must be understood) and another 15 in time (from reconnection and collision to solar cycle or longer time scales). The reaches of this science go well beyond our own solar system and Earth’s space environment to touch planetary, exoplanetary, and astrophysical domains. The history of Heliophysics has, like many sciences, been one of specialization--categorizing and separating domains and building understanding within those ever more boundaried systems. The approach has produced remarkable achievement, yet in a century in which:
\begin{itemize}
    \item sensing capabilities are revealing multi-scale (distinct phenomena occurring at different scales) and cross-scale (ways in which interactions occur across scales) behavior (e.g., field-aligned currents across scales \cite{McGranaghan_2017a}); 
    \item data analysis and computational tools are enabling cross-system and multi-scale research \cite{McGranaghan_2017b} (e.g., combined particle and magnetohydrodynamic (MHD) simulations \cite{Sorathia_2017}); and
    \item the practical demands on Heliophysics science are growing (e.g., societal dependence on space, risk to critical infrastructure due to space weather, expansion of humanity into space and the solar system--each in some respect dissolving of the boundary between Heliophysics (fundamental scientific discovery of Sun-planet connection) and humanity (the ways Heliophysics processes and phenomena affect lives and society)),    
\end{itemize}
\noindent the Heliophysics community faces the need to shift the paradigm by which it creates new scientific knowledge \cite{Kuhn_1962}--the advent of Complexity Heliophysics. We will capitalize the term in this review to draw attention to the fact that we posit it as a paradigm, a framework of assumptions, principles and methods from which the members of the community work \cite{Kuhn_1962}; a kind of generalization that characterizes the next stage of a community's work \cite{Anderson1972MoreID}. Note that we are not inventing the paradigm, merely giving it a name and outlining and connecting the varied research and research avenues that compose it with the intention of providing inspiration, helping researchers think more coherently about these ideas, and guiding future research. One of the world's premier complexity science institutions, the Santa Fe Institute (SFI), originated from a recognition that trends and challenges in chemistry, biology, and psychology revealed the need for a new paradigm. One of SFI's founding workshops was titled ``A Response to the Challenge of Emerging Syntheses in Science: A New Kind of Research and Teaching Institution'' \cite{pines2018emerging}. 

In order to usher in this new paradigm, we must first understand what complexity science is, specifically in the context of Heliophysics research. Complexity is a difficult thing to define. It is often used synonymously with `something we do not yet know' \cite{Cilliers2000}, a placeholder label for a new frontier of scientific knowledge (see \cite{Ladyman2020WhatIA} for discussion about definitions of complexity). However, this review will adopt a more principled view. We will refer to complexity science as a paradigm, drawing distinction from attempts to describe it as a discipline, ill-suited to what complexity actually is. Indeed complexity is a paradigm of scientific discovery \cite{Hofstadter1999, Mitchell2009ComplexityA, pines2018emerging, krakauer2018worlds, Ladyman2020WhatIA}. It is fundamentally distinct from `complicated.' It is the study of phenomena that emerge from a collection of interacting objects. To understand a complex system requires a plurality of frameworks and the ability to move more seamlessly between scales of the system (e.g., micro and macro). As such, complexity science spans numerous dimensions. This review reveals those dimensions through examination of a large corpus of Heliohphysics research articles. These dimensions organize this review, so we introduce them upfront along with their context in Heliophysics to provide a kind of scaffolding for the material that follows.

To understand this review readers must be aware of the central motivation: Heliophysics faces a grand challenge to unify basic and applied research. Indeed, this has been and will remain a grand challenge across the sciences \cite{Bush1945TheEF}. The challenge in the 21st century and that is made clear in this review is now one of understanding the connections between complexity science and machine learning/artificial intelligence.

\subsection{Systems science and crossing scales}
Systems science is the philosophy and methods for studying systems, sets of things interconnected in such a way to produce their own pattern of behavior over time \cite{meadows2008thinking}. It is a core of complexity thinking, whereby the pattern of behavior is not reducible to the behavior of any of the individual things. Crossing between the scales of the system is the foundation of the complexity science paradigm. Complexity science inherently expands across scale--the interacting objects may be particles in a simulation regime, flux ropes and their individual and connected dynamics on the surface of the Sun, plasma in the magnetotail subject to magnetic and electric fields, or coupling of entire systems like the magnetosphere and ionosphere. Indeed it can be any collection of interconnecting things that give rise to behaviors from their interactions--people in a research group or institution, cells in an organism, anything. 

In the context of Heliophysics, complexity science is the study of a star, interplanetary environment, magnetosphere, upper and terrestrial atmospheres, and planetary surface as interacting subsystems. Each of these subsystems can be further broken down into regions (e.g., the auroral region of the upper atmosphere) and all the way down to more elementary components such as electrons and protons. At which scale one chooses to examine the Heliophysics system determines the methods one uses and ultimately one's understanding of it. For instance, in the magnetosphere one might choose the macro-scale, dictating a magnetohydrodynamic (MHD) method, or the micro-scale, requiring a kinetic method. Indeed, one of the most vexing questions that has obstinately refused answer in Heliophysics has been, ``What level do we need to look at the system to understand and predict it?'' \cite{Denton2016PrefaceUP, Viall_2020, Borovsky2020OutstandingQI}. Complexity science is a paradigm that suggests ways of reconciling the micro and macro scales. It is the collection of methods to understand a system across scales, the smaller scale behavior in connection with the phenomena that emerge from it. 

\subsection{Emergence, self-organization, and scaling theory}
Inseparable from cross-scale understanding is the concept of emergence. Emergence is the term used to describe phenomena that are `more than the sum of their parts' \cite{Rosas_2020}. Emergence is observed in virtually all areas of inquiry, such as how large numbers of individual fish are able to behave dynamically as a school when threatened by a predator \cite{Parrish_2002}. In terms of scale, emergence is the occurrence of actions at one scale giving rise to phenomena on another level. The idea that order at some higher order or \textit{coarse-grained} level of a system arises from a number of interacting sub-systems is called self-organization \cite{Castiglione2008ChaosAC, Flack_2017, Hooker_2011}. Self-organization is a powerful toehold in complexity analyses because it reveals that emergence is observable in statistical characteristics of the system (often as power law distributions). Indeed, many fields have searched for a universal underlying generative mechanism for the ubiquitous observation of power laws in nature and self-organization has created excitement as one of the potential mechanisms \cite{Newman2004PowerLP}. Two caveats apply: 1) it is likely that there are many different mechanisms that produce power laws with different mechanisms applying in different situations; and 2) self-organization is a collective reference to many processes applicable to different systems and unsurprisingly produces behavior that is not always power law \cite{Clauset2007PowerLawDI}. However, the regularity of power law or power law-like behavior in nature and the frequent correspondence of these distributions to self-organization make the mechanism (self-organization) and this particular outcome from it (power laws) an important feature for understanding complexity. 

Power laws are a departure from assumptions of normality that has governed much traditional scientific and engineering analyses and instead involves heavy tails in the probability distributions. Normal and Gaussian distributions are `light-tailed,' meaning there is an exponential fall off moving into the tail such that the likelihood of extreme events are exponentially bounded. Heavy-tailed distributions, on the other hand, are not exponentially bounded and have heavier tails than the exponential distribution. In this review, we use the general definition of heavy-tailed as any distribution that has a heavier tail than a normal distribution. Like self-organization, power laws are powerful because they imply underlying driving mechanisms that are identical across scales of the system and produce the same statistical signature at all scales. For instance, biological organisms across a remarkable range (e.g., mice to elephants) exhibit power law scaling with a 1/4 slope such that the metabolic rate increases proportionally to the body size raised to a 1/4 power \cite{west2017scale}. So as body size doubles the metabolic rate, or the rate at which the organism consumes energy, increases only by about 75\%. Such scaling relationships could provide fundamental principles governing systems, in this case the physiology and energy requirements of living organisms. Power laws are found across systems, from cells to cities, and there is little doubt that they extend to Heliophysics. Instead, the question Heliophysics has been grappling with is whether they reveal deeper understanding of the processes driving the distribution and how to pull out principles of the physical system that they might point to. This review will cover instances where power laws are found in Heliophysics systems. It is important to note that power laws are not the only distributions characterizing scaling laws in nature, only the most widely discussed \cite{Clauset2007PowerLawDI}, so we adopt it as the category through which to examine the literature relating complex behavior to distributions it generates. The similarity across scales that a power law reveals is a property called scale-free or scale-invariant. Scale-invariance means that there is some feature or behavior of a system that does not change if the scale of the variables are multiplied by a common factor (for instance, a branch of a tree looks like the complete tree, only on a different length scale; often associated with self-similarity). It is an observed property and does not strictly imply a single driving mechanism or physical process. However, scale invariance as indicated by a power law and correspondence of the power law slope between those systems has often been the diagnostic to look for a mechanism or physical process that is acting commonly across those systems (i.e., universality in statistical mechanics). Such instances have been found in Heliophysics \cite{Freeman2000}, giving credence to the potential for these techniques to lead to new scientific discovery in space physics. 

Returning to the relationship between distributions indicating underlying generative mechanisms in the system and the notion of self-organization, \cite{Bak_1987} proposed the concept of self-organized criticality (SOC) to explain power law behavior and the correlation that extends over many orders of magnitude in complex dynamical systems. In the original SOC paradigm of \cite{Bak_1987} the system that inspired it was a pile of sand with grains slowly and continuously added and the event was an avalanche, mimicking a dynamical system with spatially complex patterns. Ultimately, SOC implies that the stability of macro-scale systems depends on the self-organization of local events into scale-invariant dynamical patterns characterized by power law probability distribution functions with certain values of the exponents \cite{Bak_1987}. Power law behavior across many astrophysical phenomena made the concept relevant to understanding space physics \cite{Aschwanden2011SelfOrganizedCI}, mirroring a similar utility across scientific domains from geophysics \cite{Smyth2019SelforganizedCI} to economics \cite{Stanley2002SelforganizedCI} to social sciences \cite{Galam2012SociophysicsAP}. A definition of SOC evolved to the broader physics context is provided by \cite{Aschwanden_2014}, ``SOC is a critical state of a nonlinear energy dissipation system that is slowly and continuously driven towards a critical value of a system-wide instability threshold, producing scale-free, fractal-diffusive, and intermittent avalanches with [power law]-like size distributions.'' Important points in this definition are that criticality is broadened to `critical point,' including almost any nonlinear system with a (global) instability threshold and that the system must be self-organizing or self-tuning without an external control parameter. 

Consider a magnetospheric substorm. The critical state is the point at which the substorm occurs and the driver is the slow and continuous accumulation of magnetic flux from the solar wind into the magnetosphere that brings the system back to its critical point after each substorm. Tying together the three components of this dimension of Complexity Heliophysics (SOC, power laws, and scaling theory), the SOC system is often recognized by power law distributions of `event' (exceedences of the critical threshold) occurrences and characteristics (e.g., size and repeat time) and a scaling law captures the relationship in the distribution that exhibits relationship across scale. In our example, the substorm is the event, power law distributions are found from empirical data for occurrence and other characteristics of the substorm, and scaling laws are determined that quantify the change of the properties of the substorm with the scale of the substorm. 

Though we will address SOC, we point readers to more comprehensive examinations of the topic and its history in space physics \cite{Aschwanden2011SelfOrganizedCI, Aschwanden_2014, Watkins_2015, Sharma201625YO, McAteer201525YO, Chang_2003}. Those excellent developments will free up space in this review for novel development of the topic.  

The study of these scale-invariant patterns is generally referred to as scaling theory, a framework that focuses on relationships between scales. The existing body of work around scaling theory offers approaches to connect small and large-scale dynamics or micro and macro states. The identification of these scaling relationships and application of scaling theory in general have been a focus of Complexity Heliophysics research. 

Related to scaling theory is the concept of coarse-graining. Coarse-graining is considering a system at a higher or coarser level at which some of the finer scale behavior has been smoothed over. Newton's laws are a coarse-graining for the physics of motion. At these scales, the laws describe the system sufficiently, though they may break down at finer scales. There are coarse-grained theories that are dynamically and statistically sufficient; aggregations that are as good predictors of their future selves as any more microscopic description is. Tools to explore self-organization and emergence provide details about when these aggregations may exist and when they do not. Such tools include cellular automata (computer simulations that apply a simple mathematical redistribution rule yet, across iterations, produce complex spatio-temporal patterns \cite{Wolfram_2002}), statistical mechanics, network science, and genetic algorithms/evolutionary programming and agent-based simulations \cite{Holland1975AdaptationIN, Holland1992, Schelling1971DynamicMO, Axelrod1997TheCO}. We discuss network science and collective behavior, a generalization of approaches such as agent-based modeling, further below as a separate, but related, dimension of complexity science. 

Coarse-graining is thus incredibly useful because it has led to ways to develop an effective theory, which is a representation that allows one to better predict the system than if the intricacies of a finer scale were considered. For instance, measuring the temperature of a system, a coarse-graining of the aggregate motion of its particles, permits more accurate predictions for a given computational capacity than if each of the individual particles' speeds and directions were measured \cite{Castiglione2008ChaosAC, Flack_2017}. 


\subsection{Information and uncertainty quantification}
Self-organization is most often described in combination with emergence \cite{Wolf2004EmergenceVS}. \textit{Nonlinearity} is a unifying characteristic between them, which \cite{glansdorff1973thermodynamic} mathematically showed has the property of \textit{auto-catalysis} or a positive feedback loop (i.e., small changes, large effect). Feedback loops connect microscale interactions to macroscale behavior, which scaling theory statistically describes. We previously introduced one particular type of scaling, power laws, and one particular mechanism by which they are generated, self-organized criticality \cite{Newman2004PowerLP}. However, the phenomena of self-organization and emergence are more general aspects of system behavior and complexity science requires more general theory to study them. Emergence is a way that order arises from many interacting parts. To analyze order mathematically, the driving principle of the complexity paradigm, one must begin with information and its counterpart entropy. Information quantifies the amount of dependency or connection between a random variable and itself at a different time or with other variables at the same or different times. 

\begin{equation}
    I(E) = -log_2(p(E)),
\end{equation}
\noindent where $E$ is an event and $p(E)$ is the probability of that event. Entropy quantifies the amount of uncertainty involved in the value of a random variable, given by \cite{Shannon1948AMT}:

\begin{equation}
    H(X) \equiv - \sum_{x \in \mathcal{X}} P(x)logP(x),
\end{equation}
\noindent where $X$ is a random variable with probability distribution $P$ over events $x$.

\cite{Hidalgo_2015} differentiates information from entropy, ``In a physical system, information is the opposite of entropy, as it involves uncommon and highly correlated configurations that are difficult to arrive at.'' It is in these uncommon configurations that mathematicians, physicists, and scientists have observed various physical systems. The mathematicians of the early to mid 1900s applied the study of order and disorder to communication systems, creating the field of information theory as the `mathematical treatment of the concepts, parameters and rules governing the transmission of messages through communication systems' \cite{Smelser_2001}. These pioneers (e.g., Claude Shannon, Florence Violet McKenzie, Warren Weaver, Alan Turing, Norbert Wiener, to name only a few) became the first information theorists or cyberneticists \cite{Wiener_1961}. It should be noted that women and minorities are often left out of the history of cybernetics, but played integral roles whose influences are still being discovered, from Ada Lovelace's pioneering work to develop the world's first complex computer program 100 years before the first computer existed \cite{Plant1995TheFL} to Margaret Mead's instrumental contributions to the field of cybernetics and new systems thinking (notably memorialized in the influential Macy Conferences alongside another often unrecognized contributor, Janet Freed Lynch \cite{Hayles1999}), from biologist Ross G. Harrison to the innumerable ways that African culture was studied by anthropologists like Mead and Gregory Bateson and used to scaffold the ideas upon which cybernetics was built \cite{Haraway1976CrystalsFA}. Subsequently, the complexity paradigm was built on ideas of information. The transmission of messages in a communication system  provides an apt analog to the transfer of energy through the physical solar system. Heliophysicists have taken the information theoretic approach to the solar-terrestrial system (e.g., \cite{Wing_2019}). Information theory provides rigorous mathematical formalisms to study the nonlinear relationships and feedbacks that characterize complex systems \cite{Thayer_CEDAR}, especially because they can go beyond linear correlational analyses, capture nonlinear relationships, and establish causalities. Long has the dilemma of correlation vs. causation been an undercurrent in scientific studies and discussions. A central contribution to the conversation was the Granger causality paradigm \cite{Granger1969}, which formulated causation in a predictive manner, asserting that a time series $X$ Granger causes another $Y$ if past values of $X$ provide statistically significant information about future values of $Y$. A core assumption in Granger's work that the cause has unique information about the future values of its effect is grounded in the notion of \textit{directed} information, an information theoretic measure. Many complementary concepts now accompany Granger causality, most of which share a grounding in causality being studied based on the number of bits of information that one process provides about another, or information theory. Information theory is a rich field with many measures that are based on probability distribution functions ($pdf$s) and can therefore capture nonlinear relationships between variables (e.g., transfer entropy). Systems sciences recognize the limitations of correlation and regression-based methods for complex systems, moving toward causal inference frameworks based in ideas and formalisms from information theory. This is notably evident in Earth Systems Science \cite{Runge2019InferringCF}. Space physics, too, has been enmeshed in the correlation vs. causation dilemma and though correlation analyses remain common and in some cases useful, there is a growing body of literature around the application of information theory. We review that literature in Section \ref{information theory section}. Researchers, in space physics as across fields, have found that information theory can describe the structures and signatures of order against the random, entropic background on which they act. Information theory thus provided an entry point into the complexity science paradigm. 


Information theory is inherently probabilistic. The equations deal with random variables and probability distributions. As apparent in the connection to entropy, implicit in information theoretical approaches is a quantification of uncertainty. Indeed the complexity paradigm requires an acknowledgement of uncertainty and uncertainty quantification becomes important within it. Progress toward information theoretic approaches and uncertainty quantification in Heliophysics leads to the advent of risk science and resilience studies as bridges between deterministic physics-based methods and empirical data-driven approaches \cite{Camporeale_2019}. The intersection or reconciliation of these two is a grand challenge for Heliophysics and one for which past work suggests the complexity paradigm may provide new possibilities for progress (see Section \ref{key challenge section}).

\subsection{Networks, network science, and collective behavior}

Information leads to another central dimension of complexity science: networks. 

\begin{figure}[h]
\centering
\includegraphics[width=\textwidth]{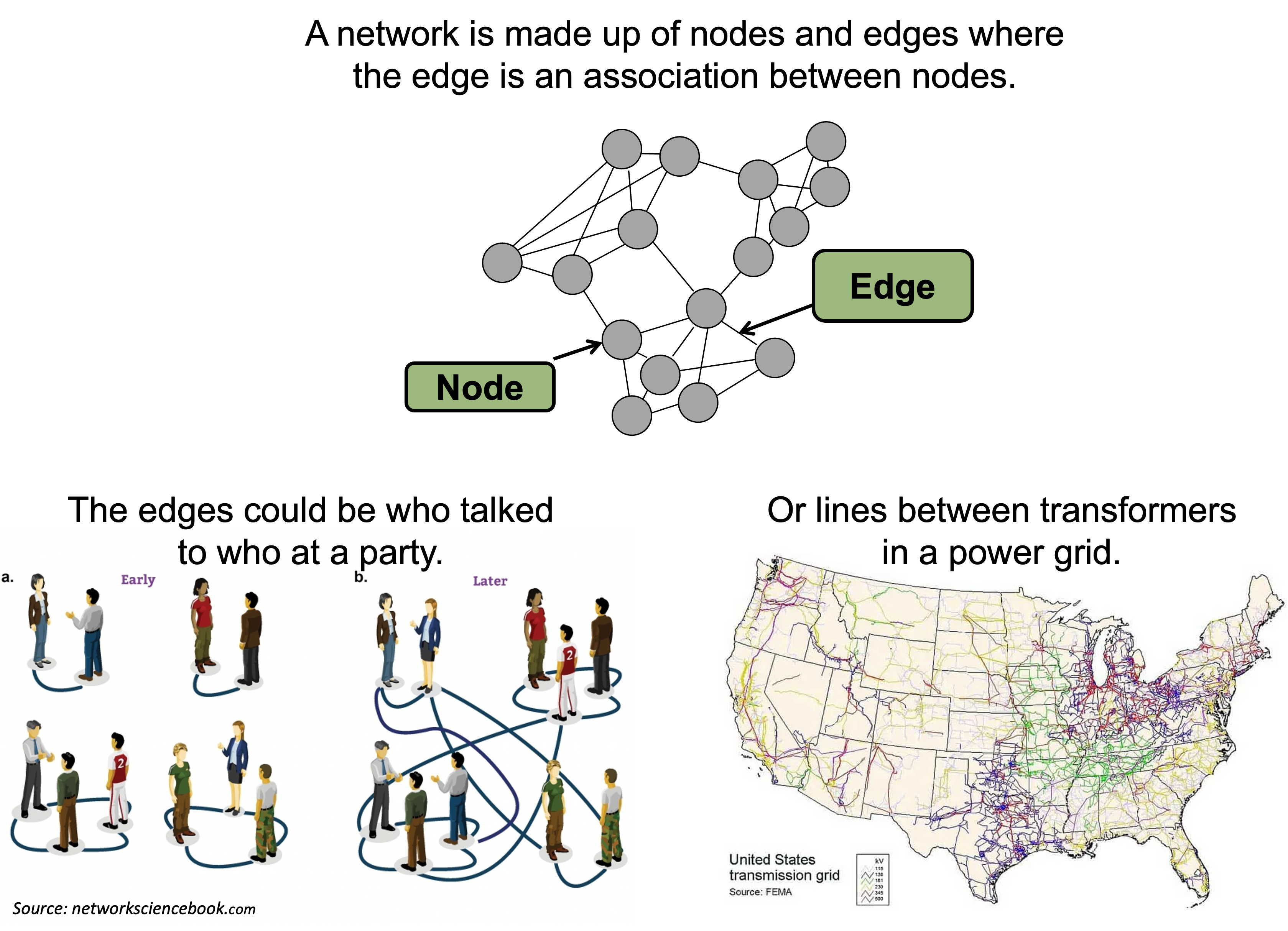}
\caption{A network is made up of nodes and edges where the edge is an association between nodes. Networks are an ubiquitous structure the world, from social ties to brain function, links across the internet to the physical power grid.}
\label{network science visual}
\end{figure}

If the complexity science paradigm is about understanding the emergence of patterns from the interactions of their parts, then networks are its specimens and network science its toolkit. A network is simply a collection of entities, or nodes, and their relationships, or edges (see Figure \ref{network science visual}. For example, in a social network the nodes are people and the edges whether they know or are friends with one another. 

Networks, also called graphs, permit the representation of a system in a way that captures more of the complexity than, say, a rigid spatial grid representation could. As the network structure is remarkably representative of the natural world \cite{Kauffman_1993}, thinking of a system in this way can lead to new and useful insights \cite{Newman_2018}. 

The 21st century has witnessed the advent of new theoretical tools to extract knowledge from networks of many different kinds \cite[and references therein]{Torres_2021}. Graph theory, dating to the 18th century, was used to represent networks and as the scale of these networks grew (e.g., internet-scale data that are inherently networked) new solutions were created for the computational and interpretive challenges. New metrics for understanding the networks at different levels were created, including various measures of centrality (\cite{Newman_2018}, Chapter 7) and community detection \cite{Porter2009CommunitiesIN}. From the mid-1900s into the 2000s, common graph structures were discovered and models developed to generate them such as the random graph and the Erd{\"o}s-R{\'e}nyi model \cite{Erdos1984OnTE}, the small-world graph and the Watts and Strogatz model \cite{Watts1998CollectiveDO}, and the scale-free graph and the Barab{'a}si-Albert model \cite{Barabsi1999EmergenceOS}. Even more recently there have been advances in understanding how networks change over time, network dynamics, and spread across networks \cite{Valente1995NetworkMO, PastorSatorras2001EpidemicDA, Gonzlez2008UnderstandingIH, Dutta2013OnTC}, notable in recent years for their use in epidemiology and the spread of disease. The availability of new tools was grounds for a concomitant dawn of network science in Heliophysics. As an objective of this review is to focus on areas of the complexity paradigm that have received less development, special emphasis is placed on networks and network science in Heliophysics. 

While there is a productive and generative body of network science research in the solar-terrestrial system (e.g., \cite{Dods_2015, McGranaghan_2017c}, the related topic of collective behavior has received relatively little attention. Collective behavior is a term used to describe approaches to understanding emergent phenomenon particularly through representing the system as a network. From the function of parts (nodes) together with their interactions (edges) collective, or community, behavior emerges \cite{Radicchi2003DefiningAI, Porter2009CommunitiesIN, Fortunato2009CommunityDI}. Collective behavior has progressed from a description of phenomena to a framework for understanding the mechanisms by which collective action emerges \cite{BakColeman_2021}. The research that composes this review suggests that areas ripe for future analyses may be identified in the language and discipline of collective behavior, especially in its treatment of networks. The areas include both the study of the physical solar-terrestrial system and the social networks or communities of Heliophysics that study it. Thus, the implications are not only for fundamental science but also for how we do science. 


Ultimately, one cannot do modern science without information theory and now networks. These form pillars of complexity science that we will use to explore the history of Complexity Heliophysics. 

\subsection{Risk science and resilience}

New frameworks are required to handle uncertainty and embody the complexity paradigm. This is acutely true in Heliophysics, which has an existential counterpart in the societal impacts of the solar-terrestrial connection known as space weather \cite{Schrijver_2015}. In order to translate the science of Heliophysics into actionable knowledge for space weather, the complexity paradigm dictates a risk science approach and an emphasis on resilience \cite{Scheffer_2001, Sobel2014SciencebasedRA, DeBruijn_2017, Angeler_2018}. Risk science is the set of approaches and research areas that bridge from the fundamental Heliophysics understanding to concrete and quantitative impacts that inform decision-making. In practice, it is the layer between Heliophysics knowledge and applications that encompasses the dimensions listed above (e.g., systems science, information and uncertainty, network science) and their downstream impacts, permitting a quantification of risk, defined as consequence times likelihood. Once we quantify risks, we can then consider the resilience of our systems, whether that system is a crewed spacecraft, the power grid, or the magnetosphere itself. Resilience emphasizes discovering mechanisms of a given system that maintain functionality under changing or uncertain environments \cite{flack_mitchell_2021}. Together, risk science and resilience outline a framework, what we will call a risk and resiliency framework, that defines a future research agenda for Heliophysics and space physics capable of responding to the grand challenge articulated above: namely navigating the intersection between basic science (e.g., complexity science) and applied science (e.g., predictive models like those from machine learning). In a risk and resilience framework a system is treated as complex and can be defined by whether or not it can accommodate changes and reorganize itself while maintaining the crucial attributes that give the system its unique characteristics \cite{Scheffer_2001}. Risk and resilience offer ways that data-driven information can be incorporated with complex systems understanding and decisions made amidst uncertainty \cite{DeBruijn_2017}. How do we define risk and natural hazards (the physical phenomena that create risk) and how might understanding Heliophysics and space weather in these terms illuminate a path toward a society resilient to their vicissitudes? This review pulls together the existing literature and knowledge to discuss this question. We draw from important sister disciplines, namely Earth Science, terrestrial weather, natural hazards, risk studies \cite{Burgess_Routledge_2016}, and disaster risk reduction \cite{Wisner_Routledge_2011}, to understand risk science \cite{Burgess_Routledge_2016}. Because the idea of Heliophysics and space weather as risk sciences is relatively novel in the published research, our treatment of the topic must be somewhat speculative and subjective. However, we ground those perspectives in the support that does exist both within and tangential to space physics. 

Although the volume of work in that treats the solar-terrestrial connection as a complex system and casts the problem as one of risk quantification and resilience-building \cite{Valdivia2005TheMA, Jonas2016, green2016building, Eastwood2017, Eastwood2018, Oughton2019, McGranaghan_2020, Mcgranaghan2022ConvergingTS} is small, it is growing, and the changes we observe in the literature motivate the need to articulate a framework for risk science and resilience. Therefore, we conclude this review with a look at just a few of the works that have considered Heliophysics and space weather as a risk science and examine the impact they have on societal infrastructure and life on Earth (resilience), providing scaffolding for creating a framework for risk science and resilience within which actions can be determined and decisions made amidst, sometimes extreme, uncertainty. The attempt to review the existing works and to abstract a framework for risk and resilience offers guidance on the topic we have identified is central to Heliophysics and space physics in the 21st century: bridging research and operations; determining the relationship between basic and applied research. The review culminates in a discussion of this grand challenge in Section \ref{key challenge section}. 



\subsection{Approach and roadmap for this review}


With the paradigm shift comes new capacities to understand the system. Heliophysics has embraced some of the dimensions, however, a number of them remain relatively unexplored. Among the nearly 400 articles manually reviewed and cited in the bibliography, and corroborated by the order of magnitude more articles included in a corpus automatically generated from natural language processing (NLP), uneven coverage of the five dimensions introduced above was discovered. `Emergence, self-organization, and scaling theory' (albeit with less emphasis on emergence) enjoyed the best coverage in Heliophysics, followed by the more general category of `Systems science and crossing scales.' `Information and Uncertainty quantification' is slightly more prevalent than `Networks, network science, and collective behavior.' `Risk science and resilience' is significantly the least well-represented. The coverage in this review inevitably reflects this representation, with the lack of coverage contributing to what is identified as gaps potentially deserving of special focus. Where other review articles have already covered a topic, we choose not to reify that material but rather to point to the appropriate review. 

The purpose of this review is to apprehend the development of the paradigm and to establish the historical trajectory for the tools of complexity science within Heliophysics. From this history, important trends emerge that will guide researchers (early and senior) and those with the responsibility to direct research resources toward directions that may be more capable of responding to the problemscape of Heliophysics in the 21st century. The structure of the review matches these goals: Sections \ref{setting the stage section}-\ref{SOC section} review the work that sets the stage for Complexity Heliophysics, taking 1996 as a pivotal year; Section \ref{1996 onward section} covers transitional years (1996-2010) when the foundation for the paradigm was being established in earnest; and Section \ref{trends section} broadens the scope of relevant literature to identify topics and trends for which Heliophysics exploration is relatively nascent or that we have not yet explored. This history is chronicled to substantiate the grand challenge confronting Heliophysics and space physics, which we make explicit in Section \ref{key challenge section}. Finally, we attempt to synthesize the history and the grand challenge into a coherent response and recommendation for the future for Heliophysics, space physics, and space weather in Section \ref{space weather as risk science section}. Sections \ref{key challenge section} and \ref{space weather as risk science section} attempt to articulate the challenges and future directions for Heliophysics. Such discussion about the interpretation of the current state of a field and about what lies ahead will be necessarily subjective. Therefore, those sections have subjective elements. However, we have grounded those opinions in literature and evidence to the extent possible.

First, a word about the philosophy of history that we adopt. The approach is one that did not assume the important dimensions of complexity \textit{a priori}, but rather reviewed the literature liberally for those dimensions to emerge from the existing work. Second, we acknowledge the importance of positionality, or where one is located in relation to their various social identities, in constructing any history. It is unavoidable to have a bias toward Heliophysics, however to address a paradigm as encompassing as complexity, one needs to step beyond the unidisciplinary perspective and bring in works traditionally considered beyond our field. Indeed this is one aspect that distinguishes this review from others on similar or related topics (e.g., \cite{Aschwanden_2014, Balasis2023ComplexSM}). 

The effect is a relatively uneven coverage of the dimensions of complexity and a reference list that is much wider than Heliophysics. We believe this paints an accurate picture of Complexity Heliophysics and provides the breadth that justifies our extrapolations in the final sections of the review.

The objectives of this work are: 
\begin{enumerate}
    \item To define the paradigm of Complexity Heliophysics in the context of the seminal works that compose it, seen in a new converging light; 
    \item To detail the network of complexity studies in Heliophysics, setting the stage for the needed research in the 21st century. The corpus of cited text will be uncommon to most Heliophysics research papers (e.g., pulling liberally from areas outside of the traditional scope of Heliophyscis research articles) and a unique resource in and of itself (e.g., serving as a hub for the network of research that all readers can use in their exploration of Complexity Heliophysics); and 
    \item To lay a foundation for how complexity science may help address outstanding questions in Heliophysics science, including the intersection of fundamental and applied research and the use of artificial intelligence and machine learning (AI/ML). 
\end{enumerate}

A function of this review is to enable further investigation of the research that was compiled and explored and the artifacts from it. Thus, we provide those resources in a navigable way that lends itself to further exploration. Artifacts that we provide include: 
\begin{enumerate}
    \item This review article; 
    \item A glossary of terms (the bedrock of information search, integration, and automated analyses \cite{pomerantz2015metadata}) that define complexity Heliophysics; and
    \item A new corpus of Complexity Heliophysics compiled using NLP where the included papers have been filtered based on Heliophysics or Heliophysics-adjacent journals and by matching terms in the papers to those in the new glossary. 
\end{enumerate}

The automated corpus contains roughly 3000 articles, which augment and grow by an order of magnitude the 400 that were manually reviewed. Given the proliferation of scientific publications, outstripping the capacity of any individual researcher or even any small group, NLP is required for comprehensive treatment or knowledge of any subject. 

All resources are provided in a Github repository for this publication\footnote{ \url{https://github.com/rmcgranaghan/ComplexityHelio_LivingReviews}}. Something we would like to see come from this work is a more collaborative examination of the corpus of articles (the collection of documents manually compiled in the references of this review along with those automatically generated through NLP methods (see Appendix \ref{corpus appendix section})). One way to accomplish this is to create a database of the papers in the corpus (our references list plus the articles compiled from NLP analysis of Heliophysics literature) on a platform that allows the community to add margin notes, annotate, and hold conversations around the papers. This would allow the insights researchers generated when reading published works to build on one another and for the conversations around those works to evolve. Inspiration can be found in the Fermat's Library\footnote{\url{https://fermatslibrary.com/}}. Altogether, the resources provided with this review constitute a \textit{library} of Complexity Heliophysics in the ethos of libraries as cultural technologies\footnote{\url{https://www.societylibrary.org/build-multimedia-libraries}} and sacred places \cite{Reynolds2019}. 

Previous works have capably reviewed complexity science up to $\sim$1996 \cite{Klimas1996}. We do not attempt to restate those works here, and instead take the Klimas review as our starting point. This enables us to give more attention to the voluminous body of work that has been created since. Parts of the discussion will, of course, reach to works prior to 1996.

We augment the literature review and synthesis with a more perspective-based portion (Sections \ref{trends section} and \ref{key challenge section}) where we attempt to describe trends perceived through creating this review and to identify key issues confronting Complexity Heliophysics. 

We acknowledge that some readers will not incorrectly read parts of this contribution as opinion; however, we have tried to support our views by quotes and references compiled across more-than-Heliophysics publication and knowledge wherever possible. 

Finally, like any review or synthesis, this is incomplete. It is \textit{not} the goal of this work to review exhaustively every paper that is related to complexity science in Heliophysics, but rather to present a useful selection, deliberately chosen to reveal the story of the complexity science paradigm in Heliophysics, and to illuminate generative areas for future thinking and research. Our purpose is to create a strong foundation to support future research, a place to build on for the inspired reader and community to respond to its inadequacies. The first step in scientific change is getting people into a space where they can acknowledge that there are alternatives to the ways that it does science at present and opening a discourse about what those alternatives are. Then you begin to build social movements and actions around those alternatives in order to achieve them. That is the hope for this review. 

\subsubsection{The use of Natural Language Processing (NLP) in this review} \label{NLP use in this review}

Given the volume and breadth of the information available to a Heliophysicist, a problem exacerbated by the pace of scientific research and publication, traditional approaches to search and discovery as well as ingestion of new materials (e.g., largely manual) will need to be augmented by new tools. Perhaps most pressing is the need to create and adopt mature natural language processing (NLP) tools to help one search through, organize, and summarize the vast literature. NLP refers to interactions between computers and human language and is often used to refer to the programming of computers to process, analyze, and respond to large amounts of natural language data. We have employed these techniques in this review. As far as we are aware, the use of NLP to augment the review is a novel element and further distinguishes it from related contemporary works. 

As a supplementary piece that intends to make this a living contribution to the state of knowledge, we provide a corpus that was generated by natural language processing methods of 33 journals (those within the NASA Astrophysics Data System (ADS) deemed most important to Heliophysics) that any reader can use freely to determine other trends not addressed here. Across the 33 journals, we compiled a corpus of nearly 125 thousand articles with their authors and abstracts. After matching words in the title and abstract with terms in our complexity glossary, we arrived at a Complexity Heliophysics corpus of roughly three thousand documents, two orders of magnitude larger than a typical journal article bibliography. The details of how the corpus was generated can be found in Appendix \ref{corpus appendix section}. It augments the bibliography, a corpus in itself, of works directly cited in this review. The author acknowledges a wealth of knowledge much greater than the papers directly cited in this review contributed to the writing. Resources related to the automated corpus generation and results are provided in an accompanying Github repository for this review\footnote{ \url{https://github.com/rmcgranaghan/ComplexityHelio_LivingReviews}}. 

We suggest that such automated corpora could perhaps even become standard for future reviews. The one provided herein should be considered a resource that complements the extensive references cited in the body of this review and contains high potential for discovering trends and knowledge about Complexity Heliophysics. It is important to note that the manual and automated corpora are not disjoint nor is the manual corpus strictly a subset of the automated corpus. Many references are shared across them, lending validation to the process of generating the automated set, but there are many references in the manual set that are not included in the automated one. This points to the flexibility of the scientist-driven discovery process, pulling in relevant references and material that might be more distant or irregularly connected to the research at hand than the necessarily more rigid automated process. This review, in particular, read widely in gathering material, many connections of which an automated approach would likely not have captured. The point is there must be an intersection of manual and automated gathering of resources, the manual approach benefiting from flexibility and capacity to range widely and be discerning and the automated approach benefiting from the volume of resources it can examine. 


The augmented way that we have approached this living review illuminates a trend in Heliophysics research and one point of this manuscript is to demonstrate the hybrid manual-automated approach. We discuss this trend in research below in Section \ref{NLP subsection}. 

Readers interested in the history of complexity science in Heliophysics will enjoy beginning with Section \ref{setting the stage section}. Those who might be interested in a more quantitative analysis might begin with the importance of NLP (Section \ref{NLP subsection}) and our provided corpus, where they can emerge their own conclusions and trends. Those with an interest in solving the key open questions and challenges will find most value in Sections \ref{trends section} and \ref{key challenge section}.

\section{Key definitions} \label{definitions section}

Any conversation or published work situated sufficiently in spaces between established fields requires a period of adjusting or coming to a more shared language. We provide a set of definitions that are important to the review that follows as a way to create common language. 

\begin{itemize}
    \item \textbf{artificial intelligence} the theory and development of computer programs able to perform tasks normally thought to require human intelligence \cite{McCarthy2006APF}.   
    \item \textbf{coarse-graining} considering a system at a higher or coarser level at which some of the finer scale behavior has been smoothed over
    \item \textbf{collective intelligence} the study of collective behavior, that is adaptive, wise, or clever structures and behaviors by groups, in physical, biological, social, and many engineered systems \cite{Flack2022EditorialTT}
    \item \textbf{corpus} a collection of documents
    \item \textbf{disaster} the state of a natural phenomenon resulting in major consequences for society (cf., `natural hazard' and `risk')
    \item \textbf{emergence} the term used to describe phenomenon that are `more than the sum of their parts'
    \item \textbf{feedback} a loop of interactions across a system; in the computational fields, feedbacks are defined by outputs of a process being put back into an input of the same process
    \item \textbf{graph/network} a collection of entities (nodes or vertices) and their relationships (edges)
    \item \textbf{information} Information quantifies the amount of dependency or connection between a random variable and itself at a different time or with other variables at the same or different times. Its counterpart, entropy, quantifies the amount of micro-states involved in the value of a random variable 
    \item \textbf{machine learning} the study of computer algorithms that allow computer programs to automatically improve through experience \cite{mitchell1997machine}. ML is a sub-field of artificial intelligence. 
    \item \textbf{natural hazard} potentially consequential phenomenon of nature outside of human control (cf., `disaster' and `risk')
    \item \textbf{natural language processing (NLP)} programming of computers to process, analyze, and respond to large amounts of natural language data
    \item \textbf{named entity recognition (NER)} subtask of NLP that seeks to locate and classify named entities mentioned in unstructured text
    \item \textbf{resilience} the property of a system to accommodate changes and reorganize itself while maintaining the crucial attributes that give the system its unique characteristics \cite{Scheffer_2001}
    \item \textbf{risk} the combination of the probability of occurrence of a natural phenomenon and the magnitude of the consequences (likelihood multiplied by consequence; cf., `disaster' and `natural hazard')
    \item \textbf{sandpile cellular automata model} a model based on adding sand to a pile and observing and quantifying the results, specifically the avalanches, that occur due to simple rules defining the evolution \cite{Bak_1987}. It is a specific example of an automaton that is a regular grid of cells, each with a finite number of possible states, and a fixed rule governing its state at the next time step given its current state and the states of neighboring cells \cite{Wolfram_2002}.
    \item \textbf{scale-free/scale invariant} property of a system in which there is similarity across scales, the same structure is observed regardless of the scale at which the system is observed
    \item \textbf{self-organized criticality (SOC)} \cite{Bak_1987} proposed the concept of self-organized criticality (SOC) to explain the power law, or scale-invariant, correlation extending over many decades in complex dynamical systems. It is the observation that, as articulated by \cite{Consolini_2002}, ``...some slowly-driven, dissipative, extended, dynamical systems can naturally exhibit a spontaneous organization towards a sort of dynamical critical point. The critical state does not depend [on] the initial conditions, does not require a fine tuning, and behaves as an attractor for the dynamics. All these systems are characterized by an intermittent dynamics, 1/f noise, and by a threshold dynamics, i.e. a local stepwise instability that occurs when the local field exceeds some critical value.'' SOC dynamics can be viewed as a sort of dynamical transition between metastable configurations near a critical point and is an explanation of the ubiquity of 1/f noise
    \item \textbf{system} A group of interacting or interrelated elements that act according to a set of rules to form a unified whole \cite{mw:systems}. A systems perspective is at the center of complexity science. 
\end{itemize}

\section{Setting the stage for Complexity Heliophysics: from 1996} \label{setting the stage section}
The world of Complexity Heliophysics touches many areas. We will focus this review on areas that are important and have received less development, including the critical linkages between the papers. For instance, we will address the topic of self-organized criticality, but defer a comprehensive examination of the topic and its history in space physics to \cite{Aschwanden_2014} and \cite{Chang_2003}, allowing space here for further development of complexity in Heliophysics.  

As mentioned in the introduction, we begin with the review article by \cite{Klimas1996}, which could be interpreted as a turning point. They provide the relevant framing, ``Earth's magnetosphere responds to the time-varying solar wind in an organized and repeatable fashion. Evidence has accumulated indicating that this organized evolution is a manifestation of low-dimensional magnetospheric dynamics. It appears that over the largest spatial scales and over substorm timescales and beyond, a relatively small number of magnetospheric state variables dominate the evolution. The identities of these variables are not known at present, and very little of the dynamical system that governs their evolution is understood. Determining these variables and understanding the related dynamics are the primary goals of this research. If these goals are reached, then a spatio-temporal framework will result within which reside the complex phenomena that are collectively called geomagnetic activity.''


Klimas et al., details the transition over the preceding several decades from an era of linear correlative studies to the beginning of a new era of nonlinear dynamical studies. Why? Because the linear filters of \cite{Bargatze_1985} across varying solar wind activity demonstrated clear nonlinear behavior through distinct peaks in the time lags of activity. ``Generally, there is a peak in the filters at lag time 20-30 min showing that there is always a response in electrojet activity to solar wind activity 20-30 min earlier; Bargatze et al. attributed this peak to the directly driven magnetospheric response \cite{Perreault_1978, Akasofu_1979, Akasofu_1980}. There is a second peak at lag time one hour that is most evident for the moderate activity filters; this peak was attributed to the unloading magnetospheric response \cite{McPherron_1970, McPherron_1973, Hones_1979, Baker_1979, Baker1981GlobalPO, Baker1981AHT}''. The Bargatze work made clear the direct and indirect (directly-driven and unloading) modes of the magnetosphere \cite{Kamide_1996}. Klimas et al., begin from this background and the implication that nonlinear explication of the solar wind-magnetosphere coupling requires nonlinear treatment. A key contribution of their review is a convergence of work to address the directly-driven and unloading modes of the solar wind-magnetosphere system and the corresponding evolution of the methods from linear correlative studies to nonlinear dynamical studies. It is an excellent starting point to understand the Complexity Heliophysics paradigm. 


The central question in the review was whether it can it be shown that the magnetospheric dynamics, represented by systems of differential equations, are effectively represented by a low-dimensional dynamical system and, if so, what is the nature of that dynamical system? The response to this question is traced through three approaches, interleaved in time: autonomous time series, analog (input-output) models, and computationally mature input-output models. The self-described summary of the work is a review of approaches to ``find a low-dimensional analogue model of the magnetospheric dynamics derived directly from data and interpreted in terms of magnetospheric physics.'' It laid out the existing state-of-the-practice toward what was a major goal of the magnetospheric community since early in the formation of Heliophysics as a discipline: to find a low-dimensional analogue model of the magnetospheric dynamics.

The first observation from the collection of articles in the Klimas review is the prevalence of using the auroral electroject index (AE and the auroral upper and auroral lower indices, AU and AL, that constitute it) \cite{Davis_1966} to be the measurable variable to reconstruct the magnetosphere. In fact, the standard or benchmark dataset from most of the work addressed is a set of 34 events for which solar wind data are collated with the AL index \cite{Bargatze_1985}. Appendix \ref{datasets appendix section} contains a list of important datasets, and their original appearance in the literature, that appear across this review to aid readers who wish to compile datasets and explore data-driven research across the datasets that have factored importantly in the Complexity Heliophysics paradigm.


\subsection{Autonomous Time Series} 
Autonomous approaches make the assumption that the evolution of the system representation (the state vector) is solely a function of the internal dynamics and independent of external factors. They are concerned with the question of whether or not the evolution of the magnetosphere is organized such that a few variables alone can describe its evolution. Studies attempt to determine whether the dissipative magnetosphere dynamically evolves into a low-dimensional `attractor' state and then attempts to describe the attractor to develop a model of the organized physical state. A measure of the attractor dynamics is the correlation dimension. The correlation dimension considers the set of state variables, or the phase space, of the system. To calculate the correlation dimension an arbitrary measurement from those variables is chosen and compared to other measurements in the neighborhood of the chosen one, defined by a sphere of radius $r$. The number of measured state vectors within the neighborhood are counted ($C_r$) and the variation of this number as $r\rightarrow0$ is determined. The importance of the correlation dimension, $D_{cr}$, is that if $C(r) \propto r^{D_{cr}}$ as $r\rightarrow0$, then $D_{cr}$ is an estimate of the attractor dimension near the chosen point. For an entire system, an average correlation dimension can be found by averaging the correlation dimension from each chosen measurement over all of the measured points. For a simple closed loop phase space, purely periodic system, in three dimensions (three state variables), $D_{cr}=1$. In a chaotic system, the attractor phase space is made up of many folded and closely packed layers and $D_{cr}$ falls between 2 and 3. A non-integer correlation dimension is a feature of `strange' attractors \cite{Ruelle_1980}. The dimension of the attractor is an important estimate of the physical system that gives rise to it. In the periodic case, the dimension is one, indicating that any two of the variables can be written in terms of the third and that the system is one-dimensional. Between two and three, the correlation dimension indicates the minimum phase space dimension that supports the attractor is three, and that is the physical dimension of the system. 

Across the literature a rather wide range of correlation dimensions between 2.4-4 have been found: from 3.6 in the first application of the technique to magnetospheric dynamics \cite{Vassiliadis1990LowdimensionalCI} to a set of papers that found low correlation dimensions \cite{Roberts1991IndicationsOL, shan1991chaotic, shan1991embedding, Roberts1991IsTA, Pavlos1992EvidenceFS, Sharma1993ReconstructionOL}. There were numerous critiques of the correlation dimension technique to determine the magnetospheric dimension, including the requirement of large databases of measurements and lack of knowledge of what volume would be sufficient, spurious periodicities in the databases as have been identified in the \cite{Bargatze_1985} data, potential presence of background or superimposed trends in the data across activity levels and unrelated to the magnetospheric dynamics, and sensitivity to delay times chosen for the analysis. Among the limitations that exist across attempts to understand the complex magnetosphere and geospace environment is the inability to measure all of the state variables or even to define them \textit{a priori}. The solution has been to substitute for the unobservable variables functions of measurable ones to constitute the state vector. Among the most readily available has been the AE/AL/AU indices and the best available datasets have been those that align the solar wind drivers with the AE indices' responses. Limitations in the use of these reconstructed variables permeate the history of complexity Heliophysics studies of the magnetosphere. 

A far-reaching critique of the approaches reviewed was the fact that the AL index time series is a colored noise output of a high-dimensional stochastic process, which would result in a misleading low correlation dimension when those data are used to proxy the complex magnetosphere \cite{shan1991chaotic, Takalo1993CorrelationDA, Takalo1994PropertiesOA}. Colored noise is stochastic noise that, like a chaotic time signal, exhibits a power law spectrum $f^{\alpha}$ \cite{Klimas1996}. In the case of colored noise, the correlation dimension is actually a measure of the fractal dimension of the system and is unrelated to the existence of an attractor. Numerous studies examined the likelihood that the AE/AL index time series are generated by a stochastic signal, with varying conclusions. Though useful methods arose from these investigations (e.g., self-affinity \cite{Osborne1989FiniteCD}, autocorrelation analyses \cite{Theiler1986SpuriousDF}, singular spectrum analyses \cite{Sharma1993ReconstructionOL}, the use of surrogate data \cite{Theiler1992TestingFN}), no consensus was established.

The conclusion from the autonomous method studies was rather that no conclusive statement could be made about the low dimensionality of the magnetosphere--contradictory results pervade the literature. The lack of resolution of the central question of the variables that govern the magnetosphere and its evolution led to the consideration of magnetosphere by nonlinear input-output methods. From \cite{Takalo1994PropertiesOA}:

\begin{quote} 
[The] magnetosphere is not an autonomous system but is continuously driven by the stochastic solar wind. It is therefore possible that, even if the magnetosphere were a low-dimensional chaotic system, we might not find it by studying just one time series. This is because the magnetosphere may not have time enough to converge to the possible attractor for times long enough to produce data with a detectable number of close returns of the trajectory. For this reason it has been suggested that the  magnetosphere should be described as a nonlinear input-output system [Prichard and Price, 1992; Price and Prichard, 1993] \cite{Prichard1992SpuriousDE, Price1993TheNR}.
\end{quote}

\subsection{Input-Output Models: Analogue Models} 
Autonomous methods assume that the system evolves solely due to internal dynamics. Loading rates or external forcing may be present, but it is taken to be a constant and a parameter of the system rather than a dynamical variable. In the autonomous methods there is a common approach to use a series expansion of a function to represent dynamics at a certain point in space and time and to discard the higher-order terms. The shortcomings of the autonomous methods raised the question of the dynamics that reside in these higher-order terms. Subsequent to the controversies and conflicting conclusions of many autonomous system approaches and studies, input-output approaches began to play a central role in the study of the magnetospheric dynamics. Input-output (I-O) approaches use both the input to the system and its output response to determine the coupling characteristics. The Klimas et al. review divides models into two categories: 1) analogue and 2) more recent nonlinear and computationally intensive approaches. Analogue modeling makes assumptions about the magnetospheric coupling characteristics (determining them from `preconceived notions of the processes that constitute geomagnetic activity') whereas the latter I-O methods reviewed determine them directly from data. The output of the latter I-O methods is a so-called phenomenological dynamical system, one that has been deduced from time series analyses of the input and output data rather than through analogue modeling. The two approaches reveal a debate central to science and that foreshadows the key challenge discussion that is introduced at the conclusion of this review (see Section \ref{key challenge section}) about the relationship between theory- and data-driven theories of discovery. 

Analogue approaches to magnetospheric dynamics assume that the magnetosphere is a coherently driven system that evolves in an manner that can be modeled by a low dimensional dynamical system. The analogue models reviewed by Klimas et al., share the assumption that the magnetospheric dynamics are low dimensional. 


To constitute the input and output time series, the analogue models shared the approach of using one of the best available observables thought to be related to the magnetospheric dynamics: the electrojet indices \cite{Davis1966AuroralEA}. The AU index is a measure of the eastward electrojet driven by reconnection dynamics in the dayside magnetosphere. The AL index, alternatively, is a measure of the westward electrojet that is connected to high-latitude magnetopause reconnection in the magnetotail. In using the electrojets as measures of the response to solar wind energy input, the assumption is that the currents flow as a result of appropriate conductances and electric fields mapped from the magnetosphere to the ionosphere through Alfv\`en waves along intermediary field lines. 

\begin{figure}[h]
\centering
\includegraphics[width=\textwidth]{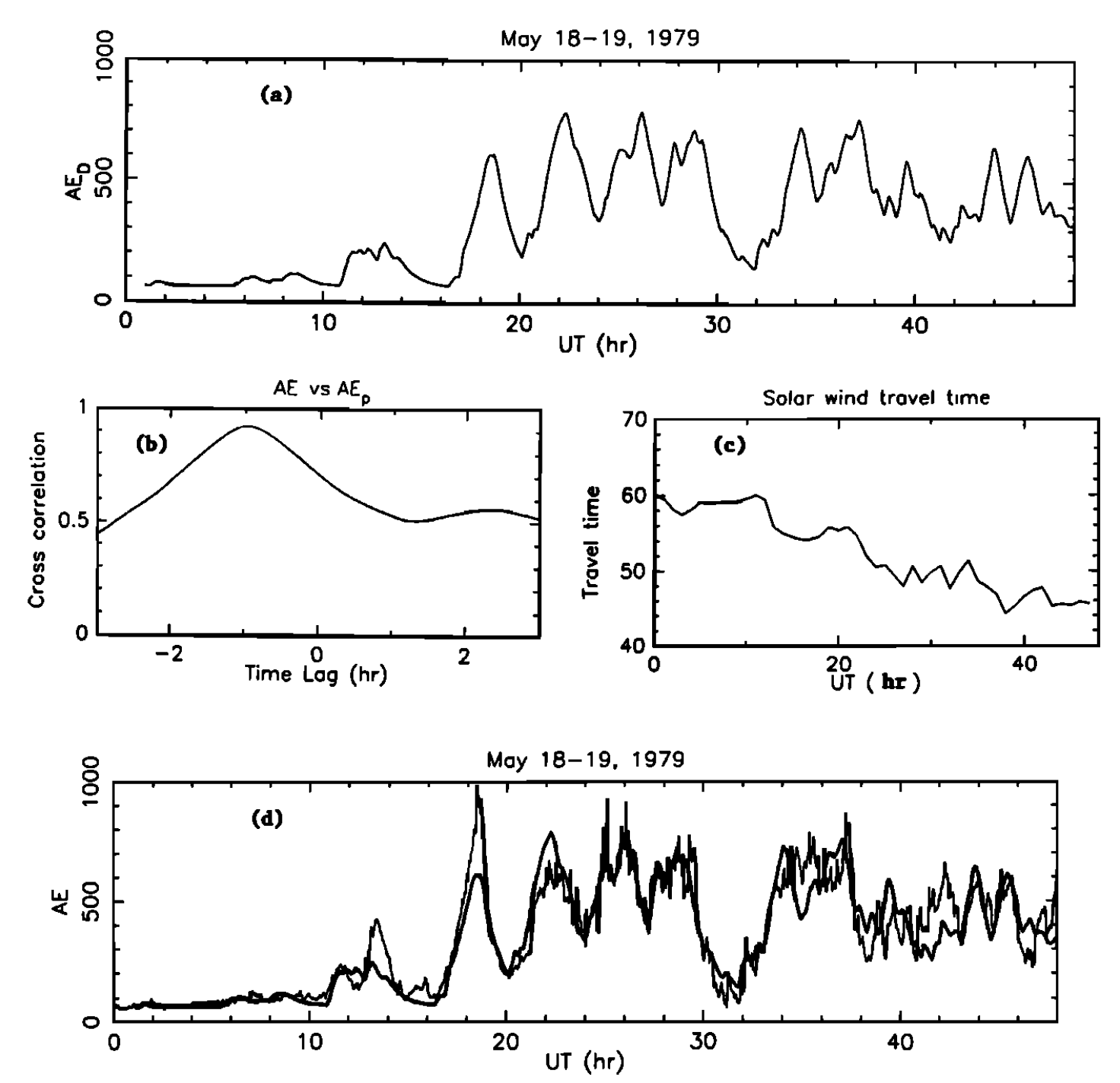}
\caption{Figure 7 reproduced with permissions from \cite{Goertz1993PredictionOG}. (a) The variations of AE as predicted from the solar wind for time delay of 55 minutes. (b) The cross correlation between the predicted variation of AE and the Gaussian smoothed observed variation of AE as a function of delay time. (c) The solar wind travel time from ISEE 3 to the subsolar magnetopause (at $x=10$ $R_E$) as a function of time. (d) The superposition of the (unsmoothed) variation of AE (light line) and the predicted variation of AE (heavy line) calculated with a variable time delay (detailed in the original publication).}
\label{Goertz1993_Fig7}
\end{figure}

\cite{Goertz1993PredictionOG} developed a model of only the directly-driven magnetosphere (responses of the magnetosphere as a result of direct energy input from the solar wind rather than unloading responses that occur at longer time lags due to magnetotail activity \cite{Akasofu_1979}) using a system of three nonlinear ordinary first-order differential equations with six independent constant parameters:

\begin{equation}
    \tau_{AD} \frac{d AU_d(t)}{dt} + AU_d(t) = A_1 E_{mpt_{max}}(t),
\end{equation}

where $\tau_{AD}$ and $A_1$ are constant parameters, $AU_d(t)$ is the eastward electrojet for the directly-driven response ($d$), and $E_{mpt_{max}}(t)$ is the maximum reconnection electric field at the magnetopause. A similar form is assumed for the westward electrojet and relationship to the electric field in the central plasma sheet, with appropriate adjustments in the parameters. ISEE 3 solar wind data were used to compute the electric field at the magnetopause, the input to the model, and measured AE (composed of the AU and AL indices) was compared with predicted AE over May 18-19, 1979. A cross correlation of 0.9 was achieved (see Figure 7 from the original publication, the central result of the paper, reproduced here in Figure \ref{Goertz1993_Fig7} with permissions).

Their model is strictly of the directly-driven response to the solar wind, and the agreement indicates little dependence on the unloading response \cite{McPherron_1970}. There were periods of marked disagreement, attributed in large part to differences in the AL index, though these were isolated in time as a result of non-adiabatic phenomena so that the overall correlation remained high. Absent a term for these non-adiabatic dynamics, the model has been shown to perform poorly on different data intervals as pointed out by \cite{McPherron1993CommentO}. 

\cite{Klimas1992AND, Klimas1994SubstormRD} produced a model that included analogues to both the directly-driven and unloading magnetospheric responses to the solar wind. Their model is a Faraday Loop for the time-dependent magnetotail convection with a superposed loading-unloading cycle. The model is built on the concept that if the loading rate on the dayside magnetosphere is above some level, then the unloading of the accreted energy through Earthward convection is insufficient to balance it and a substorm begins to grow. If loading continues, a critical point is surpassed whereafter explosive unloading follows. The resultant three-dimensional Faraday Loop Model (FLM) consists of dynamic variables for the cross-tail electric field, the flux content of the lobe, and a quantity dependent on the size and orientation of the tail. Their observable for comparison was again the electrojet indices, and they mapped the cross-tail electric field from their model to the westward electrojet in a simple way. After this study, it was implicit in the magnetosphere community that any dynamical model must account for both modes: directly-driven and unloading. Kilmas et al., cite the importance of the FLM as being a combination of numerous important statistical properties of geomagnetic activity: isolated substorms, steady loading, and time-dependent loading.

A number of studies have used the FLM model. \cite{Klimas1994SubstormRD} showed that under steady loading the FLM model evolves to a periodic attractor, with a period of substorm recurrence of roughly one hour and that the period was independent of driver strength. \cite{Farrugia1993TheEM} examined time scales of substorm occurrence for the passage of slowly varying magnetic clouds to understand substorm recurrence under various steady loading rates. They found that substorm recurrence ranged between 25 and 150 minutes, but with an average rate of approximately one substorm every 55 minutes (unloading recurrence rate). For this range of steady driving strengths the FLM model varies little from the 55 minute average unloading recurrence rate. \cite{Klimas1994SubstormRD} use of the FLM model seems to explain the \cite{Farrugia1993TheEM} observations: 1) the nearly invariant unloading recurrence rate under steady driving and 2) the distribution about one hour of recurrence rates during extended periods of loading. The results have been extended to time-dependent driving, where a Poisson distribution of recurrence rates around the most probable value of one hour was found. Note that the Poisson distribution expresses the probability of a given number of events occurring in a fixed interval of time given a known constant mean rate and independently of the time since the last event. The latter assumption has received focus in substorm research in the intervening decades under various names, among them substorm recurrence rates \cite{Borovsky_2017}, waiting times \cite{Freeman2004AMS}, and intersubstorm times \cite{Liou_2018}. Indeed waiting time statistics is used to identify Poisson random processes, self-organized criticality, intermittent turbulence, finite system size effects, or clusterization \cite{Aschwanden2010RECONCILIATIONOW, Chapman_1998, Chapman1999SignaturesOD}. 

As a side note, \cite{Farrugia1993TheEM} showed that both the epsilon parameter \cite{Perreault_1978, Akasofu1981EnergyCB} and the $VB_s$ \cite{Baker1986MagnetosphericRT} to be the most suitable measures of the rate at which energy is loaded into the magnetosphere by the solar wind. Developing some proxy of this loading rate is required to study the magnetospheric responses and has been the focus of much subsequent study (e.g., \cite{McPherron_2015, Borovsky_2013, Newell_2007}). \cite{Lockwood_2022} performed an analysis of coupling functions and established best practices in their derivation and guidance on their limitations. They find the analysis of the persistence of solar wind parameters defines how best to compile a coupling function. Further, they comment on the best metrics for testing the capability of a coupling function, revealing shortcomings in correlation as a useful metric for some applications. Finally, they provide two criteria that coupling functions must describe to quantify integrated effects: 1) the large-event tail; and 2) the core of activity distributions. Improved representation of solar wind energy input to the magnetosphere, both in the form of solar wind coupling functions and multiple signatures of energy input, have led to improved understanding of substorm evolution \cite{Haiducek2019UsingMS} and even to improved prediction of their occurrence and intensity \cite{McPherron_2015}.

\begin{figure}[h]
\centering
\includegraphics[width=\textwidth]{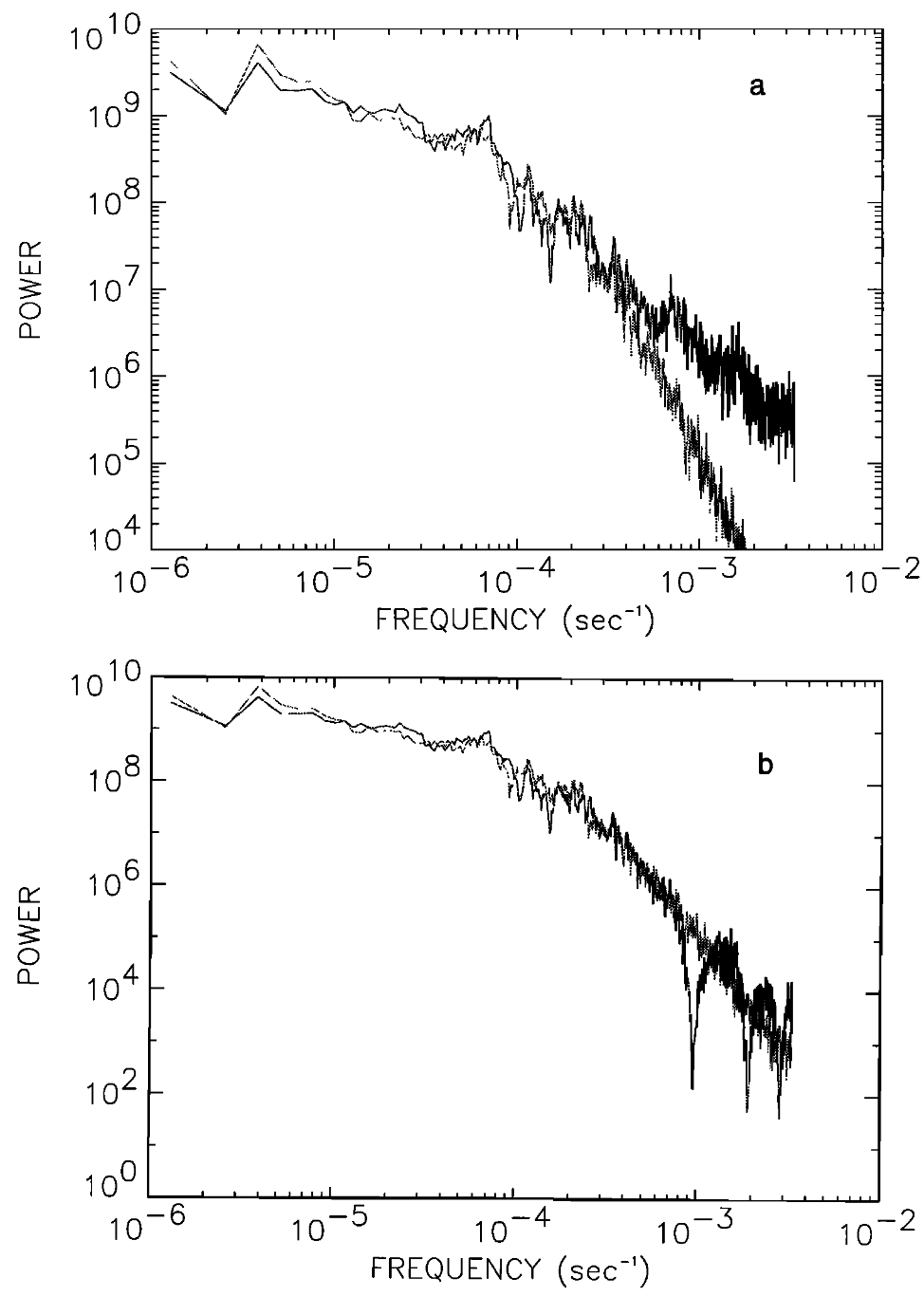}
\caption{Figure 12 reproduced with permissions from \cite{Klimas1996}. Comparisons of Faraday Loop Model (FLM, light curve) and measured (heavy curve) AL index for time periods from the Bargatze AL dataset. (a, top) The two spectra differ significantly for high frequencies due to the fractal nature of the measured time series, which is a different dissipation process from the FLM. (b) The measured time series has been smoothed to bring its high-frequency content in agreement with that of the FLM.}
\label{Klimas1996_Fig12}
\end{figure}

\cite{Bargatze_1985} calculated linear prediction filters (LPFs) to relate driving ($VB_S$) to response (AL index) for a number of intervals during a set of 1-2 day intervals. Their dataset is an important one for Complexity Heliophysics and is described in \ref{datasets appendix section}. They found a peak in these filters at $\sim$20 minutes and $\sim$one hour, which were ascribed to the directly-driven and unloading magnetospheric modes. Power spectra were created comparing FLM-modeled and measured AL index values to $VB_S$ for an interval from the Bargatze data set. They are reproduced from \cite{Klimas1996} in Figure \ref{Klimas1996_Fig12} given their importance to subsequent work on power law relationships across the Heliophysics system.

Figure \ref{Klimas1996_Fig12}a shows a break in the measured and modeled power law spectra at high frequencies, due to the fractal nature of the AL index at high frequencies. In Figure \ref{Klimas1996_Fig12}b the agreement has been recovered when the measured AL index has been smoothed over a 17.5 minute running average. The power law scaling is abundant in and indeed a hallmark of complex systems, indicating the presence of some underlying mechanism that creates self-similar, scale-invariant behavior. \cite{Klimas1996} marks an important recognition of power laws and their significance in Complexity Heliophysics.

\subsection{Input-Output Models: Computationally Mature Input-Output Models} 

The progress and shortcomings of the \textit{linear} prediction filters pointed to the next step in I-O modeling: capturing nonlinearities. The Klimas review provided an early look at two methods to specify nonlinearity: neural networks and local-linear prediction. Their advantage is that they are data-driven, meaning they determine relationships directly from the data rather than, as in the analogue approach to I-O modeling, from `preconceived notions of the processes that constitute geomagnetic activity.' In other words, they do not make \textit{a priori} assumptions and instead allow the data to determine the conclusions. With the data-driven approach comes the challenge of physical interpretation of the derived relationships. Interpretability (making physical meaning) of magnetospheric models was, of course, already important even in relatively explainable simple models, but the awareness was exacerbated as data-driven approaches used models with more parameters, more complexity, and became more difficult to explain. Will increase representativeness, interpretability often suffers. The tradeoff of representativeness and interpretability between data- and physics-driven approaches has only intensified in Heliophysics in the 21st century. A few comments related to the approaches that proved most successful of characterizing the magnetospheric behavior, local linear prediction methods and a subset from the field of ML known as neural networks, from the Klimas review articulate the tension:

\begin{itemize}
        \item ``To understand the physical content of this``model," it is necessary to understand the global structure of the nonlinear coupling surface. In this case, to extract the physical content of the local-linear predictor, it will be necessary to reconstruct the coupling surface from the many local approximations to it that are already available.''
        \item ``...it does appear that further research into extracting the physical content of the network is warranted.''
        \item \cite{Hernandez_1993} ``It is often said that neural networks yield no usable information on the physics of the system that they model. However, it appears that this prejudice may not be correct.''
\end{itemize}

Compiling these comments has proven prescient of a deeper discussion that has emerged and is being shaped in our more modern era of computation and artificial intelligence (AI). Klimas et al., summarize this trend with the statement, ``It is anticipated that in the future, a combined approach involving both analogue [physics-based] modeling and input-output data analysis will prove most effective for understanding the magnetospheric dynamics that couples solar wind input to geomagnetic activity output.''

The early work on neural networks and local-linear predictions focused on geomagnetic activity prediction. \cite{Hernandez_1993} created a neural network prediction of the electrojet index, using the \cite{Bargatze_1985} $VB_S$-AL database. They created a state-space reconstruction (SSR) network, using input solar wind information and past output of the model (e.g., previously predicted AL index values) to predict future AL index values. As a point of comparison, they also created a nonlinear prediction filter (NPF) in which the predicted AL index values are produced based solely on solar wind input information. They determined that the SSR outperformed the NPF and that the neural networks performance were dependent on hyperparameters such as the activation function used. A severe limitation of the nets were the inability to predict large values of AL at all. 

\begin{figure}[h]
\centering
\includegraphics[width=\textwidth]{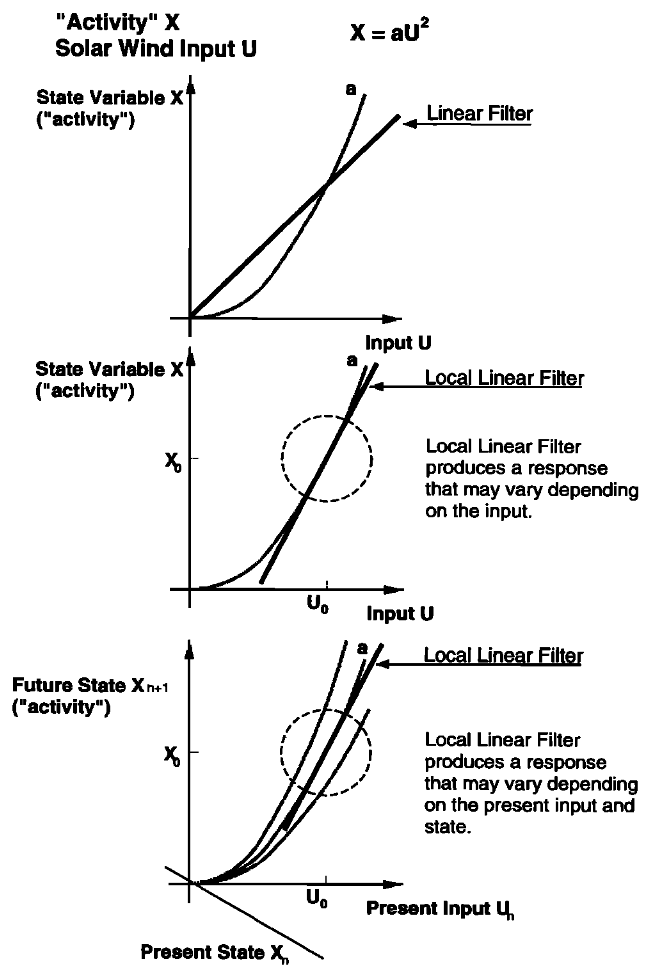}
\caption{Figure 15 reproduced with permissions from \cite{Klimas1996}. Extension of the linear prediction filter (LPF) technique to the local-linear prediction (LLP) technique.}
\label{Klimas1996_Fig15}
\end{figure}

Several works advanced the concept of a local-linear predictor (LLP) \cite{Farmer1989ExploitingCT, Casdagli1992ADS, Weigend1994TimeSP}. LLP is an extension of LPF to capture nonlinear coupling between the input and output. Figure \ref{Klimas1996_Fig15} shows the relationship.

The LLP addresses the situation where the input and output vary over a small range of values. As that variation increases, or as the curve is more nonlinear, the LPF approximation fails. The LLP fixes the the point on the nonlinear coupling curve around which input-output data samples will be used to reconstruct the relationship. The process is one of defining the viable neighborhood. The approximation is valid only within the neighborhood and thus the process must be repeated at each step in time to predict the next step. Applications to the magnetosphere discovered an important extension that both the past and present inputs to the magnetosphere and its geomagnetic response need to be used as inputs to predict future outputs. The size of the neighborhood in the method reflects the degree of nonlinearity in the system, with the relevant neighborhood decreasing as the nonlinearity increases. 

\cite{Price1993TheNR} used the approach to examine the nonlinearity of the magnetosphere, finding only weak evidence. \cite{Vassiliadis1995ADO} is perhaps the seminal early work using LLP. Using the \cite{Bargatze_1985} dataset, they found that the best fit for an LLP is low-dimensional and a local fit to a nonlinear predictor.

To apply the method, they build a large database of instances of the I-O data and use a pattern-matching approach to find the best values in each neighborhood of the case in question, varying the size of the neighborhood to find a minimum in the prediction error. As the nonlinearity of the I-O relationship increases, a smaller neighborhood for the LLP improves the prediction. Several important assumptions accompany the LLP method:
\begin{itemize}
    \item It is assumed that there is a set of variables, small in number, that adequately specifies the global state of the magnetosphere;
    \item Although not all of the variables that specify the global state of the magnetosphere may be measurable, it is assumed that an equivalent state can be reconstructed from those that can be measured;
    \item The next time step, $X_{n+1} = F(X_n, U_n)$, is differentiable, where $F_n$ represents the magnetospheric dynamics that relates previous input and output to future output. 
\end{itemize}

They suggest that the low dimensionality and nonlinearity of their LLP are characteristics of the physical magnetosphere. They provide strong evidence for nonlinearity and low dimensionality, and the results indicate that the evidence is persistent over many different local phase space neighborhoods (i.,e., over many different solar wind and magnetospheric conditions). Their work is a foundation for geomagnetic activity prediction, using past to present geomagnetic activity indicators like the AL index to predict the indicator's evolution into the future (e.g., \cite{Lundstedt1994PredictionOG, Topliff2020SimultaneouslyFG}). It is now also apparent that the Vassiliadis et al. work was a predecessor of modern data-mining approaches using larger data sets (e.g., \cite{Stephens_2019}). Finally, LLPs suggest that, ultimately, a model of the magnetospheric dynamics can be low-dimensional, nonlinear, phenomenological. 

The Klimas review summarized and may have helped sparked deeper and wider exploration of characterizing the magnetosphere as a low dimensional system. One interesting trajectory through the subsequent literature has been toward ML methods to characterize the magnetosphere: 1) specification of a magnetospheric state vector: ``the state of the magnetosphere, resulting from continuous but variable forcing of the solar wind and the interplanetary magnetic field (IMF), can be empirically specified by a magnetospheric state vector,consisting of a set of hourly-averaged magnetospheric driver and response parameters.'' \cite{Fung2008SpecificationOM}; 2) including vector correlations between the solar wind and magnetospheric state vector \cite{Borovsky2018ExplorationOA}; and 3) toward the advent of high-dimensional machine learning (ML) models of the magnetosphere and its coupling to the solar wind upstream and the ionosphere downstream \cite{Bortnik2016AUA, Chu2017ANN, Maimaiti2019ADL, McGranaghan_2021}. We will return to ML in Section \ref{key challenge section}. 




\subsection{Important Themes through 1996 that set the stage for Complexity Heliophysics} \label{themes through 1996 subsection}
The studies reviewed represent launching off points for the Complexity Heliophysics paradigm that will be useful to readers (and in some cases supported by quotes from the Klimas article or others cited therein for convenience and context):
\begin{itemize}
    \item Dimensionality of the magnetosphere and self-organized criticality; 
    \item Relative success of input-output methods whereas autonomous systems methods were inconclusive; 
    \item Local linear predictions \cite{Vassiliadis1995ADO} as a foundation for geomagnetic activity prediction (uses past to present geomagnetic activity indicators like AL to predict the indicator's evolution into the future); 
    \item The wide-reaching utility of neural networks. ``Little seems to have been done using neural networks to predict substorm indicators such as the electrojet indices. Nevertheless, Hernandex et al., [1993] \cite{Hernandez_1993} indicate that this is a promising direction for prediction. Further, in view of [the Klimas review] concerning the equivalence of neural network internal parameters to Volterra kernels, it does not appear that further research into extracting the physical content of a network is warranted.''; 
    \item Explainable AI (e.g., this review already raised the question of interpretability of the input-output models that proved most successful of characterizing the magnetospheric behavior (primarily neural networks and local linear prediction methods). These questions reach into modernity; 
    \item Converging the autonomous and the local linear prediction filter methods (gaining the benefits from each: interpretability and the success of the data-driven approach). ``It is anticipated that in the future these local-linear predictor models will be studied carefully with the goal of organizing these bits and pieces into a global nonlinear predictor model. It may be advantageous to cast these predictor models as analogue models in order to maximize their physical interpretation.''; and 
    \item The assessments of the computationally mature input-output models point to a component of a risk formulation for Heliophysics models: By perturbing the input and witnessing the change one can create a measure of the sensitivity and thus a component of the resilience of the prediction technique. 
\end{itemize}

Additionally, the Klimas review has numerous connections to the dimensions of complexity science discussed in the introduction and reveals Heliophysics-specific items that will be thematic, including: self-organized criticality of the magnetosphere, system dimensionality, the variability of the solar wind, bi- or multi-modal behavior of the magnetosphere/geospace, implications of various observables or metrics for analyzing the system (e.g., the sufficiency or insufficiency of the auroral electrojet indices), nonlinear I-O modeling such as neural networks, and representativeness vs. interpretability of physics-based and data-driven modeling approaches.

\section{Emergence of the connection between Self-Organized Criticality and the Magnetosphere} \label{SOC section}

The concept of self-organized criticality (SOC) provided Klimas et al. with an entry point for assessing the complexity science paradigm in Heliophysics. It's purpose is similar here, yielding a foundational concept from which we branch to other dimensions of complexity that are important to the history of Heliophysics, including power laws and scaling theory \cite{west2017scale}, fractality \cite{song2006origins}, network science \cite{Newman_2018}, emergence \cite{holland2000emergence}, coupling between domains of the solar-terrestrial system, observational considerations, and machine learning. SOC has been well chronicled \cite{Aschwanden2011SelfOrganizedCI, Aschwanden_2014, McAteer201525YO, Sharma201625YO, Aschwanden2019SelforganizedCI} and we will not attempt to recapitulate those excellent reviews, but will provide the necessary background for this review to be self-sufficient and point readers to the most valuable resources to discover SOC research in Heliophysics in more depth. 

\cite{Bak_1987} provided the world with the concept of SOC, an explanation for the ubiquitous $1/f$ power spectra characterized by a power law function $P(\nu) \propto \nu$ that holds for any power spectra that is not purely random white noise ($P(\nu) \propto \nu^0$). The significance is that white noise represents traditional random processes with uncorrelated fluctuations, and the $1/f$ spectra is something else--an indication of non-random structures with long-range correlations in a time series. Bak used a sandpile as the example. Imagine that grains of sand are dropped onto a table. A cone-shaped pile will build up until a grain causes an avalanche. The longer a pile avoids an avalanche the larger the avalanche will be when it occurs. Bak attempted to determine the conditions under which these avalanches occur. He found that avalanches were unpredictable, dependent on the interactions between the individual grains of sand. Eventually the pile reaches a critical point at which the pile transforms into something more complex and properties emerge that are not part of the individual grains themselves. The tendency of the pile toward this critical state (self-organized criticality) was a new way of viewing nature \cite{bak1997}--out of balance, but in a poised state. 

Bak's avalanches are the non-random time structures represented by the $P(\nu) \propto \nu^0$ function. That seminal work sparked a new avenue of research, beginning first with numerical simulations of avalanches and cellular automata, iterative applications of a simple mathematical redistribution rule that yields complex spatio-temporal patterns \cite{Wolfram_2002}, and subsequently to widespread application in the physical sciences. \cite{Charbonneau_2001} provides a review of cellular automata in the field of Solar Physics. 

Nearly ten years after and a continent away from where Klimas provided a backdrop for Complexity Heliophysics in large part beginning with the concept of the magnetosphere as a self-organized system, \cite{Valdivia2005TheMA} offered a synthesis of the field's thinking that self-organization is an explanation for magnetospheric behavior. They addressed this history of the magnetosphere as a self-organized critical system, beginning with the premise that a complex system is one that is characterized by multiscale spatio-temporal (S-T) behavior. Their central question was how could the magnetosphere exhibit complexity at small-scales, but coherent and repetitive behavior at global scales? The canonical example is how there could exist self-similar turbulent behavior in the plasma sheet simultaneously with repeatable and coherent substorm phenomena. A key question in their history is what the threshold state between the predictable and unpredictable might be. Self-organized criticality provided the answer. 

\begin{figure}[h]
\centering
\includegraphics[width=\textwidth]{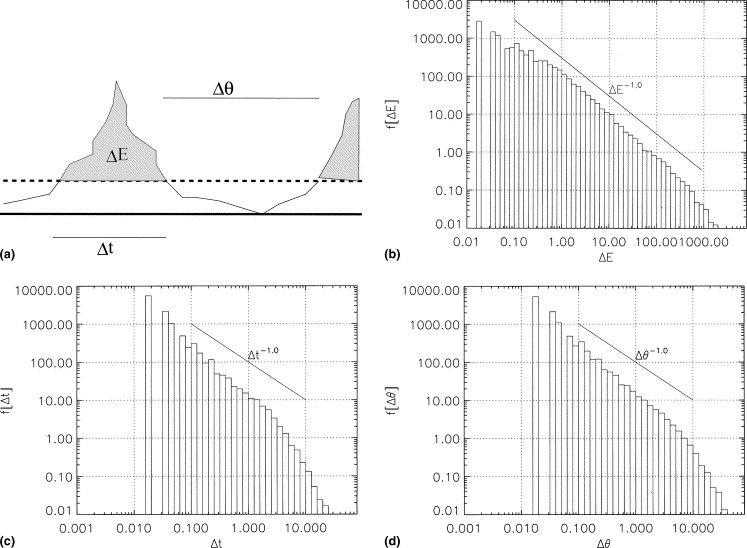}
\caption{Figure 3 reproduced with permissions from \cite{Valdivia2005TheMA}. (a) Schematic AL time series with the three characteristics of self-organization labeled: $\Delta E$ (energy dissipation; area under the AL curve), $\Delta t$ (time duration), and $\Delta \theta$ (time separation between events); Distributions of two years of AL index data for (b) energy dissipated, $\Delta E$, during an event; (c) event duration, $\Delta t$; and (d) time between events, $\Delta \theta$.} \label{Valdivia2005_Fig3}
\end{figure}

Valdivia et al., brought the SOC equations into the magnetospheric context. Taking the AL index as an indicator of global magnetospheric energy dissipation, they describe three characteristics used to indicate self-organization: $\Delta E$ (energy dissipation; area under the AL curve), $dt$ (time duration), and $\Delta \theta$ (time separation between events). Figure \ref{Valdivia2005_Fig3} details each characteristic in AL index data. Their distributions cover two years of AL time series data between 1986 and 1988, taking the threshold of -100 nT to define an event (for a total of N=10365 events). The statistics show clear power laws in all three characteristics.

Finding evidence for a self-organized state, they posit that it may provide a key for understanding substorm onset and derive a 1-D general dynamical model for the magnetic field in the diffusion region of the magnetotail to attempt to explain magnetospheric observations. Using their simple model they compile event statistics for the collective effects of many interacting instability sites (assuming a simple parameterization of the dissipation, derived from observations and data analysis), in a manner not dissimilar from microphysics and particle kinetics simulations. They identified ranges, a phase diagram of sorts, for the $(\mathbf{U} \times \mathbf{B})_y$ term and a certain range in which robust critical behavior occurs. Their model cannot describe the details of the microphysics of the magnetotail, yet it serves to indicate that `the statistical behavior of many complex distributed systems is more a property of their self-organized state, if it is achieved, than the details of the physical processes that allow such state. This is a general characteristic of systems that are close to criticality where many systems belong to the same universality class, suggesting that it is probable that the statistics of substorms, pseudobreakups, and even the evolutions of the growth and expansion phases, are unrelated to the details of the dissipation process (Shay et al., 1998) \cite{Shay1998StructureOT} other than that dissipation allows for the establishment of a self-organized state.'

Subsequently, a quite comprehensive review of SOC as applied to solar physics and astrophysics was created during two week-long meetings at the International Space Sciences Institute (ISSI) \cite{Aschwanden_2014}. They reviewed self-organized criticality across these fields from 1989-2014, highlighting trends, open questions, and future challenges. 

The importance of SOC systems is punctuated by its application across domains, including: ecology \cite{Kauffman1991CoevolutionTT, Halley1996EcologyEA, Milne1998MotivationAB}, evolutionary biology \cite{Langton1991ArtificialLI,Kauffman_1993, Sneppen1995EvolutionAA, Holland1995HiddenOH}, geology \cite{Bak1989EarthquakesAA}, cognitive science \cite{Plenz2021SelfOrganizedCI}, computer science \cite{Wolfram_2002}, the social sciences \cite{Axelrod1997TheCO, Miller2009ComplexAS, Newman2002RandomGM}, economics and finance \cite{Stanley2002SelforganizedCI}, political science \cite{Brunk2001SelfOrganizedCA}. These diverse systems share common features that are linked through SOC: driven, dissipative, and far from equilibrium and releasing energy in a bursty intermittent manner on multiple scales with numerous routes to instability that lead to the energy release and reconfiguration \cite{Watkins_2015}. The importance to this review is that SOC systems are inextricably connected to the statistics of nonlinear processes, which is signaled by power law-like size distributions. The exposition of SOC in Heliophysics was a hallmark of the complexity paradigm in the field in the years after 1996. 

\section{Beyond 1996: Complexity Heliophysics} \label{1996 onward section}


\subsection{Power laws in Heliophysics}

To set the stage for Heliophysicists' adoption of the concept of SOC, we begin with the origin of the idea itself. \cite{Bak_1987} documented peculiar features of the sandpile cellular automata in the discovery of the concept of self-organized criticality. The peculiarity is that the system responds to external perturbations by dissipating the stored energy in an avalanche, where the size of the avalanche is described by a power law distribution with 1/$f$ noise. Nature seems to love power laws -- they appear widely in physics, biology, earth and planetary sciences, economics and finance, computer science, demography and the social sciences \cite{Newman2004PowerLP}. The origins of power law behavior, the mechanisms that can cause it, have been and remain a point of debate in the scientific community. Power law behavior in investigations of the solar wind-magnetosphere system have been taken to be strong indicators of complexity and SOC in the magnetosphere’s dynamic evolution. It is important to mention that appropriate detection of a power law distribution is non-trivial and care should be taken with the methods and power law interpretation applied to empirical data \cite{Clauset2007PowerLawDI}.

\begin{figure}[h]
\centering
\includegraphics[width=\textwidth]{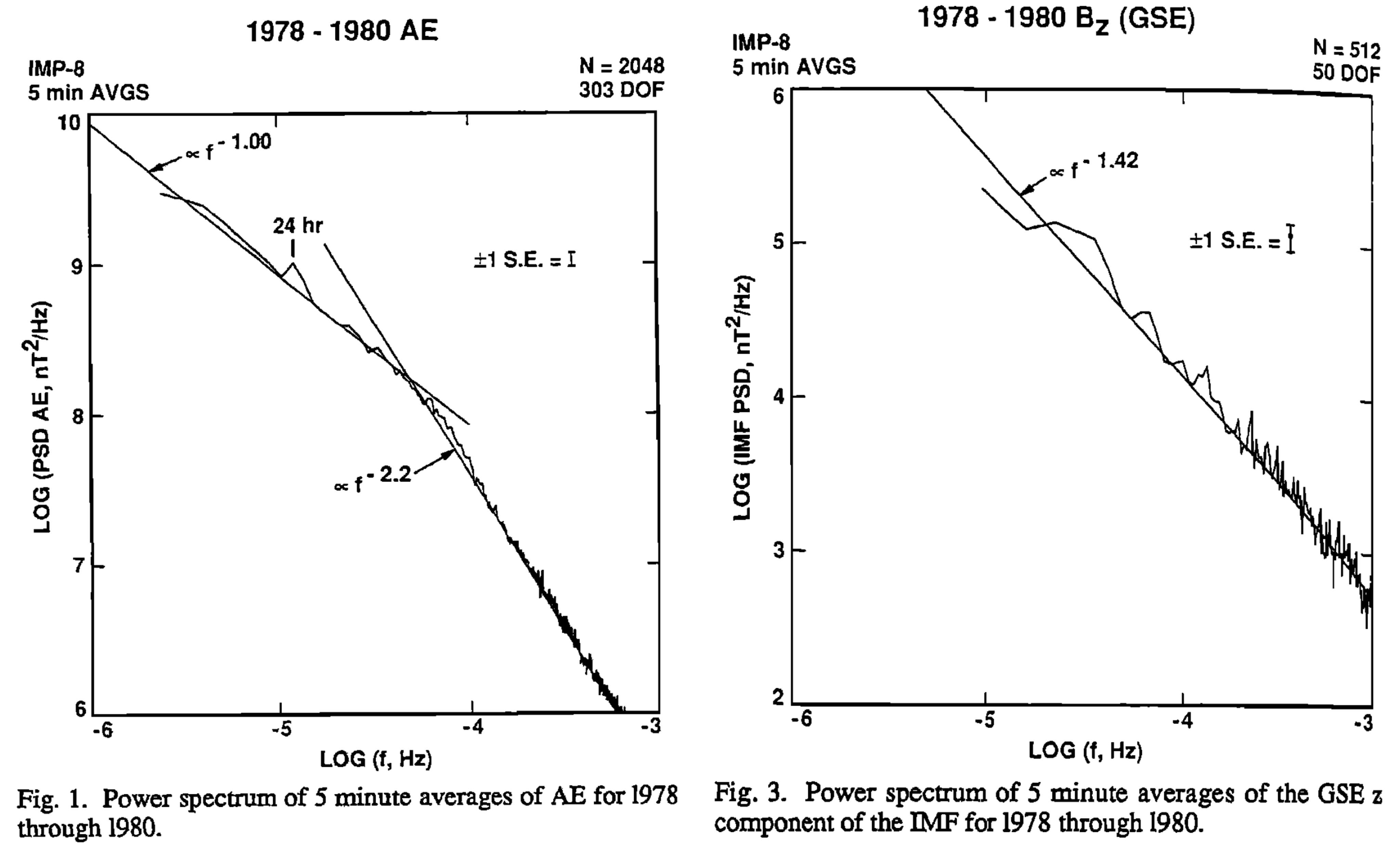}
\caption{Figures 1 and 3 reproduced with permissions from \cite{Tsurutani_1990}. Power spectrum of 5 minute averages of AE (left) and the interplanetary magnetic field (IMF) north-south component ($B_{Z}$) (right) for 1978 through 1980. The figure for the IMF Geocentric Solar Ecliptic (GSE) coordinate frame is reproduced, though the authors analyze and visualize the Geocentric Solar Magnetic (GSM) frame, too. PSD=power spectral density. DOF=degrees of freedom (twice the number of time segments). IMP=Interplanetary Monitoring Platform} \label{Tsurutani_1990_Fig1_3}
\end{figure}

\cite{Tsurutani_1990} examined the power spectra of the AE index and the interplanetary magnetic field (IMF) north-south component ($B_{Z}$) over frequencies between 17 minutes and 28 hours using five minute averages between 1978-1980 (Figure \ref{Tsurutani_1990_Fig1_3}).

Notably, the power spectrum revealed a peak at 24 hours and a break in the spectrum at $\sim$ five hours. The spectra on either side of the break are fit using power laws with $f^{-1.00}$ at lower frequencies and $f^{2.2}$ at higher frequencies. The spectral break was found to be independent of the choice of data interval and averaging length. Overlapping the AE dataset that was used to create Figure \ref{Tsurutani_1990_Fig1_3}(left), they computed the power spectrum for the IMF $B_{Z}$ component, finding an unbroken power law that roughly follows a $f^{-1.42}$ slope (see Figure \ref{Tsurutani_1990_Fig1_3}(right)). Thus, the break in the AE index behavior at around 4.7-5.2 hours was not explained by the solar wind IMF $B_{Z}$. They showed the ratio of the power of AE to the power of the IMF $B_{Z}$ as a function of frequency and found a clear break at $\sim$4.6 hours, below which the ratio is independent of frequency and above which the ratio decreases at a rate of $f^{-0.5}$. The $\sim$5-hour break in the AE power spectrum is longer than substorm time scales of around 30 minutes for expansion phases and a couple of hours for total length. However, their results indicated no preferred period of $B_{Z}$ (no break in the power spectrum) and no preferred period of substorms (no break in the spectrum below roughly five hours). The authors suggest a number of potential explanations as to why the AE index power diminishes with increasing $B_{Z}$, citing saturation mechanisms, but do not draw firm conclusions. 

Research to resolve the question of whether the magnetosphere is an SOC system was reinvigorated, perhaps as a result of the Klimas review, employing power law techniques as the prominent mechanism of investigation. 

\cite{Consolini_1997} (and later extending the statistics in \cite{Consolini_2002}) used the AE index and power law distributions to attempt to explain the intermittent nature of the magnetospheric dynamics (e.g., the rapid fluctuations in the AE index even in periods when the solar wind parameters were relatively smooth). They evaluated the distribution function for the intermittent burst behavior of the index. Defining the quiet time AE index level as $L_{AE} = \left[45\pm15\right]$ nT, they derived the strength of a burst as $s=\int_{\Omega} \left(AE(t)-L_{AE}\right)dt$, where $\Omega$ is the time interval over which the AE index is greater than the quiet time level. The distribution ($D(s)$) for a period of AE data covering around 3000 burst events was found to follow the power law form $s^{-\tau}$ with $\tau\sim1$ over more than four decades ($10^{1} - 10^{5}$ nT$\cdot$minute). They suggested an interpretation as an absence of characteristic length or time for the magnetospheric system. Their result related the magnetosphere to the clear demonstration that SOC systems exhibit a spontaneous organization towards something like a dynamical equilibrium. They extended the analysis to study the second important component of these systems that there are often various scaling regimes when the system is not quasistatically driven. To determine the relevance of the magnetosphere as an SOC system they explored the presence of a $1/f$ regime. They constructed the distribution of the full power spectral density (PSD) of the AE index, corroborating Tsurutani et al.'s results that the PSD divides into two power law regimes ($1/f$ and $1/f^{1.89}$) with a spectral break around $5.5\times10^{-5}$. The low frequency regime represents the random superposition of single burst events while the high frequency regime is the result of interaction among the bursts and is the regime of the SOC state of the magnetosphere. The high frequency regime is associated with minutes to hours time scales of magnetosphere dynamics, which are those associated to magnetic storms and substorms. 

\cite{Consolini_2002} reifies many of the points made by \cite{Consolini_1997}, recentering the questions of whether the magnetosphere is an SOC system, what evidence there is, and what can be measured to resolve the hypothesis. They note that magnetohydrodynamic (MHD) modeling is incapable of describing the highly intermittent and multifractal character of the magnetospheric dynamics during magnetic substorms and storms. As a result of the lack of a physical model through which to study these dynamics, there has been perhaps an over-reliance on the most readily available information: the AE index. They well-capture the background on this reliance. However, the index remained the best available diagnostic, and they used it to extend the statistics of \cite{Consolini_1997} examining the burst size power and burst lifetime distributions of AE, with the goal to discriminate the impulsive dissipative events in the AE index from the enhancements that result due to convection (\cite{Kamide_1999} directly-driven and unloading modes of the magnetosphere). They found power law scaling in the power distribution functions of the form $D(x) \approx x^{\tau}$ with exponents 1.0 $\leq \tau \leq$ 1.5 ($\tau = [1.35 \pm 0.06]$, and $\tau = [1.5 \pm 0.1]$ for the burst size and lifetime distributions, respectively). These relationships were shown to hold over four and two decades, respectively, falling off only once the magnitude of the burst size or lifetime exceed points where the method is likely able to distinguish between the impulsive unloading and the convective dynamical modes of the AE index. The main result of the analyses is that AE index time behavior seems to be the possible occurrence of criticality in the Earth’s magnetospheric dynamics, again meaning that no characteristic scale or time represents the magnetotail dynamics and instead scale-free behavior is exhibited. These findings support previous ones (e.g., \cite{Tsurutani_1990}) that the magnetotail may be an open, dissipative dynamical system at a critical state.

Further, their analysis of the power spectral density (PSD) function of the AE index data revealed two distinct regions characterized by scaling exponents of -2 at high frequencies and -1 at low frequencies with a spectral break at f$\sim$70 $\mu Hz$. 
To clarify the origin of the 1/f-noise region at lower frequencies, the authors explore the relationship between the AE index and simultaneous solar wind parameters, attempting to unravel the how the solar wind behavior may be driving the magnetospheric response from the magnetosphere itself as an SOC system. The work to understand the relationship between the solar wind parameters and the auroral indices is more conclusively taken up in \cite{Freeman2000}, reviewed below. A final word of significance from this study. They issued a key warning that is prescient of future directions in Complexity Heliophysics: ``to find scale-free distribution functions does not mean that the system is in a self-organized critical state. As a matter of fact, while SOC systems display scale-free distribution functions, many other physical mechanisms may produce scale-invariant distributions. In order to address this issue, we must investigate in great detail the physical mechanisms of this scale-free avalanche process in the magnetotail dynamics.''

Tsurutani and Consolini et al. provided strong evidence through broken power law forms of the power spectrum of AE and then its burst lifetime and size distribution that the magnetosphere behaves as a system near its critical point.

If the magnetosphere is a system near its critical point, it challenges the ability to predict its evolution as those dynamics are random. However, a system near its critical point may be confined to a sub-space characterized by a few dimensions and therefore could be well represented by a few parameters. \cite{Chang_1999} review the concepts and mathematical techniques for examining the deterministic chaos of low-dimensional nonlinear systems with fractal characteristics for the magtnetosphere. Chang begins from the idea that systems near critical configurations may exhibit low dimensionality: a dynamical system connected to a reduced number of relevant parameters. In the Heliophysics context this is a possible framework for the explanation of bursty bulk flows, low-dimensionality, and power law magnetic field spectra in the magnetotail, postulating the magnetotail to be an open, dissipative dynamical system near ``forced- and/or self-organized criticality'' (FSOC) \cite{Consolini_2002}. Readers will recognize their postulate--it is a foundation of the arguments of \cite{Valdivia2005TheMA} (reviewed at the end of Section \ref{SOC section} above) in describing the magnetosphere using SOC. The magnetotail plasma being near the point of criticality and causing a substorm onset is like a fluid being at the critical point for equilibrium liquid/gas phase transitions--both are FSOC systems. They first emphasize that the magnetosphere is inherently multiscale and place focus on the mathematical tools to address the interplay of the kinetic, intermediate, and magnetohydrodynamic (MHD) scale fluctuations. In describing the merging of coherent magnetic structures in the magnetotail, they were perhaps the first to suggest that this process is the explanation of ``bursty bulk flows,'' a topic that remains an active area of research and more recent work seems to corroborate \cite{Gabrielse2014StatisticalCO, Wiltberger2015HighresolutionGM, Merkin2019ContributionOB}. Of further note is that they open the possibility that the magnetic reconnection events leading to bursty bulk flows likely come from many if not all of the suggested microscopic instability mechanisms such as the collisionless tearing instability, or the cross-field two-stream instability. In such nonlinear systems there are likely overlapping and interacting mechanisms driving observed behavior and suggest multiple explanations rather than attribution to a single cause.  Complexity thinking is open to multiple explanations over the traditional central and singular explanation. 

The Chang paper is a good introduction to the nontraditional mathematical techniques of dynamic merging of coherent structures, nonclassical nonlinear instability, path integrals, the theory of the renormalization-group, low-dimensional chaos, self-similarity and scaling, fractals, coarse-grained helicity and symmetry breaking and an excellent complement to this review. An important contribution is the introduction of powerful techniques for quantitatively and computationally studying dynamical systems far from equilibrium, such as the renormalization-group transformation procedure. Each of the ideas in the paper play an important role in Complexity Heliophysics. Coarse-graining (the representation of a physical system in which some of the fine-grained structure has been smoothed over without introducing external details, or in other words remaining true to the microscopic details\footnote{\url{https://www.edge.org/response-detail/27162}}), especially, emerges from this review as a central concept in Complexity Heliophysics. A rigorous understanding of coarse-graining is important to build compact representations of a system and thus to bridge the gap from complexity research in Heliophysics to decision-making based on the knowledge, which we further develop in Section \ref{trends section}. 

\begin{figure}[h]
\centering
\includegraphics[width=\textwidth]{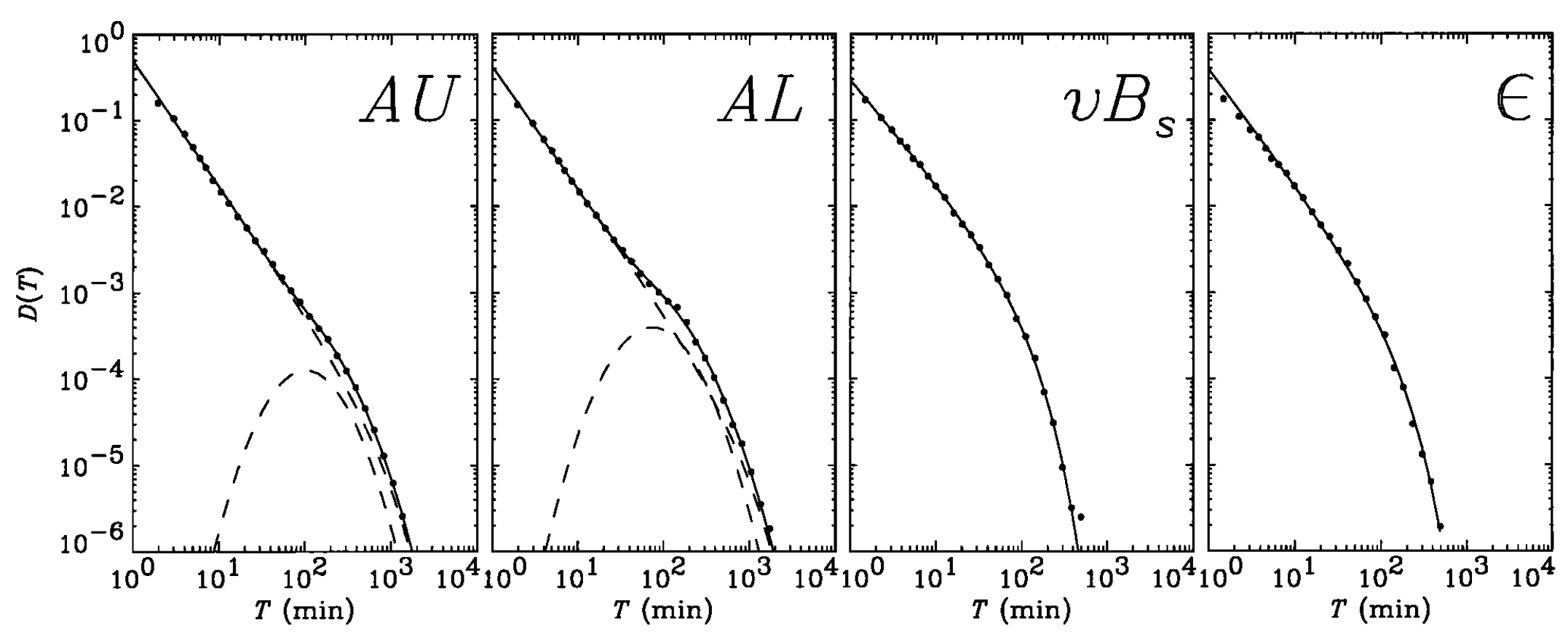}
\caption{Figure 2 reproduced with permissions from \cite{Freeman2000}. Functional best fits (solid lines) to the burst lifetime probability density functions (dots) of AU, $|$AL$|$, $vB_s$, and $\epsilon$. D(T) is the burst lifetime (T) probability density function. The power law and lognormal components of the fitted function are shown by the dashed lines for AU $|$AL$|$ while $vB_s$ and $\epsilon$ are fit using only an exponential.} \label{Freeman2000_Fig}
\end{figure}

Subsequent to the knowledge of broken power laws in the auroral indices and the enumeration of the techniques to study them, particularly in defining measures of energy bursts \cite{Consolini_2002}, were numerous studies that identified additional characteristics or anomalies in the distributions. Power laws still governed the distributions of burst magnitude and duration, but Consolini (in submitted work in the year 1999 that was not published entitled ``Avalanches, scaling and 1/f noise in magnetospheric dynamics'') identified small `bumps' with characteristic values of magnitude and duration and suggested that a better fit to these distributions was a power law with an exponential cut-off plus a lognormal distribution. \cite{Freeman2000}, studying AU and $|$AL$|$ data 1978-1988 paired with additional observations from the WIND satellite's \cite{Szabo2014} particle and magnetic field instruments between January 1995 and December 1998, showed a power law component of the burst lifetime distribution ($P(T)$) in two measures of solar wind-magnetosphere coupling, Akasofu’s epsilon and velocity$*$southward IMF ($vB_s$), but absent a bump. This close correspondence is illustrated in Figure \ref{Freeman2000_Fig}, comparing curves for AU, $|$AL$|$, $vB_s$, and $\epsilon$. 

They found that AU and $|$AL$|$ distributions were fit well by power law with exponential cut-off plus lognormal distributions, and that there was no evidence of the log normal component in the solar wind variables ($vB_s$ and $\epsilon$). They examined each component of the AU and $|$AL$|$ distributions and described the physical implication. First for the power law component: The power law exponent of the solar wind variables did not significantly differ from those of the AU and $|$AL$|$ indices. This similarity of the input component (solar wind) to the output component (AU and $|$AL$|$) points to the system being `directly driven' (quasilinear relationship) by the solar wind at short ($\sim$20 minutes) time lags. The similarity between AU and $|$AL$|$ points to the fact that this component of the magnetospheric output acts throughout the auroral oval because the AU and AL indices rely on contributing magnetometers from different local times. Finally, because of similarity to the solar wind and the global distribution of these relationships, this power law burst lifetime component in the AE indices may be attributable to the Disturbance Polar type 2 (DP2) convection electrojets \cite{Nishida_1966}. Next, the lognormal component: It was most prominent in the $|$AL$|$ index for which the contributing magnetometers are concentrated in the post-midnight sector and acted over the characteristic magnetospheric timescale of 2-5 hours. Both were considered evidence that the lognormal lifetime component is the substorm unloading component associated with the DP1 electrojet, the `unloading' current system \cite{Obayashi1968LargescaleEF}. This component is not scale-free, but rather does have a characteristic timescale (2-5 hours).

The work of Freeman et al. extended the exploration of the connection between the solar wind driver and magnetospheric response distributions (i.e., relating the spectral density of the output of the magnetosphere (e.g., AU and AE) to the input/driver (e.g., $vB_s$)) as a means to understand the magnetosphere's dynamical nature. A key conclusion was that the scale-free burst lifetime of AE is not conclusive evidence that the magnetosphere is an SOC system, and that additional observations are needed. This motivated work that utilized network analysis and imagery data to attempt to unravel what could not be unraveled from the time series alone. Their key comment that would lead to seminal work in the following years is, ``...whilst scale-free behaviour in the system output is a feature of SOC systems, recent SOC models have been developed in which the scale-free behaviour is in the local or internal system output and not in the global or system wide output measured by the AE indices [Chapman et al., 1998] \cite{Chapman_1998}. Thus in order to assess whether these models are an appropriate description of the magnetosphere, attention should turn to other observables that include spatially localised as well as global phenomena.''     

\cite{Chapman2000AvalanchingAS} took up the uncertainty in many of the previous authors' minds about the inability of the AE index to unambiguously determine the type of dynamical system that characterizes the magnetospheric behavior. They identify three main classes of `SOC' system relevant to the magnetosphere: 1) the original definition of Bak et al.; 2) forced SOC (F-SOC, \cite{Chang1992LowdimensionalBA}; and 3) a phenomenological definition based on observation of some or all of a set of possible SOC diagnostics (e.g., bursty time series, 1/f power spectra, avalanche distributions). Responding to previous observational evidence and inquiry, they ask how does one can reconcile the low dimensionality observed in the magnetosphere with the observed robust bursty evolution. Key is the knowledge that systems at criticality can also be low dimensional, and that therefore SOC as well as competing explanations for these systems must be distinguished by other means than dimensionality. For instance, avalanche models (strictly SOC models) must have fixed points around which the low dimensionality is observed. They argue that the complication stems from distinguishing SOC from SOC-like, ``it is critical to understand to what extent measures of the system dynamics such as auroral indices also measure the solar wind driver directly and hence to quantify their appropriateness for such studies.'' Systems that appear to exhibit SOC through some parameters used to proxy the state (e.g., the AE indices) may not be able to distinguish between a system with an internal attractor (SOC) from those that are driven to some state (F-SOC or SOC-like). The authors attempt to establish whether the idealized SOC state is in fact needed to account for the observed burstiness, self similarity and low dimensionality associated with magnetospheric dynamics. 

\begin{figure}[h]
\centering
\includegraphics[width=\textwidth]{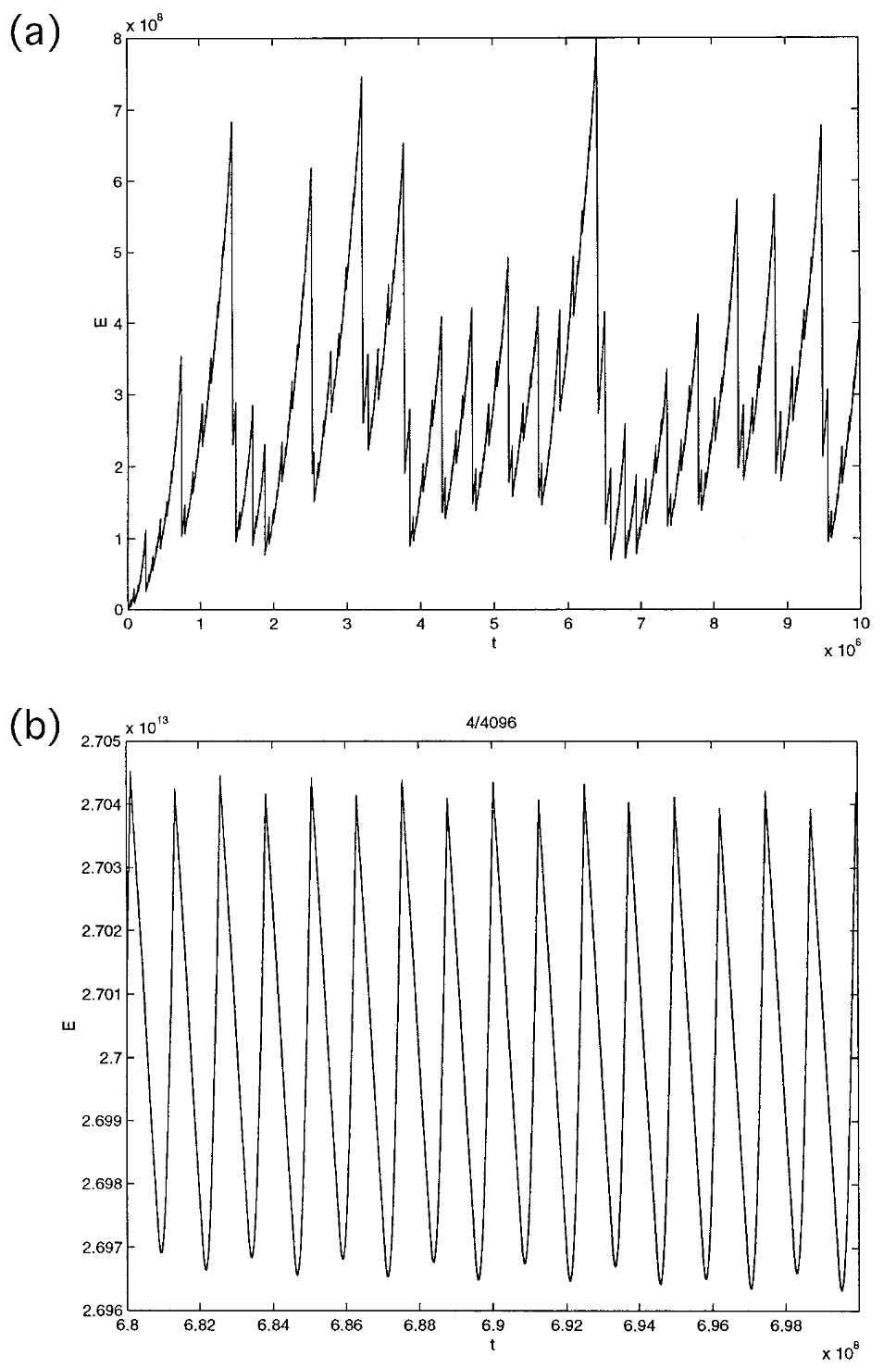}
\caption{Figure 1 reproduced with permissions from \cite{Chapman2000AvalanchingAS}. Timeseries of energy released during avalanches with flattening back length (a) $L_f$= 2000 and (b) $L_f$=4.} \label{Chapman2001_Fig}
\end{figure}

In exploring this distinction they elucidate the descriptions available for turbulent and other high-dimensional systems related to the different classes of SOC description. These are numerical models available to study avalanching and intermittency, including avalanche models and Coupled Map Lattice (CML) \cite{Kaneko_1993}, which is an approach that consists of decomposing the processes underlying the phenomena of interest into potentially nonlinear independent components (e.g., convection, diffusion), and then reducing each of these to simple parallel dynamics on a lattice. They present one result in an attempt to explain how systems at criticality can also be low dimensional, an attempt to understand the fact that observational evidence exists for low dimensionality in the dynamic magnetosphere while bursty evolution is also robust. Importantly, avalanche (sandpile) models, ``...have robust emergent phenomenology that produces bursty time evolution with power law burst statistics as required but these systems are by construction high dimensional, in the same sense as CML. If in addition these systems exhibit fixed points, then close to the fixed points, that is, close to criticality, the behavior is low dimensional.'' So for avalanche models to explain the magnetospheric observations they must have fixed points. They then demonstrate that avalanche models can be altered to exhibit low-dimensional behavior by introducing a `fluidisation parameter,' $L_f$, which is a fixed distance behind the leading edge of an avalanche that is flattened back, effectively moving the system away from a repulsive fixed point. Figure \ref{Chapman2001_Fig} is a reproduction of their Figure 1 that reveals behavior of an avalanche model with varying $L_f$. Their model is one in which sand is redistributed when a critical gradient is exceeded locally. Redistribution in their model occurs across all sites within an ongoing avalanche by construction. The fluidisation parameter has the effect of flattening back the sand behind the leading edge of an ongoing avalanche for a fixed distance, $L_f$. They illustrate the central point that, under certain conditions, an originally high-dimensional sandpile model can exhibit low dimensional dynamics. 

When $L_f$ is on the order of the system size, the behavior is that of the sandpile model--evolution is bursty and burst statistics are power law (Figure \ref{Chapman2001_Fig}a). Reducing $L_f$ significantly, the evolution becomes quasiregular, exhibiting a distinct loading-unloading cycle (Figure \ref{Chapman2001_Fig}b). Statistics in this case are power law only over a restricted range. Thus, by changing certain conditions of the avalanche model, the high-dimensional sandpile model can exhibit low-dimensional dynamics. The implication for systems, the magnetosphere for instance, is that low dimensionality can signify a system close to criticality or certain classes of avalanching systems whose specifci parameters produce either intermittent, or quasiregular, time evolution. 

The complication of distinguishing SOC from SOC-like reveals the importance of understanding to what extent measures of the system dynamics such as auroral indices also measure the solar wind driver directly. Systems that appear to exhibit SOC through e.g., the AE indices as proxies for the state may not be able to distinguish between a system with an internal attractor (SOC) from those that are driven to some state (Forced-SOC or SOC-like). Chapman et al. ultimately conclude that auroral indices are not effective at distinguishing the internal dynamics of the magnetosphere from that of the intermittent solar wind driver. The statement from the article that resounds across Complexity Heliophysics is, ``Of principal concern in the magnetosphere is the variability of the driver and the extent to which any given observable yields the output of the system, the system's internal dynamics, or a mix of these with the driver superimposed.'' Raising the implications of this paper to a broader level, the authors write that dealing with real observations exhibits complications that one must be aware of in studying the phenomenology of SOC for a given dynamical system (e.g., the magnetosphere). 

Encompassing works surrounding the beginning of the new millennium, \cite{Consolini_2002} called SOC a new paradigm for magnetospheric understanding, implicitly naming power law analyses a vital diagnostic. 

Thus, numerous studies identify power law behavior as evidence of self-organized criticality in the internal magnetospheric dynamics (especially in connection with phenomena like substorms based on the AE index spectrum), though `the observed characteristics of the spectrum are also amenable to alternative interpretations' \cite{Chang_1999}, as will be seen in \cite{Lui_2000} and \cite{Uritsky_1998}. The openness of interpretation, the advent of computational capability, and the availability of observations from new missions and sensors spurred investigation of potentially richer and less ambiguous data, such as imagery. 

\cite{Riley2012OnTP} explored distributions for phenomena occurring across the solar-terrestrial system: solar flares, speed of coronal mass ejections, Dst index, and $>$ 30 MeV proton fluences as inferred from nitrate records, for the purpose of estimating the likelihood of rare extreme events. Their method is that of extrapolation, assuming that the range over which the events are well observed can be reliably extended to regimes where they are rarely, if ever, observed. They argue that power laws represent the phenomena studied and from which extrapolation of probabilities is trivial. Their empirical data meet the criteria they spell out for power law distributions and they are able to estimate likelihoods of space weather events not observed in the space age in the next decade.

\subsection{From time series to imagery} \label{image analysis section}


Like \cite{Riley2012OnTP}, some studies have looked beyond the AE indices and their intrinsic limitations for observables/data to understand the magnetospheric, and ionospheric, complexity. A predominant source of observation for magnetospheric output is auroral optical activity or imagery. Imagery is a more direct measure of energy output from the magnetosphere than ground-based indices \cite{Lui_2000}. The intensity, color, and location of the aurora contain information about the magetospheric particles that cause them. The size, shape, and extent of the auroral region enables inference about the size and shape of the magnetosphere and the fluctuations in the solar wind driver. These data provide capabilities that indices or in-situ observations do not, e.g., observing over large spatial areas of the high-latitudes, but also require additional care in preparing and interpreting the data as we will see. There is rich and wide literature on the use of auroral imagery to study the magnetosphere and geospace. We will focus on those studies that have used these data to inquire about the nature of magnetospheric dynamics specifically.


\cite{Lui_2000} is one of the early examples of examining auroral imagery data to infer the complex adaptive behavior of the magnetosphere. Motivating their study was the seeming distinct behavior of internal high-dimensional plasmasheet dynamics (e.g., burstiness) that drive small-scale auroral structures and global dynamics (storms and substorms) and the fact that the causal dynamics of the two modes are not understood. A number of studies demonstrated that these modes are inherently connected \cite{Chang1992LowdimensionalBA, Chang1992PathID, Chang_1998,Chang_1999, Consolini2001MagneticFT, Klimas2000OnTC}. Principle among them and already discussed above is \cite{Chang_1999} who discussed forced self-organized criticality whereby global dynamics can be a consequence of high-dimensional SOC behavior and the high-dimensional plasmasheet behavior (driving small-scale aurora) can be an artifact of the global loading-unloading magnetospheric dynamics. 

\begin{figure}[h]
\centering
\includegraphics[width=\textwidth]{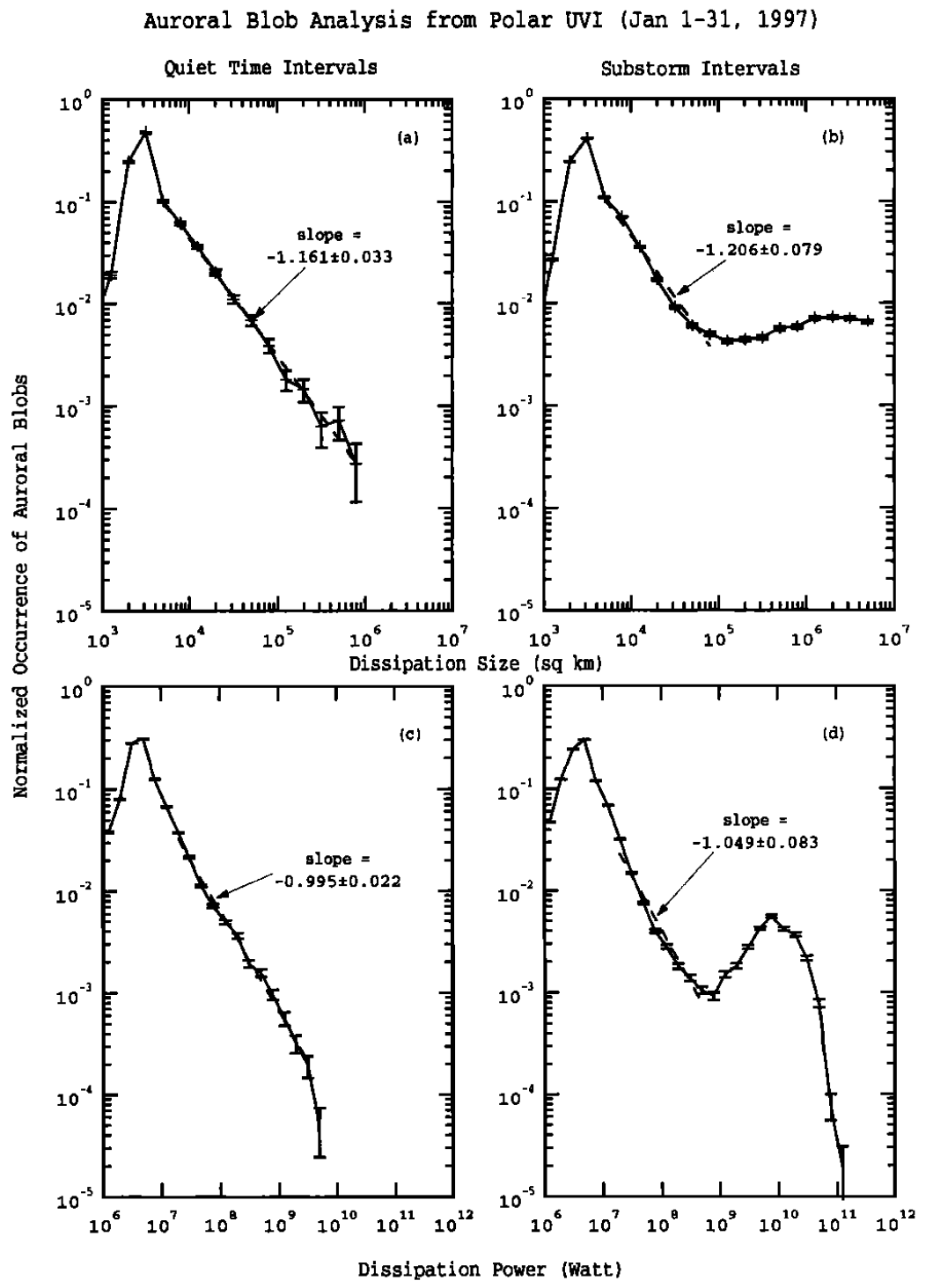}
\caption{Figure 3 reproduced with permissions from \cite{Lui_2000}.
Probability distributions of size and power output of individual auroral regions. (a) Size distribution during quiet times. (b) Size distribution during substorms. (c) Power distribution during quiet times. (d) Power distribution during substorms.} \label{Lui2000_Fig3}\end{figure}

Lui et al. attempted to use auroral imagery to monitor the total energy output of the magnetosphere across scales (small auroral arc-like, and global). They produced the first probability distributions of the power and spatial size of the magnetospheric energy output across scales from auroral emission regions. To determine the distribution of power and sizes of these auroral activity region they identified individual auroral blobs in the ultraviolet (UV) imager on the Polar spacecraft \cite{Torr_1995}, calculated the power in the intensity of the blob in the image \cite{Germany_1997} and the area of each blob, and compiled these values into statistics. Each image was manually inspected and classified as substorm or non-substorm to separately examine the statistics for these global and non-global cases. The distributions are shown in Figure \ref{Lui2000_Fig3} (Figure 3 reproduced from \cite{Lui_2000}). 

Quiet time distributions (left column in Figure \ref{Lui2000_Fig3}) display power law behavior across roughly four decades of dissipation size and power. Similar power laws (with slopes matching those of the quiet times within uncertainties) are found for the substorm intervals, but a peak above $\sim10^5$ km$^2$ and $\sim 5\times10^8$ Watt exists in the size and power, respectively. They interpret these power law regions to mean that there is an ever-present component of auroral activity which exhibits the scale-free behavior of an avalanche system and that this behavior exists regardless of the presence or absence of substorms. They interpret the behavior that is independent of the level of activity as `bursty, internal (localized) relaxations of the system.' On the other hand, the peaks noticed in the substorm intervals, but not in quiet times, are interpreted as global reconfigurations. Put another way, their interpretation was that there was a scale-free (power law) component of auroral activity that was always present (i.e., regardless of whether or not there is substorm activity), but that global events during substorm intervals superimpose on the scale-free behavior well-defined peaks in emitted power and size of emission regions. Their ultimate conclusion was that the magnetosphere acts as an avalanche system. 


There remained questions about the peaks found by Lui et al. Prominent among them was whether the avalanching magnetospheric system could exhibit power law (scale-free) behavior in the energy due to internal relaxations/burstiness while having a characteristic mean in the energy released with global reconfiguration that scales with the system size (e.g., the global extent of magnetospheric activity) \cite{Chapman_1998}. \cite{Uritsky_2002} took up those questions and the idea that, ``understanding the complexity in the magnetospheric behavior associated with critical phenomena appears to be necessary for a correct description of geomagnetic activity as a response to the solar wind driver,'' suspecting that the absence of the temporal dimension might have contributed to the strange peaks. They extended Lui et al.'s method from a static spatial analysis to a spatio-temporal analysis using the same UV Imager data from Polar. This spatio-temporal representation was found to be vital to describe SOC dynamics in a strongly driven system \cite{Watkins_2015}. 
It begs the question of why the spatio-temporal perspective is necessary? In avalanche models the main characteristics of an avalanche are its size and energy. These characteristics are found by integrating over both spatial and temporal coordinates. While in theoretical or laboratory avalanche experiments the driving can be closely controlled, this is not true of forcing in the real world. With a natural and uncontrollable source, the task to verify power law statistics requires more elaborate techniques to identify individual events. They cite two limiting cases permitting the use of laboratory SOC inference techniques (i.e., separating temporal and spatial domains) to physical systems: 1) the avalanching event lifetime is much shorter than the sampling time of the dataset and spatial propagation characteristics of the event are well known, then a purely spatial analysis is warranted; and 2) in the absence of spatial information but one can be certain that there are not multiple avalanches evolving simultaneously, then avalanche distributions can be calculated from time series of the output characteristics of the system. Neither case is true for the magnetosphere and auroral emissions. For instance, the low driving rate condition in effect requires that only one reconnection site in the plasmasheet be active at any one time given that the frequency of reconnection in the plasmasheet is high relative to the driving of the magnetosphere. However, it is well-established that there are often multiple reconnection sites over an extended spatial region \cite{Angelopoulos_1999}. This leads to a key statement of the text, ``Therefore, the low driving rate assumption is not satisfied in the magnetosphere and so the results of previously reported time series analyses related to the hypothesis of SOC in the magnetosphere \cite{Consolini_1997, Takalo1999ACM, Freeman2000, Uritsky2001ComparativeSO} are insufficient for obtaining correct avalanche distributions in terms of a rigorous SOC approach. Moreover, since the lifetime of many auroral activations is longer than the sampling time of the Polar UVI image series, the static spatial analysis reported by Lui et al. [2000] \cite{Lui_2000} is also inappropriate.'' 

\begin{figure}[h]
\centering
\includegraphics[width=\textwidth]{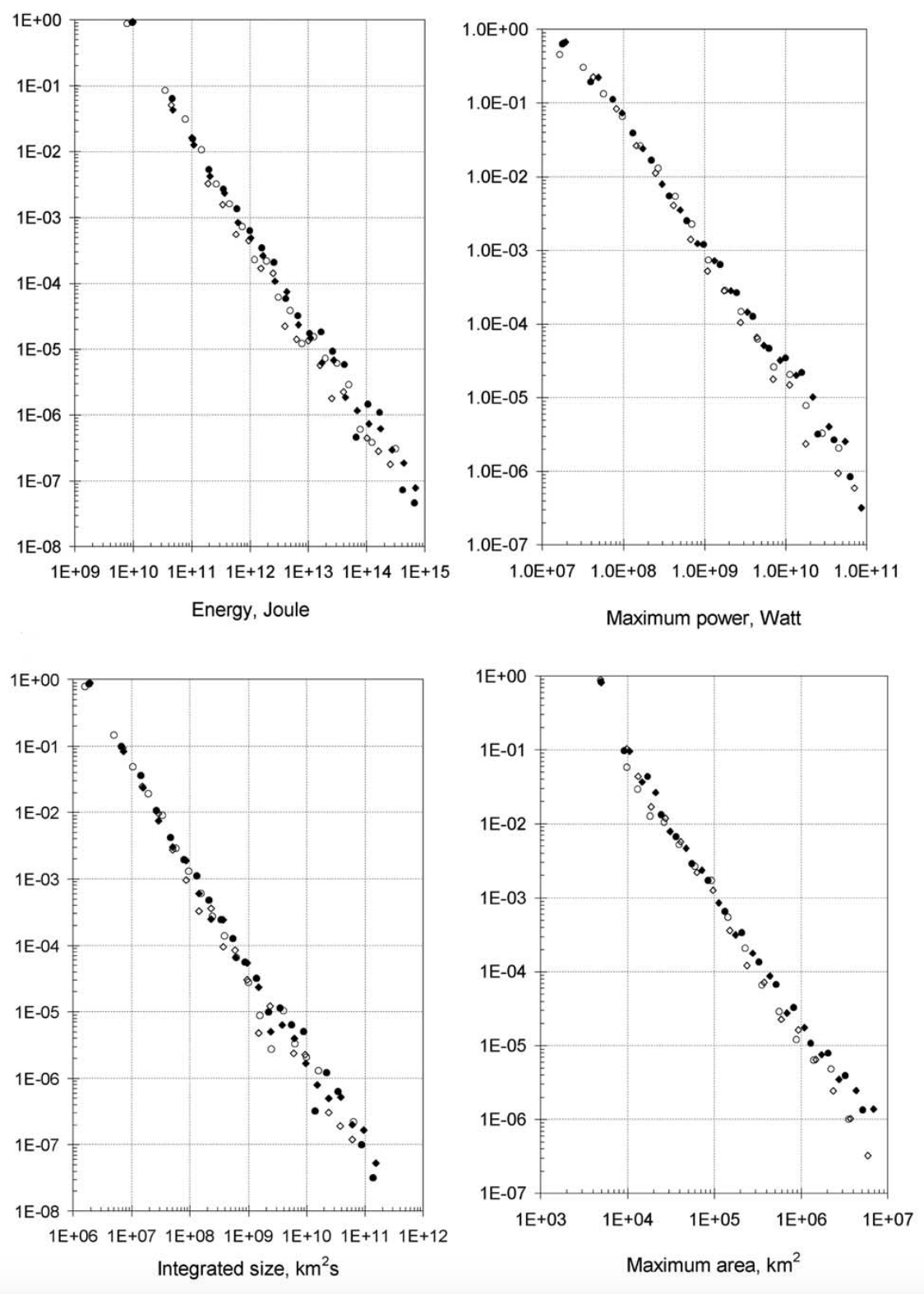}
\caption{Figure 3 reproduced with permissions from \cite{Uritsky_2002}. Text from their caption: Normalized occurrence of spatio-temporal auroral perturbations as a function of maximum area A, maximum power output W, integrated size S, total energy deposition by auroral electrons E (resolution 184 s). Specification of months are: January 1997 (solid circles), February 1997 (solid diamonds), January 1998 (empty circles) and February 1998 (empty diamonds).}
\label{Uritsky_2002_Fig3}
\end{figure}

The key to their spatio-temporal approach is to identify and distinguish multiple simultaneous avalanches (auroral emisions) and calculate their individual properties \cite{Hwa_1992}. They analyzed more than 30,000 Polar UVI images from January–February 1997 and January–February 1998 and largely underwent a similar preprocessing as the Lui et al. data. The sampling time of the images was 184 seconds, also including a period where the sampling occurred at a higher rate (37 seconds). They compiled statistics for active auroral regions that persisted longer than the sampling time and thus across two or more consecutive images, tracking that region through the images up to five hours. They treated events that split but had a unique source as a single event, and merged events with spatially distinct sources as separate events, following \cite{Becker_1995}. For each event, they calculated lifetime T, integrated size S, and integrated energy E as well as maximum active surface area A and maximum energy deposition rate W. Figure \ref{Uritsky_2002_Fig3} reproduces their principal result. 

The central finding is that, across UV images and a variety of solar wind conditions, they found no characteristic time, size, or energy scales within the entire available range of studied parameters. The auroral events exhibited well-defined power law statistics over a broad range of scales. They did not find the distribution peaks that Lui et al. reported, concluding that the peaks are an artifact of an incomplete avalanche detection methodology that missed the temporal component. Thus, auroral emissions exhibit statistical properties of avalanches in SOC models. In relating the magnetosphere dynamics through auroral observation and the behavior they observed, they were able to draw analogies to the avalanche model itself: auroral activations are the avalanches while reconnection events in the plasmasheet are the internal dissipation of SOC models.



In addition to addressing methodological issues that may have produced the peaks in Lui et al., Uritsky et al. contributed several other key findings, including: 
\begin{itemize}
    \item The power law distribution, and thus SOC-like behavior, was exhibited over many orders of magnitude for the duration, power, and size of auroral activations and therefore the magnetosphere behaved as an SOC system across wide ranges of geomagnetic activity. It is worth noting that not all Heliophysics research claiming power law distributions cover the same observational range as Uritsky et al., exhibiting this kind of relationship over only a small number of orders of magnitude;
    \item The dynamics comprising all levels of magnetospheric activity remains scale invariant; and 
    \item The magnetosphere as represented by the spatio-temporal evolution of auroral emissions operates in a self-organized state. The dynamics of auroral perturbations corresponds well to avalanche dynamics at criticality.
\end{itemize}
The lasting discovery is that one can expect cross-scale coupling effects to play a significant, if not crucial, role in the development of geomagnetic disturbances and that large-scale properties of the magnetotail plasma sheet depend critically on the statistical hierarchy of small- and intermediate-scale perturbations associated with sporadic localized magnetic reconnections, current sheet disruptions, and other localized plasma instabilities \cite{Klimas_2000b} 
A question raised by this work for the community is whether statistics of mesoscale magnetosphere simulations match observed statistics from the SOC paradigm. 

A general comment can be made from the Lui-to-Uritsky development: the spatio-temporal domain is required for identifying SOC dynamics in a strongly driven system. To make appropriate comparison, identification of the auroral activations had to be conducted in spatio-temporal space. Uritsky et al. found that the spatial-only analysis of Lui et al. produced `bumps' in the power law distribution that were entirely methodological and were incorrectly interpreted to be unique behavior during periods of high perturbation.
 
In order to truly understand the system, data across the full spectrum of system activity were critical. 

Following the demonstration of the importance of the spatio-temporal perspective in concert with the advent of imaging platforms, Complexity Heliophysics began to use imagery data more regularly, a trend that persists into the 2020s and is expected to continue \cite{Kozelov2004PowerLP, Uritsky2007CoexistenceOS, Golovchanskaya2008ScalingBO, Klimas2010MultiscaleAE, Aschwanden2014GlobalEO, Longden2014MagneticLT}.

Indeed the NASA mission intended to resolve longstanding fundamental questions about the nature of substorms, the Time History of Events and Macroscale Interactions during Substorms (THEMIS), included All Sky Imagers (ASIs) in the ground observatories that accompanied the magnetospheric spacecraft \cite{Donovan2006TheTA}. Imagery, as well as multi-modal observational systems, will continue to be a vital component of unraveling the complexity of the solar wind-magnetosphere-ionosphere system, and the capabilities of these systems will grow (e.g., see the University of Calgary's Transition Region Explorer (TREx) sensor web \cite{Spanswick2018FirstLightOF}). 

Of course, imagery data have relatively shorter histories and pose their own processing and analysis challenges, so time series analyses remain important sources of new complexity studies \cite{Chapman2004RobustnessAS, Consolini2008OnTE, Borovsky2019CompactingTD, Meng_2021}. Prominent among the continuing body of research are results related to the complexity of the geospace system made possible by the Swarm mission, which we do not provide a detailed review of in this manuscript but point readers to several important works \cite[and references therein]{deMichelis2015MagneticFF, Papadimitriou2020DynamicalCO}. As far as processing challenges, more recently groups have been bringing tools from artificial intelligence and machine learning (AI/ML) to bear (on auroral imagery \cite{syrjasuo2002analysis, Syrjsuo2004DiurnalAO, Clausen_2018, Kvammen_2020, Nanjo2022AnAA} and solar imagery \cite{Galvez2019AML, Armstrong2019FastSI, Upendran2020SolarWP, Brown_2022}). Section \ref{key challenge section} discusses the intersection between complexity science and AI/ML, positioning it as a key challenge for 21st century Heliophysics and indeed all of science. 

%
%

\section{Emerging literature: Topics and trends} \label{trends section}
The following section examines emerging literature (largely drawn from publications after the year 2010) and extracts topics and trends. These perceived topics and trends are organized by section. This section is a departure from the previous ones in that along with the literature review, subjective assessment of areas that might be important to Heliophysics in the coming years are provided. These could be interpreted as predictions and should thus be treated with a degree of uncertainty or openness to interpretation. We also draw extensively from complexity science literature outside of the field of Heliophysics to establish trends. 

\subsection{Metrics and Diagnostics of Complexity} \label{metrics and diagnostics subsection}
The first observation is that the complexity paradigm in Heliophysics is shared across disciplines (physics, in general, biology, social sciences, etc.), and understanding the topics common to each of these versions of the complexity paradigm informs the future research avenues for each of them. The first has to do with how we quantify and make legible complexity. The means of 'metric-ing' complexity will be a good transition into this section as it will call back to themes of the literature review above and point to trends we identify below. 

There is no single metric of complexity, just as there is no single metric suitable to understand the capability of a model. \cite{Lloyd_2001} gives three dimensions along which to measure complexity: 
\begin{itemize}
    \item How hard is it to describe?
    \item How hard is it to create?
    \item What is its degree of organization?
\end{itemize}

Many metrics have been proposed for assessing and quantifying a complex system (see \cite{Lloyd_2001} for an informative, but non-exhaustive enumeration), and the list continues to grow. A complex system is by definition multi-faceted. No single measure could describe it adequately. Complexity pioneer and Nobel laureate physicist, Murray Gell-Mann, noted as much ``A variety of different measures would be required to capture all our intuitive ideas about what is meant by complexity'' \cite{Gell-Mann_1995}. Complexity is a collection of features, not a single phenomenon \cite{Ladyman2020WhatIA}. Therefore, a science of complexity must understand the various metrics available to quantify aspects of complexity, their capabilities and shortcomings. We have already encountered numerous measures of complexity in the literature review above, namely self-organization and power laws. Here we will review several other measures, filtered for inclusion based on their relevance to Heliophysics. The literature suggests that the degree of organization dimension perhaps dominates in the domain. We will not address the more computational/computer science metrics such as logical depth and algorithmic complexity. \cite{Mitchell2009ComplexityA} (Chapter 7) provides a more general development of the topic of complexity metrics. 

The most basic complexity metric is numerosity, or counting entities and interactions between them. Numerosity is common to all science. Next are measures of order/disorder in a system. Disorder is mathematically represented through probability distributions and their measures of dispersion such as variance. Related to these measures of disorder and one of the important core measures of complexity science is \textit{Shannon entropy}, measuring the amount of uncertainty in a probability distribution \cite{Shannon1948AMT}:
\begin{equation}
    H(X) \equiv - \sum_{x \in \mathcal{X}} P(x)logP(x),
\end{equation}

\noindent where $X$ is a random variable with probability distribution $P$ over events $x$. Shannon entropy quantifies the difficulty of predicting an actual outcome given possible outcomes ($x$) or, in a temporal context, predicting future outcomes given past events ($x$). Shannon entropy is inextricable from uncertainty quantification. In some domains, `diversity' is used as opposed to disorder. Shannon entropy is a part of most measures of diversity in these domains (e.g., ecology) \cite{Page_2011}. Numerosity and entropy allude to statistical physics, more generally, as an approach to quantifying complexity \cite{LpezRuiz1995ASM, Sethna2021StatisticalME}. 

The next feature of complex systems that requires measurement is feedback. Feedbacks are loops of interactions across a system. Feedbacks are created when a change in a component of a system affects the rate of change of that same component \cite{meadows2008thinking}. There is no measure of feedback. However, feedbacks support the persistence or disappearance of a behavior over time such that their signatures are present in the system. In the computational fields, feedbacks are defined by outputs of a process being put back into an input of the same process. Feedbacks reveal themselves in physical systems through patterns, structures, and nonlinearities such that measures to quantify the effects of feedback are those of structure formation and nonlinearity. Fractals themselves, well studied in Heliophysics as indicated by the volume of publications into their statistical signature--the power law--in the solar-terrestrial system, are a result of repeating a simple process over and over in a feedback loop. 

A widely used computational tool for studying feedback is the agent-based model, a simulation of collections of `agents' of entities of the system in which the agents follow certain rules for interactions and their evolution is studied. The closest example in Heliophysics are test particle simulations (e.g., \cite{Sorathia_2017}). Attempts to understand the societal impact of Heliophysics research, i.e., space weather, might be an arena for agent-based models (ABMs) of human behavior that might help understand preparedness to respond to space weather storms, especially in the context of potential compounding effects such as terrestrial weather or related system failure \cite{Mcgranaghan2022ConvergingTS}. This would be one way to integrate Heliophysics systems models with human behavior models. ABMs, capaciously defined, is an intriguing future direction for Heliophysics and space weather sciences. 

The results of ABMs are multidimensional data characterized by interactions between agents. Their structure is inherently a graph or a network. Therefore, making sense of those data depends on graph theory, which enjoy well-defined and mathematically rigorous formalisms. Section \ref{network science subsection} introduces graphs and discusses important quantitative measures to understand them. In short, graph theory provides a way to study the geometry and evolution of a network through centrality measures, community structure, and modularity \cite{Newman2010NetworksAI}. These measures offer deeper insight into a system and should be considered core metrics of complexity. 

Finally, among the most essential ideas in complexity science is self-organization, that order can spontaneously arise from many uncoordinated interactions \cite{Ladyman2020WhatIA}. Self-organization is measured through the order of the system. This is perhaps the measure best explored in Heliophysics to this point. Correlation measures, from linear Pearson correlation to covariance to information theoretic calculations like mutual information and transfer entropy, reveal order in a system. Nonlinear order is often studied with power law relationships. 

A way of measuring complex systems that has not been as widely explored is robustness and resilience. Robustness refers to the ability of a system to maintain its structure or function in the presence of perturbation. Sans the requirement that structure is maintained, resilience refers to the property of a system to accommodate changes and reorganize itself while maintaining the crucial attributes that give the system its unique characteristics \cite{Scheffer_2001}. Tools to study robustness and resilience include dynamical systems theory and theories of phase transitions \cite{Ladyman2020WhatIA}, such as stability analysis \cite{DEMIREL2019573}, critical slowing down \cite{Scheffer2009EarlywarningSF}, and tipping points \cite{Scheffer2009CriticalTI}. Resilience offers a way that decisions can be made based on complex systems understanding, perhaps permitting a new framework for bridging research and operations in Heliophysics and Space Weather. 

The measures identified above have in some manner been used in Heliophysics. However, the metrology of complexity is a young field, and therefore rapidly changing \cite{WOOD198660, Lloyd_2001, Zurek1990ComplexityEA,Gregersen2002FromCT, Mitchell2009ComplexityA, krakauer2018worlds}.  Paying attention to the advent of complexity metrics might lead our research to tools for better representing the phenomena we observe. 

The lesson from this brief enumeration of measures of complexity is that the multi-feature nature of Heliophysics now requires multi-faceted approaches to measurement and evaluation. The geospace community more recently has recognized this, advocating \cite{Liemohn2018ModelEG, Liemohn2021RMSEIN} and providing frameworks \cite{McGranaghan_2021} for more robust evaluation of our models across numerous metrics and levels. This review suggests that a similar approach must be taken for future work to quantitatively evaluate complexity in the system. 

\subsection{Coarse-Graining} \label{coarse graining subsection}

\cite{Chang_2007} describe `coherent structures' as a characteristic of dynamical complexity, phenomena that result from the nonlinear interactions of their constituent parts and are dramatically different than the behavior of those parts. The truism alluded to is that the whole is more than the sum of the parts. What the authors describe explicitly is present across the complexity Heliophysics literature, albeit often implicit and unnamed: there is a process of `coarse-graining' to identify the relevant macrostates of a system. \cite{Flack_2017} defines a coarse-grained description of a system as one in which some of the fine microscopic behavior of a physical system has been smoothed over. She emphasizes that this is a principled smoothing, not arbitrarily reducing the granularity, but instead based on whether information remains important to a descriptive or predictive task at hand and does not introduce outside information. She writes that this is a `lossy, but true' process. Coarse-graining is a process of integrating over parts, results in a compact representation of a system, and provides the basis for an effective theory. The preeminent example is temperature: a macroscopic description of a fluid that is an integration of microscopic behavior of particles. Averages are but one method of coarse-graining and there are many more, some much more complicated. The concept of coarse-graining is relevant across domains. In physics, renormalization theory is one remarkable example \cite{GellMann1954QUANTUMEA}. As early as the 1970s in biology, coarse-graining has been used in molecular modeling of biomolecules \cite{Levitt1975ComputerSO}. Even in art it has a long history in drawing attention to the scale at which one witnesses the world. The artist Piet Mondrian painted series of representations of trees, displaying the trees as increasingly geometric and abstract until the tree itself could scarcely be recognized \cite{Coppes2022BeyondCO}. His work, like representations in science, explore the inherent patterning and ordering of nature. The prevalence of coarse-graining in science and society suggests an essential role of this process and that a review of complexity in a domain of science should address. 

Despite the utility of macroscopic effective theories like thermodynamics and statistical physics, there exists long-standing debate in physics and biology about the right level at which to describe a system, or in how far `down' we need to go. Associated with this debate are questions related to mappings between microscopic to macroscopic states. Though much of this debate has been staged in fundamental physics and biological domains, it speaks to unresolved issues in Heliophysics \cite{Lui_2001, Denton_2016, Denton_2021, Viall_2020}, conversations punctuated by a perceived need to reconcile particle-level behavior with system-level phenomena and multiscale quandaries. The debate has shown up in contrasting modeling approaches, e.g., particle-in-cell and magnetohydrodynamic ways of describing the solar and magnetospheric systems. On the multiscale understanding side, questions abound about the `right' level at which to look at the system and how to study the relationships between scales. The literature suggests that what is needed are hybrid models (e.g., note progress made in unifying particle and magnetohydrodynamic modeling \cite{Sorathia_2017}) and new methods for conducting multiscale analyses (e.g., explore the advances made in comparing scales and studying relationships between scales \cite{McGranaghan_2017a, Nishimura_2021, Consolini_2021, Nishimura_2022}). It is quite possible that multiscale understanding and ideas about the right level at which to represent the Heliophysics system will bear on the outstanding desire to bridge the gap between research and operations \cite{Mcgranaghan2022TheEO}. Research and operations function at different scales. Where research may need to look at the finest scales technologically possible to advance the boundaries of knowledge, operations requires an efficient representation of the adequate knowledge to make a decision. In some sense, the research to operations (and operations to research) gap is the problem of developing an effective theory or a compact representation that permits moving between scales. Presaging a discussion that concludes this review, this is the same tension that exists between scientific understanding and prediction. These separate regimes of knowledge discovery and science dictate concomitant separate regimes of representation. In Section \ref{key challenge section} we center this discussion in terms of fundamental science vs. prediction-oriented science (e.g., basic science vs. applied science; physics-based modeling vs. artificial intelligence/machine learning) and suggest this is inextricable from the future of Heliophysics research. 

In the sense that coarse-graining is capturing underyling structure and pattern in complex systems, two forms of coarse-graining are particularly important to Heliohpysics: information theory and network science. We discuss them in sequence next.


\subsection{Disentangling drivers and parameters amidst nonlinearities: Information theory} \label{information theory section}

Like coarse-graining, information theory is often used to simplify a complex system, to understand the more parsimonious description of its functioning. Indeed, in the context of entropy, the two concepts are quite similar. There is a growing body of research within Heliophysics that suggests that information theory is a useful form of coarse-graining for our field. 

When a system's drivers are nonlinearly correlated, and the parameters of the system that they effect are numerous, it is a challenge to untangle their relative effects. There are an immense number of frameworks with which to find and investigate causality relations among different time series \cite{Runge2019InferringCF}; information theory has proven powerful for the detection of causal information flows in complex systems. Information theory provides a mathematical framework for quantifying nonlinear flow of information from drivers to system parameters and between system parameters. It assumes that a given domain of interest can be described by coupled subsystems that interact (exchange information) with one another. Information theoretic approaches then use a number of possible measures to extract the direction of the information flow and to infer causality. 

Recent literature has revealed promise for information theory to provide observational constraints that can help guide the development of the theories and physics-based models and for feature selection to create more accurate data-driven models \cite{Wing_2019}. 

\cite{Stumpo2020MeasuringIC} discusses a method for quantifying the strength and direction of the coupling between the solar wind and the magnetosphere-ionosphere system. The authors introduce a new measure of information transfer to the solar wind-magnetosphere-ionosphere domain called the Transfer Entropy (TE), which is useful for the nonlinear analysis of the relationship between two time series. TE is based on transition probabilities between two random processes, $X$ and $Y$, obtained by inserting the Markov condition into the conditioned Kullback-Leibler distance such that the information flow from $X$ to $Y$ is accounted for. TE is a quantitative way of determining if the past history of $X$ is predictive of the future $Y$. TE also provides a means to distinguish bidirectional information flow, thus providing evidence about feedback processes, mentioned above as a difficult thing to measure in complex systems. They show that TE is a useful measure for capturing relationships between the solar wind and magnetosphere. They showed a strong information transfer from the vertical component of the interplanetary magnetic field $B_Z$ into the geomagnetic indices, with time delays of about 30 to 60 minutes. Further, they inferred that substorms drive geomagnetic storms from a strong observed information flow from the AE index into the SYM-H index (analogous to the Dst index). 

Similarly, \cite{Manshour2021CausalityAI} were interested in measures from information theory that could illuminate the relationship between the solar wind and magnetosphere-ionosphere system. Their measure was Granger causality. They found a tighter temporal relationship between $B_Z$ and the AE index, a delay of only 10 minutes, and a 30 minutes delay with SYM-H commensurate with \cite{Stumpo2020MeasuringIC}. They, however, did not find any relationship between AE and Sym-H, casting uncertainty on the claim that substorms might drive geomagnetic storms. 

In addition to solar wind-magnetosphere coupling applications, two areas of Heliophysics have demonstrated the physical discovery potential of information theory: radiation belt dynamics \cite{Wing2016InformationTA} and solar cycle dynamics \cite{Wing2018InformationTA}. Researchers in the decade following 2010 have been applied the theory to understand the influence and the timing of the solar activity on the near-Earth environment, extending its utility to the space weather domain \cite{Materassi2011PredictiveSW}. 

A caution and a difficulty of information theoretic analyses is that they require robust support in the available data. Care must be exercised in assessing the data in numerous ways, including: 1) data sufficiency: the available data must be statistically representative of the system, sampling the full space; 2) relevant variables: the data must include all relevant variables that affect the transmission and reception of information in the system; and data diversity: the data must be diverse enough to capture the full range of possible transmission scenarios and conditions, which often requires integrating data from multiple platforms and sources. 

Information theory is promising, but realizing its potential suggests that Heliophysicists develop or strengthen certain literacies such as probability, statistics, noise quantification, and systems science.

\subsection{Network Science: A future for Heliophysics and Space Weather} \label{network science subsection}

Networks surround us. Network structure, or a high number of dynamic interacting units, characterizes much of our society, from molecules that constitute biological organisms to our communication infrastructure like the internet to the power grid. Networks even describe the structure of how we interact with one another--social networks. We have been grappling with the prevalence of networks in our society for many years \cite{Milgram1967TheSW, Granovetter1973TheSO, Erdos1984OnTE} and more recently recognizing their capacity to represent the complex world around us \cite{Boccaletti_2006, Watts1998CollectiveDO, Barabasi2002, Newman2010NetworksAI}. 

The reason for this is that networks are how complex systems are represented \cite[and references therein]{Torres_2021}. Networks are the lingua franca of complexity, so to speak. The profound importance of this discovery is that network analysis rests on a well-defined domain of discrete mathematics known as \textit{graph theory} (graph, in this context, is synonymous with network), dating back to Leonhard Euler's solution to the K{\"o}nigsberg bridge problem \cite{biggs1986graph}. Thus, if one can figure out how to encode a system as a network, then a robust domain of mathematics can be applied to study it and discovering important properties such as community structures, key nodes in the flow of information in the network, and the type of network that reveals properties of its functioning (e.g., random networks \cite{Erdos1984OnTE}, small-world networks \cite{Watts1998CollectiveDO}, and scale-free networks \cite{Barabsi1999EmergenceOS}).  

First, a very brief primer on graphs and networks. We will hereafter use the term \textit{network} to represent both--indeed they are synonymous, with the only difference being that some communities prefer graph (e.g., mathematics) while others prefer network (e.g., social sciences). Networks show connections between things. The `things' are called nodes or vertices and can represent people as in a social network or any other definition of the entity for a domain. The connections are called edges and they represent a relationship between nodes. In a social network the vertices might indicate if two people know one another. The quantitative means that one chooses to define a vertex is central to the construction of the network. 

Another key concept is the adjacency matrix. The adjacency matrix is a square matrix with rows and columns corresponding to every node in the network. The corresponding matrix element will be either 1 or 0 according to whether those nodes are connected or not. This is one form of visualizing or understanding the network. Other means to derive important sub-networks (collections of a few of the nodes of the full network and their connections) and aggregations of the network are available and important given the complexity that most real-world networks display. The purpose of these aggregations are to permit an understanding of the network topology or geometry.

\cite{Torres_2021} provide a review of methods to encode a system as a network. Graph theory applied to complex systems, possessing irregular structure and dynamically evolving over time, gave rise to the term `complex network analysis.' In this review, we use the parsimonious `network analysis' to encompass network studies of any sort. The methods for encoding and subsequently analyzing systems as networks have been empirically, if nonlinearly and cobbled together from across disciplines, determined and were accelerated in the 20th century. The 1920s witnessed a sophistication of social network analysis--networks of relationships among social entities (e.g., trade between nations, communication between members of a group) \cite{Milgram1967TheSW}. It has found more recent application in biological, engineering, and geophysical systems \cite{Tsonis_2006, Donges_2009,  Steinhaeuser_2011, Malik_2011}.

The power of the network representation is that understanding and approaches from the mathematical field of graph theory can be used to explore and understand the networks. Indeed network measures exist to define the geometry and evolution of the network. Responding to Section \ref{metrics and diagnostics subsection}, network measures are being used as new metrics for complexity, in effect means to aggregate the system in less lossy ways (e.g., better coarse-graining). One could consider three levels of measuring a network: 1) micro: the level of nodes and edges; 2) macro: distributions aggregating quantities; and 3) mesoscale: a large class in between \cite{Porter2009CommunitiesIN}. 

At the micro level, one considers the nodes and edges individually or in small groups. A node's \textit{degree} is the number of edges it has. At the macro level, one aggregates the entire network and studies statistical distributions that attempt to describe it. Common macro measures include diameter (the length of the longest geodesic path between any pair of nodes in the network for which a path actually exists), average path length (average shortest path between all pairs of nodes in the network), degree distribution (the frequency distribution of the degree of the network nodes), and clustering coefficient (average probability that two neighbors of a vertex are themselves neighbors). Together, these are ways of understanding the \textit{geometry} of a network, which is the complement of it's size, connectivity, efficiency, and homogeneity/heterogeneity. The characteristics of the degree distribution, including its higher order moments, are a fundamental way that networks of different types and behavior are distinguished and it conveys information about the functioning of the system. 

The mesoscale level covers everything in between. A topic that has received much attention in the network science literature is \textit{centrality}, which is the attempt to quantify the most important or central nodes in a network. There are numerous different ways to think about and thus to calculate centrality, some quite simple like degree centrality (looking at the degree of a node with respect to others in the network), and others more involved and considering neighborhoods around a node such as eigenvector centrality, betweenness centrality, and closeness centrality \cite{Newman2010NetworksAI}. Another mesoscale structure that receives much attention is the \textit{community} (a group of nodes relatively densely connected to each other but sparsely connected to other dense groups in the network \cite{Fortunato2009CommunityDI, Porter2009CommunitiesIN}). Identifying communities in networks has significant implications for making discoveries about systems, though there is no consensus on technique for detecting them. It is perhaps in the mesoscale where undiscovered insight into Heliophysics systems and processes lays. Indeed, this is reflected in the Heliophysics network science literature reviewed below. 




The ability to capture multiscale relationships and behavior using a network structure or representation provides an exciting opportunity to improve multiscale understanding of the Heliophysics system \cite{McGranaghan_2017c}. 

Beginning in the mid-2010s a few pioneering works began to realize the potential of network analysis for space physics, Heliophysics, and space weather applications. Traditional approaches to systems analyses in Heliophysics and space weather have attempted to track energy flow from the Sun to the Earth or other planet through a collection of time series. This approach does not generalize to multi-event studies nor to extracting statistics. Networks liberate the systems approach from the limitations inherent in this approach. They allow one to track energy flow and dynamic changes across a system for multiple events in a principled manner and to quantify them using well-established measures from graph theory and network analysis. Not only do network approaches reveal new parameters by which to understand the system, they enable quantifying the likelihood of those measures that can become the basis of a risk quantification system \cite{simpson2021framework} and ultimately understanding how to create technologies and society resilient to the threats of space weather. Thus, we believe it important to provide a brief review of those works here. We anticipate that this field will evolve rapidly in the coming years and intend to provide a basis for researchers to become oriented and trace some of the important history here. 

\begin{figure}[h]
\centering
\includegraphics[width=\textwidth]{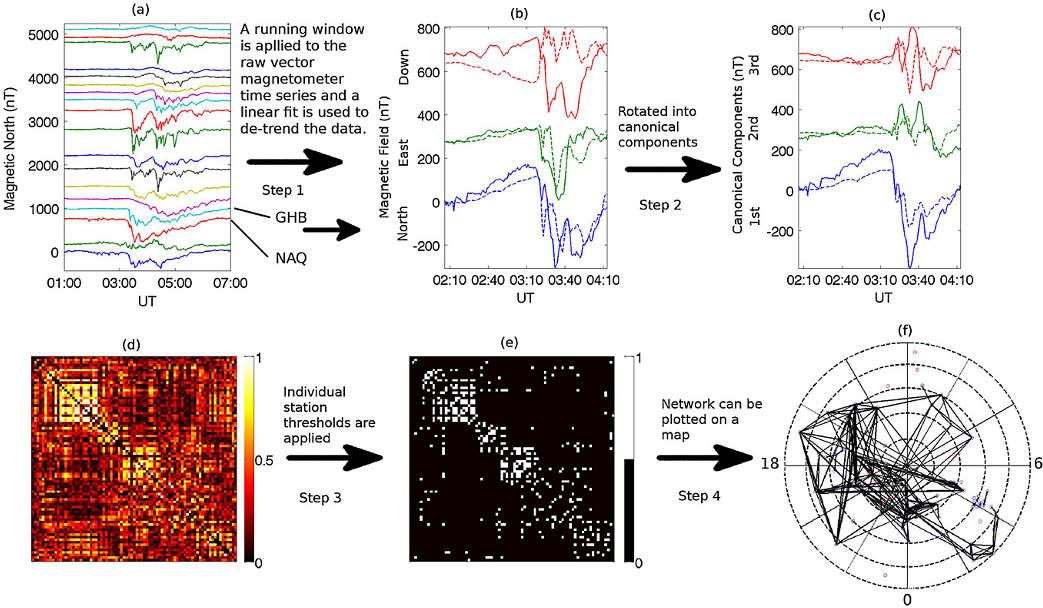}
\caption{Figure 1 reproduced from \cite{Dods_2015}. (reproduced figure caption) Illustration of SuperMAG network construction: (a) Stack plot of magnetic north time series for stations centered around magnetic midnight during the substorm event. The time series are ordered by magnetic latitude. (b) A comparison of a 128 min segment of the north (blue), east (green), and down(z) (red) component of the magnetic field for two stations: GHB (dashed) and NAQ (solid). A linear fit was used to detrend the data within the window. (c) Canonical correlation was used to form new rotated components that maximize the correlation in the first canonical component for this time window. The rotation is unique for each station pair and time window. Correlation between different canonical components is zero as both the cross and autocovariance for the canonical components matrices are diagonal. (d) The canonical correlation process is repeated for all station pairs, and a correlation matrix, $\mathbf{C}_{ij}$, can be formed. The matrix contains the correlation coefficients for the first canonical component. (e) Station-dependent thresholds are applied to $\mathbf{C}_{ij}$ to form the adjacency matrix $\mathbf{A}_{ij}$. The white squares indicate a connected station pair. Connections can be visualized on a MLAT-MLT map. (f) The magnetic north down view of the Northern Hemisphere, the blue circles indicate active stations, and the red circles are stations for which there are no data at this time. The dashed lines are contours in MLAT, at 50$^{\circ}$, 58$^{\circ}$, 66$^{\circ}$, 74$^{\circ}$, and 82$^{\circ}$.}
\label{Dods2015_Fig1}
\end{figure}

\cite{Dods_2015} applied network analysis to $>$200 distributed ground-based magnetometers that are indexed in the Super Magnetometer Initiative (SuperMAG) \cite{Gjerloev_2009}. Already a debate in the Heliohpysics community whether ground-based indices such as the AE and Disturbance storm-time (Dst), themselves aggregates of small numbers of ground-based magnetometers, were capable measures for understanding magnetospheric dynamics, Dods et al. studied whether a network representation of ground-based magnetometer data can quantitatively extend our qualitative understanding of magnetospheric substorms, creating the first application of network analysis to SuperMAG data and one of the early applications to space weather in general. The observations are vector magnetometer time series data at 1 minute cadence from the SuperMAG database (all stations from the northern hemisphere). Canonical correlation \cite{Brillinger2001TimeS} was used to establish similarity between the pair of vector time series as a function of time. To construct the network, they defined magnetometers as nodes and correlation above a threshold between the vector magnetometer time series from pairs of stations within a running time window as edges. Figure \ref{Dods2015_Fig1} reproduces their visual explanation of the network construction. They investigated four substorms. For each event, they form dynamical networks of connected stations in magnetic local time (MLT)-MLAT space in the Northern Hemisphere. 

Importantly, the correlation threshold between any two stations might be different, so they created a global threshold that effectively normalizes the likelihood of being connected to the network. This is an important step of network analyses for Heliophysics and Space Weather applications where observations are distributed and baselines are commonly distinct. 

They are able to then construct networks for any given event. Networks are reconstructed in the paper for four selected substorm events (defined according to \cite{Gjerloev2014TheLC}) and one steady magnetospheric convection (SMC) event. From each network dimensionless parameters were obtained that quantitatively characterize the network and by extension, the spatio-temporal dynamics of the substorm under observation. They found several typical signatures of the isolated substorm:
\begin{itemize}
    \item Before onset, the network has few connections; 
    \item Connectivity rapidly and clearly responds to the onset, characterized by high-latitude connections, but not without low- and cross-latitude connections; 
    \item In the recovery phase, connection structure switches from high-latitude dominant to low-latitude dominant; and 
    \item The normalized total number of connections and the average geodesic connection distance (physical distance) of the substorm period networks are greater than those for the SMC event and much greater than during quiet times. 
\end{itemize}

Thus, they discovered that network responses to substorms, SMCs, and quiet times are quantitatively distinct. They conclude also that their technique may have applicability to other magnetospheric phenomena. 

That supposition was examined by \cite{Dods_2017}, who apply the same methodology to the response of the quiet time (no substorms or storms) large-scale ionospheric transient equivalent currents to north-south and south-north IMF turnings. The calculation of the correlations between station pairs is identical to \cite{Dods_2015} but they also map the network onto a regular grid and aggregate the network responses over more than 350 events (between 1998–2004) to obtain an averaged response as a function of geomagnetic location (MLT-MLAT) and of the time delay since the occurrence of the IMF north-south and south-north turnings. 

For both north-south and south-north IMF turnings they examined short-range (station-pair connections with geodesic separation $<$4000 km) and long-range ($>$4000 km) connections. Their results indicate magnetometer correlation network responses are distinct for north-south and south-north turnings and the sign of IMF B$_{\textrm{Y}}$ component. Demonstrating the potential of network analysis, they provided new information on two competing concepts for reconfiguration of ionospheric currents in response to a change in the north-south component of the IMF: a fast initiation of the transient currents associated with ubiquitous and near-simultaneous response at the high-latitudes (e.g., \cite{Ridley1997IonosphericCD}) and a gradual reconfiguration spreading from an initial response on the dayside followed more gradually by the nightside (e.g., \cite{Lockwood_1986}). The network response shows near-simultaneous responses ($\sim$8-10 minutes) between magnetopause impact and network response, consistent with \cite[and references therein]{Ridley1998ASS}. They discuss tentative evidence for a two-step process: fast initiation of change in ionospheric equivalent currents between day and night followed by a more gradual reconfiguration that first appears on the dayside and then the nigthside. 

Using spatial maps of the edges of the network, they found that turnings are associated with increases in connectivity (correlation) in the areas known to be associated with the two-cell convection pattern \cite{Dungey_1961}. Ultimately, this was one of the first studies to reveal that dynamic correlation networks can characterize the spatio-temporal ionospheric response observed in large numbers of ground-based magnetometers. 

\cite{Orr_2019} advanced the analysis of the spatio-temporal evolution of substorm ionospheric current systems in SuperMAG data with networks by introducing lags in the canonical cross correlation. Considering lags, rather than zero lag correlation as Dods et al. had done, permitted construction of a \textit{directed} network that captured not only the formation of coherent patterns observed by magnetometers but also the direction of information propagation of those coherent structures. They used the directed networks to test different proposed methods for how the ionospheric current system evolves during a substorm. 

To assess the direction of propagation, they divide the nightside auroral ionosphere (18 MLT to 6 MLT, passing through midnight; 60-75$^{\circ}$ MLAT) into three zones of six hours MLT, a typical extent of the substorm current wedge (SCW) \cite{Gjerloev2007StatisticalDO}.  

They conclude that the magnetic perturbations are consistent with the SCW formation during substorm onset and westward expansion into a coherent current system in the premidnight (MLT) sector (see their Figure 2). Subsequently, a coherent correlation pattern emerges that spans the entire nightside ionosphere. 

\begin{figure}[h]
\centering
\includegraphics[width=\textwidth]{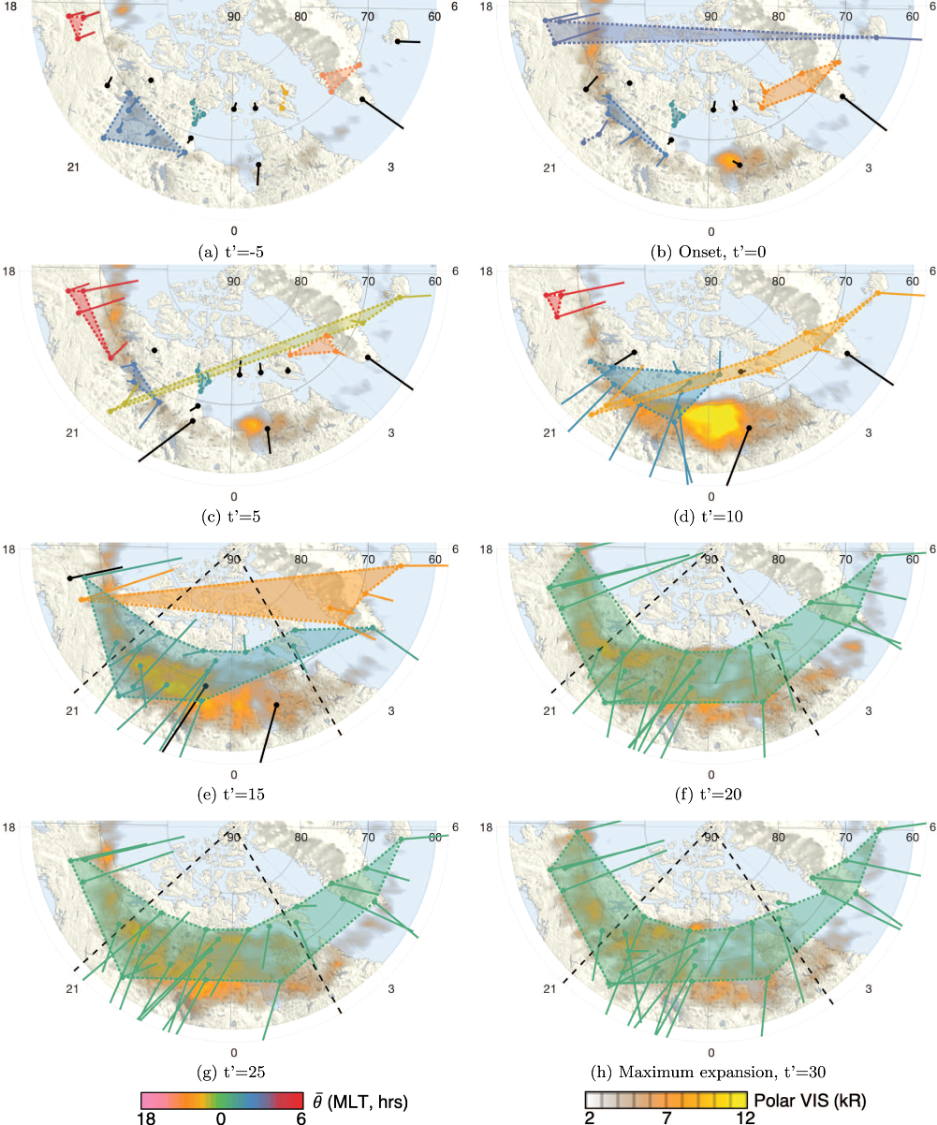}
\caption{Figure 2 reproduced from \cite{Orr_2021b}. Snapshots revealing evolution of community structures from numerous small-scale communities to coherent large-scale community across normalized substorm time. (following text reproduced from caption in\cite{Orr_2021b}) plots are in magnetic coordinates entered at the magnetic pole, where magnetic local time (MLT, h) increases clockwise, with midnight located at the bottom (MLT=0 h). Maps show the nightside from dusk (MLT=18 h) to dawn (MLT=6 h) and 60-90$^{\circ}$ magnetic latitude. Plotted are the magnetic field perturbation vectors (North and East components, B$_{N,E}$, measured in nT) for a substorm on 16/03/1997. The colorbasrs at the bottom of the figure represent the MLT of the centroid ($\overline{\theta}$(t')) of each community, and polar VIS data from left to right, respectively.}
\label{Orr2021b_Fig2}
\end{figure}

\cite{Orr_2021b} took the analysis a step further, taking advantage of a wider set of network science analysis techniques to understand the properties of a network. Namely, they studied community structure in the SuperMAG networks in which a community is defined consistently with \cite{Porter2009CommunitiesIN}: an area of a network more densely connected to one another than to the rest of the network. They detect communities in SuperMAG networks across 41 isolated substorms with 1-min resolution data. Primary results are illustrated in Figure \ref{Orr2021b_Fig2}.

In the networks, multiple discrete current systems exist prior to onset (see Figure \ref{Orr2021b_Fig2}a-d) and progressively transition into a coherent SCW (see Figure \ref{Orr2021b_Fig2}f-h), notably a transition to a coherent large-scale spatially extended structure rather than flux accumulation of incoherent small-scale wedgelets. The same pattern is observed across numerous algorithms for community detection. Thus, the SCW is a characteristic part of substorm evolution, potential resolution for long and ongoing controversy in substorm science. The spatially extended communities they observed cannot be obtained by having many, small, spatially localized wedgelets, which are internally correlated, but lack cross correlation with each other. 

Of immense societal relevance for ground-based magnetometer observations of space weather activity are the corresponding potential hazard to grounded technology like the power grid. The threat to the power grid is quantified by geomagnetically induced currents (GIC).  A pair of studies have explored network analysis for GICs, both providing insight into the GIC hazard and revealing network analysis as important for bridging from Heliophysics insight to space weather risk assessment and societal relevance. \cite{Hughes_2022} produced networks connecting SuperMAG magnetometers to newly available GIC data collected by power utilities through the the Electric Power Research Institute (EPRI) SUNBURST project. They calculate probability multipliers for all pairs of magnetometer-GIC sensors, information that would be useful to using magnetometer observations to determine risk the power grid. Overall, there is a factor of 1.83 increase in the GIC increase given magnetometer changes. On a sensor-to-sensor comparison, however, the magnetometers that provide the most information for a given GIC sensor are often not those in closest geographic proximity, meaning networks reveal non-intuitive relationships. 

\cite{Orr_2021GIC} used SuperMAG ground magnetic perturbation measurements as input to a model of the high-voltage (HV) power grid in the United Kingdom (UK), which output GICs at the grid transformers. They quantified the spatio-temporal response of the GICs in a manner similar to \cite{Orr_2019}. A number of conclusions were drawn, including:
\begin{itemize}
    \item the entire physical power grid is spanned by coherent connections with long-range correlations at intense storm times;
    \item the GIC networks are not a simple response to the rate of change of the magnetic field; 
    \item during storms, networks have intermittent quiescent periods in which distinct sub-networks form; and 
    \item GIC networks are distinct from the physical networks of the HV grid, exhibiting characteristics unlike the exponential and small-world nature of the physical grid. 
\end{itemize}

Their work offers a direct connection to space weather risk assessment: ``The GIC response networks that we have determined here have significantly different properties to that of the physical HV grid. This is important since it implies that previous studies that focus on stability of the physical grid to the failure of individual network connections may not fully inform the assessment of space weather risk.''

\begin{figure}[h]
\centering
\includegraphics[width=\textwidth]{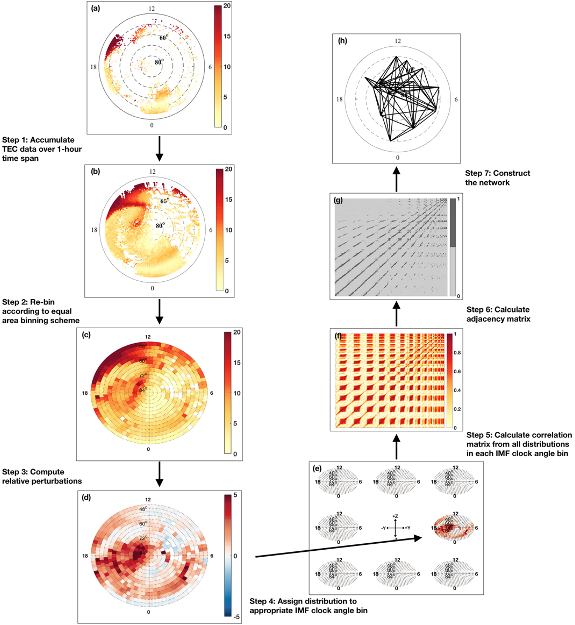}
\caption{Figure 2 reproduced from \cite{McGranaghan_2017c}. (reproduced figure caption) An illustration of the network construction process. (a) Sample 5 min 1$^{\circ}$ latitude × 1$^{\circ}$ longitude total electron content (TEC) distribution from the worldwide system of ground-based receivers used by the Madrigal upper atmospheric science database. These data are shown in magnetic latitude (MLAT)-magnetic local time (MLT) Altitude Adjusted Corrected Geomagnetic Coordinates (AACGM) with noon MLT to the top of the polar plot and a low-latitude limit of 45$^{\circ}$. The data plotted are TEC units (TECU), 1 TECU =10$^16$ el/m$^2$. (b) AACGM TEC data accumulated over a 1 h period. (c) TEC data distribution rebinned onto an equal area grid (see text for details). (d) TEC relative perturbation data (equation 1). All figures in the remainder of the manuscript use the same 45–90$^{\circ}$ MLAT range shown in Figures 2a–2d. (e) Visual showing the assignment of a given relative perturbation distribution to a particular interplanetary magnetic field (IMF) clock angle bin. For all figures in this paper the IMF clock angles increase in 45$^{\circ}$ increments in the clockwise direction. (f and g) Example correlation and adjacency matrices. (h) Visual illustrating the resultant network where lines have been drawn to indicate that two grid points are connected based on the spatio-temporal correlation between them.}
\label{McGranaghan2017c_Fig2}
\end{figure}

Concurrently and drawing inspiration from Dods and Orr et al.'s revelations about the utility of networks in geospace science, \cite{McGranaghan_2017c} applied network techniques to another important data set: total electron content (TEC). Global, high-latitude response of TEC is the result of numerous complex geospatial processes, each with unique spatial and temporal scales \cite{Mendillo2006TotalEC, Shim2009AnalysisOT, Emardson2013SpatialVI}. Despite being rich with information about the Earth's space environment, their characteristics at high latitudes are not well understood, and the complex nature of the processes in this regime requires innovative and sophisticated approaches to (1) understand the information content of these data and (2) gain the most scientific utility from them. These were motivation to attempt to understand the spatio-temporal characteristics of TEC in the high-latitude regime. In their application, nodes were defined by the magnetic coordinate system grid points (physical locations) and edges by spatio-temporal correlations exceeding a threshold between them. It was the first application of such techniques to TEC data. Their data are hourly averages of TEC data compiled from the worldwide system of ground-based GPS receivers binned into a geographic 1$^{\circ}$ latitude $\times$ 1$^{\circ}$ longitude grid (rebinned in this work into equal area magnetic coordinates). Data from winter and summer seasons in 2016 were used. Data were studied separately for the northern and southern hemispheres and all data were separated into interplanetary magnetic field (IMF) clock angle bins, the angle between geocentric solar magnetic (GSM) north and the projection of the IMF vector onto the GSM Y-Z plane, to determine dependence on the solar wind forcing. Figure \ref{McGranaghan2017c_Fig2} visually details the network construction steps. 

Using predominately the network measures of degree centrality, median geodesic separation distance, and local clustering coefficient, their analyses suggested that the Northern Hemisphere exhibits correlations over shorter distances but is more spatially uniform while the Southern Hemisphere correlations typically extend over larger distances but that the hemisphere as a whole is more spatially fragmented. Their resultant maps indicate the scale sizes important to characterize the ionosphere during geomagnetic activity depend on season and hemisphere. The proof of concept study was exciting and pinpoints several ideas for follow-on inquiry.

These studies provide a framework through which to analyze the complex magnetosphere-ionosphere-thermosphere system free of the limiting assumption that phenomena can be described by interpolating distributed observations onto a grid. As \cite{Dods_2017} write network analysis enables the characterization of, ``... the spatio-temporal correlation pattern for [any] event directly from the spatially nonuniform original observations, then aggregate many such patterns onto a single grid to give a complete spatial coverage.''

How does network analysis relate to risk and resiliency? Risk is probability times likelihood. Risk is a distribution. In the terms of the insurance industry, hazard is actual cost incurred from a risk. To effectively quantify risk for space weather we must use the most informative parameters and calculate their likelihoods. The important parameters will be hazard-specific (e.g., the relevant parameters for power grid risk will be different from those for communications systems risk). Traditionally, and as a result of observational limitations, we have relied on indices like Dst and AE as the important parameters. Yet we know them to contain inadequacies (e.g., see Section \ref{setting the stage section} and the discussion of the use of AE to represent the magnetosphere). Network measures are potentially much more directly related to the physical phenomena relevant to a given risk and thus an exciting pathway to better risk quantification, upon which resilience is built. 

Finding new applications for Heliohpysics is a trend that continues, particularly in connecting fundamental research with applied outcomes as evidenced by \cite{Orr_2021GIC, Hughes_2022}.





\subsection{The role of Natural Language Processing} \label{NLP subsection}

Natural Language Processing (NLP) is concerned with programming computers to process, analyze, and respond to large amounts of natural language data. The importance of augmenting human research and activities with automated analysis of text is now undeniable, given the sheer volume of relevant scientific literature. It is beyond the capacity of an individual researcher to understand all of the relevant scientific information about a topic. The growth rate in scientific literature is exacerbating and compounding the problem \cite{Bornmann2020GrowthRO}. For inter- or trans-disciplinary science, the kind that complexity demands,  the amount of relevant information grows exponentially simply through the need to incorporate more than one domain's knowledge. To some extent the history of disciplinary science has been to put up boundaries on the science one attempts to answer as a means of reducing the seemingly infinite amount of information that must be considered. Complexity science exposes those boundaries as artificial. Therefore, it requires new methods for handling more voluminous information. This review has already demonstrated some of those methods for data analysis, but NLP is vital for textual analyses. 

There are many applications of NLP that have already or are poised to have an impact on scientific research and the process of science. Common tasks include, each applied to a given piece of natural language (e.g., a publication):
\begin{itemize}
    \item Named entity recognition (NER): identify and locate entities in unstructured text such as person, organization, location; 
    \item Information retrieval: searching for information contained in a document;  
    \item Keyword generation: extract or identify the most relevant words, phrases, or ideas in a document; 
    \item Summarization; 
    \item Sentiment analysis: identify the affective state and the subjective foundation of a piece of text; and 
    \item Question and answer: return an answer to a natural language question a user poses based on a document or collection of documents. 
\end{itemize}

Many of these `downstream tasks' \cite{Bommasani2021OnTO} rely on a language model. A language model is a model that assigns a probability to a sequence of words \cite{Jurafsky2000SpeechAL}. Language models take a collection of words and conditioned on that information assign the probability of another word or collection of words. For a simple example, perhaps you want to predict the next word in a sentence given the ones that precede it. A language model can fulfill that task. However, this basic functionality can extend to much more complicated tasks such as taking an entire document and predicting descriptive keywords or creating a summary. 

For a number of reasons modern language models have been changing rapidly. First, the growth of textual data on the internet is growing exponentially, providing vast volumes of data for training these models, which typically have millions of parameters that need to be determined and require heretofore unimaginable volumes of data to constrain. Second, computational power is making it possible to process those data. Finally, AI research produced new modeling approaches such as recurrent neural networks \cite{Rumelhart1986LearningRB}, transformer architectures \cite{Vaswani2017AttentionIA}, and self-supervised learning (e.g., \cite{Baevski2020wav2vec2A}). Together, these influencing factors produced step changes in performance of language models on many tasks \cite{Tamkin2021UnderstandingTC}. Improvements in capability coupled with wider awareness owing to much greater accessibility of language models through services like ChatGPT\footnote{\url{https://chat.openai.com/chat}}, early 2023 has become a cultural moment for NLP and language models (e.g., \cite{Klein2023}). 

In 2018 Google developed the Bidirectional Encoder Representations from Transformers (BERT) language model \cite{Devlin2019BERTPO}. Judging the general language models incapable of deeply contextual scientific support, scientists subsequently recognized the opportunity to tailor the baseline models like BERT for their domains or narrower applications. The first result was SciBERT, a fine-tuning of the BERT model that focused on scientific papers from the Semantic Scholar corpus\footnote{\url{https://www.semanticscholar.org/}} \cite{Beltagy2019SciBERTAP}. SciBERT is part of a growing list of models that adapt BERT to specific domains and tasks, for which perhaps the most relevant to this review is that created for the astrophysics and astronomy domain, astroBERT \cite{Grzes2021BuildingAA}. One of the most sophisticated examples, likely due to the availability of large volumes of training data in the biology and biomedical domain and a more sophisticated baseline language model from a family of models known as generative pre-trained transformers (GPT) \cite{Radford2018ImprovingLU}, Bio-GPT has demonstrated improvement across six biomedical NLP tasks \cite{Luo2022BioGPTGP}. Progress in Heliophysics is nascent \footnote{see `SMD Knowledge Graph Discovery` at \url{https://www.calameo.com/read/0055032805743f9fd8bf6}} and most of the NLP work being done is on downstream tasks using existing language models rather than training our own. Despite not yet realizing the potential that some have stated exists for these domain-specific models, the concept of \textit{foundation models} \cite{Bommasani2021OnTO} remains enticing. Capable language models for Heliophysics research are far from settled and available, yet may be a core component of searching and discovering vast amounts of literature and research artifacts available. 

Encompassing the diverse disciplinary integration needed for complex systems analyses will require sophisticated NLP capabilities, including perhaps entirely new approaches to building language models. This review combined traditional literature review processes with NLP to create a hybrid review article. The NLP used was relatively simple, mining the NASA Astrophysics Data System (ADS) for articles, two-fold filtering based first on selected Heliophysics journals and second on a manually-created Complexity Heliophysics glossary, but the approach created a much larger corpus and discovered articles that were not included in the hand-selected corpus for this review. Thus, the hybrid approach produced a richer coverage of the literature. Appendix \ref{corpus appendix section} describes the NLP approach. 

It is useful to future Complexity Heliophysics work to chronicle briefly the outcomes and open questions from the application of NLP to augment this review. A more detailed taxonomy of uses for NLP in scientific research has emerged. For encoder-like language models, tasks fall generally under five categories:
\begin{enumerate}
    \item Question answering;
    \item Text classification; 
    \item Semantic equivalence; 
    \item Named entity extraction; and 
    \item Knowledge extraction. 
\end{enumerate}
For generative models, language models support tasks related to conversational artificial intelligence, conversion of data to text, and text summarization \cite{Ramasubramanian2020SurveyingTM} (and \textit{personal correspondence: Muthukumaran Ramasubramanian; March 2023}). 

There are also many open questions. A few of importance to Heliophysics are: 1) given that Heliophysics is inherently a systems science, how might we extract relationships from scientific literature that guide us to incorporate new knowledge in our science?; 2) what role might NLP play in improving our information search and discovery processes?; 3) to what extent can NLP support efforts to better represent Heliophysics knowledge in human- and machine-readable ways (e.g., support building semantic technologies and capabilities) \cite{Biffl2016SemanticWT}?



The advent of large language models (LLMs) and the discussion of their use in scientific research is pointing to a larger conversation that is unfolding in the future of science: what is the intersection of complexity science with AI/ML? We deliberate on this question in Section \ref{key challenge section} with the hope of seeding important conversations Heliophysicists need to have, placing it in the context of conversations all scientists are grappling with.







\subsection{Areas of complexity science that have not yet been widely explored in Heliophysics} \label{unexplored subsection}

There are tools viewed by the complexity science community as necessary to understand complexity \cite{krakauer_parallax, Hobson2018RethinkingAS} that have not yet been widely employed in Heliophysics. Two areas conclude this section on emerging literature. 

First, agent-based modeling (ABM). ABM is a model that simulates the actions and interactions of agents, most traditionally autonomous, individual elements with properties and capable of actions. ABMs have been used extensively in the social sciences, biology, and ecology \cite{Niazi2011AgentbasedCF} and their efficacy owes to their combination of elements of game theory, sociology, evolutionary programming, and emergence. Their operating principle is to give agents relatively simple operating rules, simulate their interaction, and study the emergent collective phenomena. Though ABM is often associated with modeling behavior of living organisms, perhaps more important to Heliophysics it is also a technique for modeling the behavior and interactions of things such as particles in the magnetosphere. Particle simulations, therefore, can be understood as a form of agent-based model and the connection might allow computational approaches to Heliophysics learn from a rich domain of research. There may also be application agent-based modeling for human responses to events such as space weather storms, though outside of `simulation game' activities \cite{Mcgranaghan2022ConvergingTS} this is virtually unexplored. However, understanding human behavior in disaster situations is an important component to quantifying risk and establishing resiliency in our societal systems. 

Second, collective intelligence. If ABMs are the tools, collective intelligence is the study of their output data. It is a nascent field of study to understand collective behavior, that is adaptive, wise, or clever structures and behaviors by groups, in physical, biological, social, and many engineered systems \cite{Flack2022EditorialTT}. Discovery in disciplines as diverse as biology and ecology to psychology and economics point to cross-disciplinary utility. At present there is exists essentially no conversation about the use of methods and techniques from collective intelligence. We urge researchers to think capaciously about how collective intelligence might impact Heliophysics, perhaps even providing new solutions to long-standing questions. Two areas, in particular, might be fruitful: 1) interpreting particle simulations and 2) studying responses to natural hazards as a way of more accurately predicting societal impacts of extreme space weather events. Anticipating the final section of this paper, there is a growing literature that exists at the intersection of collective and artificial intelligence \cite{Berditchevskaia2022ADA}, a trend indicative of most sciences. 

\section{Frontiers of inquiry and investigation emerging from complexity science} \label{future section}

This review has covered much ground. What follows is a synthesis of the discovery and insights from the history of complexity science in Heliophysics and space physics into a path forward for its research and its community. The path involves three elements: 
\begin{itemize}
    \item Articulating a key challenge for Complexity Heliophysics that is shared with numerous scientific domains (Section \ref{key challenge section};
    \item Defining a scientific framework capable of responding to it (Section \ref{risk science subsection}; and
    \item Discussing the socio-cultural dimension that must be addressed (Section \ref{convergence research subsection}). 
\end{itemize}

The sections that follow will be necessarily more subjective and perspective-leaning that the previous ones. This is deliberate. However, those perspectives are grounded in literature and scholarly artifacts. Because these are `frontiers,' coverage within Heliophysics and space physics is as yet inadequate and we must borrow liberally from sister and even more distant domains. The intention for the following material is that it is not only provocative, but suggestive of generative new thinking and inquiry. 

\subsection{Key challenge for 21st century science: the intersection of complexity and artificial intelligence and a framework to explore it} \label{key challenge section}


Artificial intelligence (AI) and machine learning (ML) are not new. They are traced as fields of study to the 1950s, when Alan Turing worked on the concept of intelligent machines and how to create them \cite{Turing1950ComputingMA} and the conference that many credit with coining the term artificial intelligence was held \cite{McCarthy2006APF}; and their origins as areas of thought predate even that by many decades (in science fiction writing \cite{AsimovRunaround} as in philosophy and the mechanical manipulation of symbols \cite{Descartes1968-DESDOM}). In the 1950s, hyperbolic beliefs about nearness to genuine artificial intelligence led to an `AI winter' when those hopes failed to materialize and lasted from the 1970s to the 2000s. However, in the past several decades the rise of internet-scale data, computing power consistent with a doubling in the number of transistors in an integrated circuit every two years (Moore's Law \cite{Moore1998CrammingMC}), and improvement in algorithms have brought a renewed zeal and concomitant advance to AI/ML. A quick note that ML is a sub-domain of AI where AI is the ability to accomplish complex goals \cite{Tegmark2017Life3B} while ML is leveraging data to improve computer performance on some task or set of tasks \cite{mitchell1997machine}. Since existing studies in Heliophysics do not truly address AI, but the broader concept is important in this section, we will adopt the shorthand AI/ML to refer to the full set of methods. 

Heliophysics and space weather have been a part of the passionate exploration, adopting and applying AI/ML advances coming out of industry. \cite{Camporeale_2019} review the state of AI/ML in Heliophysics research, including prominent applications in forecasting: geomagnetic indices, relativistic electrons at geosynchronous orbits, solar flares occurrence, coronal mass ejection propagation time, and solar wind speed. Their synopsis of the field led them to conclude that there is a need to shift forecasting in Heliophysics to probabilistic approaches centered on the reliable assessment of uncertainties, and the combination of physics-based and machine learning approaches. Their discussion echoes a long-standing conversation that has been staged across the sciences (e.g., \cite{Mazzocchi2015CouldBD}) and indeed across culture more broadly, becoming a common subject of popular science writing, science fiction literature, and futurism e.g., \cite{Anderson2008TheEO, Chiang2000CatchingCF, Kelly2016TheIU, Finn2017WhatAW, ottino2022nexus}. The two sides of this debate have been given different names over time: hypothesis-driven vs. empiricism, deductive vs. inductive reasoning, theory-driven vs. data-driven. 

Our review, too, leads us to the need to find common ground between these poles of approach to growing scientific knowledge, now perhaps with the language of AI/ML vs. complexity science. Words create worlds \cite{Heschel1989MoralGA} and it deserves some inquiry into what new perspectives these new words for the old debate may create.

\cite{Mcgranaghan2021SpaceWR} and \cite{McGranaghan_2017b} taxonomized AI/ML as a part of the broader field of data science, defining the latter it as scalable architectural approaches, techniques, software and algorithms which alter the paradigm by which data are collected, managed and analyzed and communicated. They point to a similar need to integrate knowledge of the physical domain with data-driven approaches.

The trend in the literature around AI/ML and data science in Heliophysics is clear: the future of Heliophysics must explore the intersection between data-driven approaches with theory-driven science \cite{Karpatne2016TheoryGuidedDS, Pankratius2016ComputerAidedDT}. \cite{Klimas1996}, where we began this review, points to this key challenge for Complexity Heliophysics: converging the autonomous and the local linear prediction filter methods, merging the benefits of interpretability with the success of data-driven approaches. \cite{Klimas1996} writes, ``It is anticipated that in the future these local-linear predictor models will be studied carefully with the goal of organizing these bits and pieces into a global nonlinear predictor model. It may be advantageous to cast these predictor models as analogue models in order to maximize their physical interpretation.'' Klimas et al. are pointing to a reconciliation that is at the heart of a debate that spans the sciences in the 21st century: between first principles, physics-based models and data-driven, artificial intelligence or machine learning AI/ML algorithms. They represent different reasoning approaches: inductive and deductive, and correspond to different capabilities to explain the result found. Physics-based models are inherently explainable--the behavior arises from understood laws stitched together in traceable logical reason. AI/ML models discover patterns directly from data, but are less clearly interpretable. This review has chronicled the fact that complexity introduces, sometimes extreme, uncertainty to physics-based understanding and precludes predictability \cite{Wolfram_2002}. However, the advent of AI/ML (and other data-mining approaches) and the requisite computation has produced, in some cases, capabilities to represent complicated systems more accurately than physics-based equations \cite{Camporeale_2019, Stephens_2019}. These models may be capturing some as yet unknown physical properties. This quality is similar to the way that power laws in complexity science capture some underlying mechanism that acts across scales \cite{west2017scale}. Reconciling the first principles with data-driven approaches, physics and complexity with artificial intelligence, is a grand challenge for the 21st century. 

Indeed, the consilience of physics with data-driven approaches is important to all areas of inquiry \cite{Wilson1998ConsilienceTU}. Domains tend to embrace this call to integrate the two when the fundamental and applied components of their science come into contact: understanding pathologies of diseases in medicine vs. predicting if and when a disease will occur; understanding the Earth system vs. predicting natural disasters; understanding the solar-terrestrial system (Heliophysics) vs. predicting its consequences for our technological systems (space weather). In science there is a `tangled relationship between prediction and understanding' \cite{Krakauer_aeon}. 

\subsection{Space weather as a risk science} \label{space weather as risk science section}


Recasting the debate about theory- and data-driven science as a tension between AI/ML and complexity science, the body of literature in this review suggests that risk science coupled with an emphasis on resilience can provide a new framework. Definitions of commonly confused terms help approach the subjects of risk science and resilience. 
    
Risk is likelihood of occurrence of a natural hazard multiplied by the consequence of that. It is therefore distinct from a natural hazard, which refers to the physical phenomenon, and disaster, which is the particular occurrence of a natural hazard that results in major consequences. If risk is the focus of a science that spans physical understanding to decision-making, resilience is the applied goal of that science. Resilience is the capacity of a system to recover from a disturbance \cite{Lent2017ThePI}. Risk requires specification or prediction of events; resilience requires understanding the systemic impacts of those events, including the physical system along with the elements of preparing for and responding to the risk (change anticipation, exposure, mitigation, damage minimization). Together, they are a way of formulating grand scientific and societal challenges in a way that AI/ML and complexity science converge. 



Other fields have demonstrated this convergence: Dynamical Systems: \cite{Fischer2022TowardsAD}; Climate: \cite{Hultman_2010}; Ecology: \cite{Walker2004ResilienceAA, gunderson2000ecological}; Socio-Ecological Systems: \cite{Carpenter2001FromMT}; Disaster Research: \cite{Paton2000DisasterRR}; Medical Anthropology: \cite{panter2014health}; Health: \cite{Promislow2022ResilienceIC}; among others: see \cite{Bhamra2011ResilienceTC} for a review of the concept of resilience across disciplines. 

\subsubsection{Risk science} \label{risk science subsection}

In the context of climate science, \cite{Sobel2014SciencebasedRA, Sobel2022} argued for the necessity and the great intellectual opportunity of creating a discipline of \textit{climate risk science}. They qualified such a science as being a layer between modeling and applications, requiring probabilistic approaches, and being related to adaptation. The similarities to the dimensions of complexity science and the grand challenge articulated in this review are striking and suggestive of how we might affect the trajectory of space weather research. 

The innovation is to study space weather as a risk science and develop a framework for evaluating and quantifying space weather risk. It is important that space weather follow the examples of other natural hazards in adopting a risk formulation not only to overcome disconnects between science and decisions for space weather itself, but also for eventual incorporation into multi-hazard risk studies (e.g., understanding the power grid when space weather acts contemporaneously with terrestrial weather changes), permitting multi-hazard, compounding hazard, cascading hazard, and systemic risk research \cite{Helbing2013GloballyNR}).

To develop a risk science framework, we can and should begin with the fields of risk studies \cite{Burgess_Routledge_2016} and disaster risk reduction (DRR) \cite{Wisner_Routledge_2011}. Drawing on the framework developed for DRR in \cite{Wisner_Routledge_2011}, there are five elements: 
\begin{itemize}
    \item Environment: the system under study, including its overlaps and interconnections to other systems; 
    \item Hazards characteristics: understanding the details of a natural hazard (and if a multi-hazard analysis, then the collective, co-occurrence, connected details of the hazards) such as location, intensity, frequency, probability as well as quantifying the uncertainty of statistically averaging; 
    \item Vulnerability: differential impact from a natural hazard;
    \item Capacity: resources and assets available to resist, cope with, and recover from natural hazards \cite{Wisner2004}; and
    \item Exposure: the situation of people, infrastructure, housing, production capacities and other tangible human assets located in hazard-prone areas\footnote{\url{https://www.undrr.org/terminology}}. 
\end{itemize}

\begin{figure}[h]
\centering
\includegraphics[width=\textwidth]{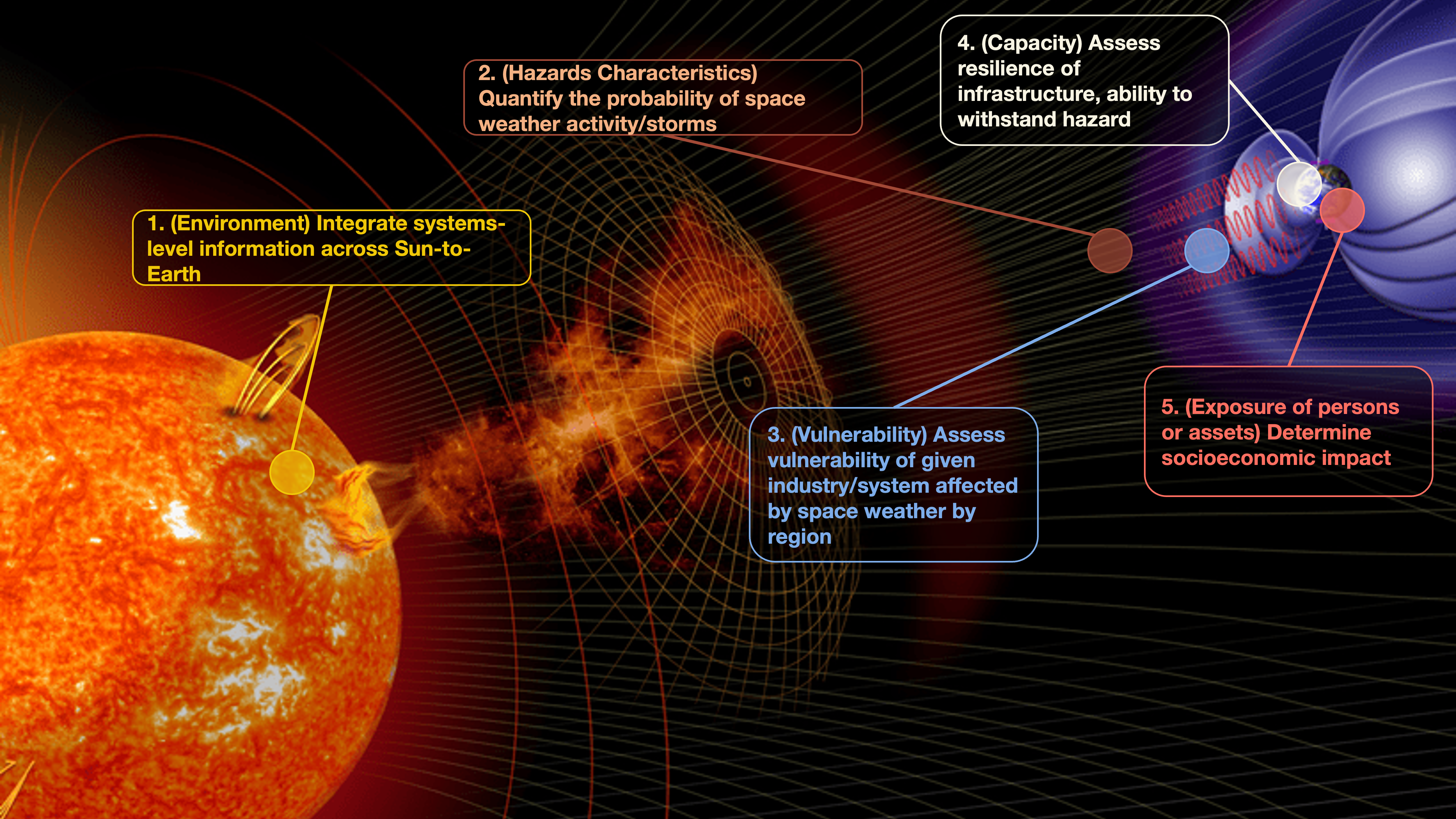}
\caption{A space weather risk science framework, covering the five elements risk adapted from the field of disaster risk reduction: Environment, hazards characteristics, vulnerability, capacity, and exposure.}
\label{risk science figure}
\end{figure}

There is a sixth element that we will not address in this review: \textit{recovery}. Figure \ref{risk science figure} relates each of the five elements to space weather. 

The framework is built on three important principles: 1) Consideration of the holistic Sun-to-society system; 2) Quantification of the uncertainty that arises from coarse-graining and statistical simplification \cite{Mcgranaghan2022TheEO}; and 3) relating space weather hazard to societal impact. 

Much of this review has already been focused on efforts to quantify the characteristics of Heliophysics phenomena (e.g., burst statistics and scaling relationships for geomagnetic storms and substorms, as well as a not insignificant body of research on quantifying probabilities for extreme space weather events that we have not reviewed \cite[and references therein]{Jonas_2018}). Here we will review articles that address the final three elements of the framework, to the extent that such research exists. 

\cite{Schrijver_2015} adopted an economic perspective to understand the relative potential impacts of extreme events and less extreme, more frequent events. He determined that societal impacts of both common severe and of rare extreme space weather are substantial, concluding that quantifying the characteristics of both kinds of event are vital to preparing for them, including creating mitigation strategies where possible. While the analysis of \cite{Schrijver_2015} was for general space weather disturbance, a majority of risk studies for space weather have been conducted around the impact to the power grid. 

\cite{Oughton_2017} explored the potential costs associated with failure in the electricity transmission infrastructure in the United States due to extreme space weather, focusing on daily economic loss, exploring both direct and indirect consequences, and commenting on the implication for cost-benefit analyses of space weather forecasting and mitigation investment. Their key contributions include creating a foundation for the quantification of the economic impact of space weather, identifying the structural relationships that tie space weather impacts to the power grid to supply chains, and methodology to connect regional and national and direct and indirect impacts. Key among the results was the finding that the direct economic cost incurred from disruption to electricity  was only a fraction of the total cost for the space weather scenarios explored. Space weather impacts are not merely of the direct kind, but systemic. 

\begin{figure}[h]
\centering
\includegraphics[width=\textwidth]{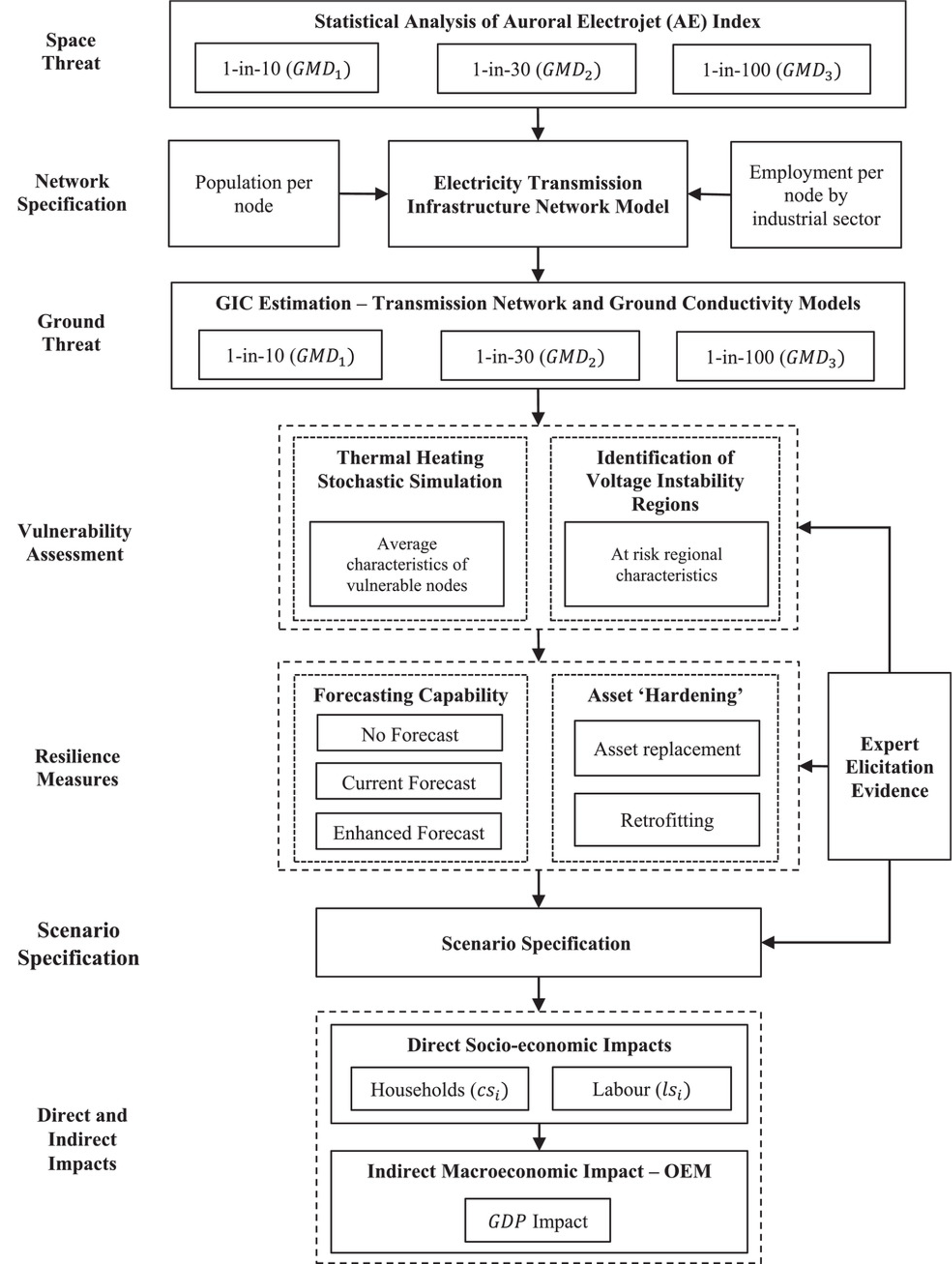}
\caption{Risk assessment framework of \cite{Oughton2019}, reproduced with permissions. The framework includes: 1) estimating the probability of intense magnetospheric substorms (space weather natural hazard), 2) exploring the vulnerability of electricity transmission assets to geomagnetically induced currents (GICs; vulnerability), 3) assessing the resilience of the infrastructure (capacity); and 4) quantifying socioeconomic impacts (exposure). We have related their figure to the language we develop in this review to aid in mapping between frameworks, which are indeed quite close.}
\label{oughton framework figure}
\end{figure}

\cite{Oughton2019} in many ways culminates these threads of space weather as a risk science in the context of the power grid, establishing a framework for risk assessment. Their framework includes quantifying general geophysical risk, asset vulnerability, and the network structure of critical infrastructure systems. Figure \ref{oughton framework figure} is a reproduction of their three-part risk assessment framework. In the description of the figure we have related their work to the language we develop in this review to aid in mapping between frameworks, which are indeed quite close.

Concurrently, \cite{Eastwood2018} explored a similar framework, also focusing on the impact of substorms on the power grid. They delineate the components of economic impact (to any hazard, not limited to space weather): spatial and temporal extent, vulnerability of the technologies/infrastructure, extent of mitigation, capacity to maintain production and support consumption across firms and consumers. Like \cite{Oughton2019} they map the physical phenomena to the impact through the intensity of the space weather event, resilience of the power grid, capability of the forecast, and socioeconomic models. The differences between these two similar exploratory studies to create risk assessment frameworks are illustrative of the open research questions in space weather risk science. They include the determination of the likelihoods of extreme storms, incorporation of forecast information, evaluation of the vulnerabilities of the infrastructure, details of recovery times, and socioeconomic models. A key challenge both studies articulate is the inhomogeneity of available information across the globe, making global impact calculations difficult. 

Across space weather more broadly, \cite{Eastwood2017} assessed the existing knowledge available to quantify the economic impact. Their survey included the phenomena that represent space weather hazards (environment), the existing research to calculate occurrence and intensity statistics (hazards characteristics), and documented impacts across industries affected (vulnerability). Thus, their work aligns to the framework that we suggested above. Where it fell short of assessing the complete risk science framework reveals important research gaps to fill. It also points to structural issues in Heliophysics research where some of these gaps exist because of a lack of collaboration between the involved communities and others due to unavailability of vital data. There is clearly much that needs to be learned to make space weather a risk science. 

A theme in this literature is that of interconnections, of disciplines, of physical phenomena to socioeconomic impacts, of spatial regions, and of critical infrastructure; it reiterates the complexity paradigm as the needed response, and risk science as a means to create general systemic understanding. A type of risk not often addressed in the literature, perhaps in large part due to the sheer challenge associated with it, is interconnected risk--risk due to externalities acting co-temporally and/or co-spatially to space weather such as extreme terrestrial weather. 

If space weather is approached as a risk science, the domain could share a common framework with other natural hazards such as extreme terrestrial weather, hurricanes, earthquakes, wildfires, and floods.

Given the potential link that risk science provides between physical understanding and data-driven specification and prediction, it is important to understand the ways that risk is studied and used as a framework for research in various scientific disciplines. That is left as a call for Heliophysicists to understand approaches of `sister' disciplines and how they might be impactful to Heliophysics research. In concluding this review, we guide that effort by relating a few of the ways the risk science framework might relate to and be used in Heliophysics. They fall generally into three categories in order from general to specific: 1) systemic risk; 2) resilience research; and 3) critical transitions. 

\subsubsection{Systemic Risk} \label{systemic risk section}

\begin{figure}[h]
\centering
\includegraphics[width=\textwidth]{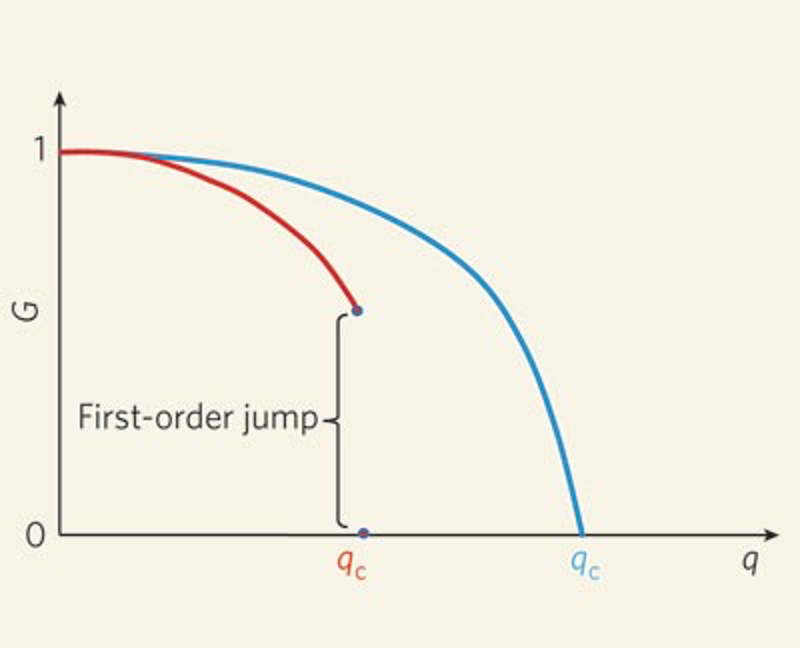}
\caption{Figure 2 from \cite{Vespignani2010ComplexNT}, reproduced with permissions. $G$ is the largest number of connected nodes in a network (as a fraction). $q$ is the fraction of nodes removed from a network. A critical fraction of nodes $q_c$ is the point at which complete fragmentation occurs. The behavior of the isolated network (blue) and the interdependent network (red) is fundamentally different, falling off at a faster rate and exhibiting a `first-order' transition.}
\label{vespignani figure}
\end{figure}

Existing understanding of natural hazards is predominantly from the perspective of a single hazard in isolation. Yet the behavior of systems is not observable by considering each part in isolation. In short, emergence only occurs in the interconnected systems.  \cite{Buldyrev2010CatastrophicCO} laid out a framework to study critical transitions (what they call catastrophic failures because of their focus on interdependent power grid and internet networks implicated in a 2003 blackout) in interdependent networks. They present details of the critical fraction of nodes that, on removal, will lead to a failure cascade and to a complete fragmentation of the interdependent power grid and internet networks. They study the functional integrity of the composite network using the largest connected set of nodes, $G$, in the system as a proxy. Network nodes are progressively removed and the effect on $G$ observed. The general finding is that interdependent networks exhibit not only a smaller critical threshold than isolated networks, leading to different levels of disruption, but also a different nature of abrupt `first-order' transitions in system breakdown (see Figure \ref{vespignani figure}) \cite{Vespignani2010ComplexNT}. There is a relationship between failure of nodes in one network and failure of nodes in the other. Buldyrev et al.'s results can be interpreted as general, exemplifying complexities and fragilities arising from network interdependencies. Relating the findings to the study of natural hazards, it is quite possible that understanding hazards in isolation will be inadequate to understanding the system and its resilience. Extreme events may be a result of the inherent interdependent system dynamics rather than of unexpected individual external events. 

\cite{Helbing2013GloballyNR}, too, points to limitations in uni-hazard/uni-disciplinary analyses. They envision a `Global Systems Science' that begins from the recognition that systemic failures and extreme events are consequences of highly interconnected systems and networked risks due both to natural connections between physical systems and connections created by the human and built world. 

The Heliophysics community has been an active participant in shaping systemic risk research. Heliophysicists helped lead the geophysical monograph \textit{Extreme Events and Natural Hazards: The Complexity Perspective} \cite{Sharma2012}, whose contributions collectively established the set of techniques (drawn from complexity science) and state of research at the time of writing for studying the extremes of natural hazards encompassed by the upper tail of the probability distribution. That monograph understands natural hazards as multidisciplinary phenomena, requiring concomitant multidisciplinary research, and recognizes that risks are only appropriately quantified if rare events across domains are coupled. They introduce the collection, ``Like most of the major scientific challenges in the Earth and space sciences, there is increasing recognition that an integrated approach involving multiple disciplines will be needed to advance the science underlying extreme events that lead to natural hazards...The distributed nature of the components and the strong interaction among them is [a] feature common to systems exhibiting extreme events.''

The conclusions arrived at from this review's examination of the history of complexity Heliophysics, those from \cite{Helbing2013GloballyNR} in establishing `Global Systems Science,' and the contributions to \cite{Sharma2012} are too similar to ignore. With consensus guidance on what needs to be done, it seems time for adoption across the research community from policy-makers to research scientists. The following two sub-sections are areas that offer pathways into systemic risk research. 


\subsubsection{An emphasis on resilience} \label{components of resilience subsection}

Systemic risk enables the study of resilience, the ability of a system to maintain specific functions in the face of change \cite{Scheffer_2001, baggio2015boundary}. 

Returning to \cite{Vespignani2010ComplexNT}, the key finding is that understanding resilience, and ultimately designing more resilient systems, requires consideration of the interconnected system, the mutually dependent network properties. Indeed, the notion of resilience has been a powerful way into risk and systems science. 

Resilience is a systemic phenomenon. At the highest level resilience is defined as a system's accommodation of changes and reorganization of itself while maintaining the crucial attributes that give the system its unique characteristics \cite{Scheffer2009EarlywarningSF, Scheffer_2018}. 

Despite evidence to its importance and vast literature about it, there exists no consensus framework for understanding and managing resilience, a fact not unrelated to the challenge of quantifying resilience \cite{Scheffer_2018}. Yet, a set of generic dynamic indicators of systems near critical transitions observable in time series data (see Section \ref{critical transitions subsection} below) has emerged that, alongside proliferation of time series data across domains and indications of corresponding spatial indicators of resilience \cite{Dakos2010SpatialCA}, enable new progress to understand resilience.


The literature on resilience is wide and multidisciplinary. Immediately meaningful in this review are seven principles of resilience that, although written for social-ecological systems, share a common basis with Heliophysics and Space Weather as a risk science in the need to build capacity to deal with unexpected change: 
\begin{enumerate}
    \item maintain diversity and redundancy; 
    \item manage connectivity; 
    \item manage slow variables and feedbacks;
    \item foster complex adaptive systems thinking; 
    \item encourage learning;
    \item broaden participation; and
    \item promote polycentric governance systems.
\end{enumerate}

An emphasis on resilience speaks to understanding sought by Complexity Heliophysics and Space Weather as a risk science:
\begin{itemize}
    \item \textbf{Both predictive and responsive}: Resilience acknowledges that all responses are dual, including both the pre-emptive actions possible and those that must be responsive to the existing conditions). One could think about these dual responses as the pre-emptive actions accommodating the more or less deterministic signals while the responsive actions are those taken in the face of uncertainty and are inherently unpredictable.  
    \item \textbf{Translational}: Resilience requires translation between the science and the responses available (thus requiring understanding of system capability and capacity);
    \item \textbf{Multi-level}: Resilience also requires a multi-level understanding of the system and the different responses for each level \cite{Sober2021UntoO}. For instance, a power grid operator monitoring the grid in Washington, DC and an individual at the Department of Energy tasked with the health of the country's grid as a whole will have distinct responses to a National Oceanic and Atmospheric Administration (NOAA) warning; 
    \item \textbf{Interdependent}: Resilience connects a system to the external systems that may amplify or attenuate a particular effect \cite{Levin2021GovernanceIT}, which reveals the final component;
    \item \textbf{Semantic}: To understand interconnections, relationships between domains must become first class citizens to enable agents (whether a human or an intelligent machine) to navigate between them and integrate information from across them \cite{NAROCK2012248, BentleyHelio, Shimizu2020TowardsAM}. This requires understanding information flow, in the physical, socio-cultural, and technological senses \cite{Mcgranaghan2022ConvergingTS}. 
\end{itemize}

These are sites of future research that must draw from both the complexity paradigm and AI/ML.

\subsubsection{Critical transitions} \label{critical transitions subsection}
Resilience is complicated by and inextricable from the notion of critical transitions, or tipping-points. Tipping-points that represent critical transitions are points at which a dynamical system abruptly shifts from one state to another. In dynamical systems language, these are points at which small changes in a parameter can lead to qualitative change in the behavior of the system, or bifurcation points \cite{Strogatz_2018}. Under enhanced resilience, a system is more likely to accommodate changes without undergoing a critical transition into a qualitatively different state. Critical transitions are notoriously challenging to predict. However, exciting research points to the existence of generic properties (and the way that they can be measured) for systems near critical transitions \cite{Scheffer2009EarlywarningSF}: 
\begin{itemize}
    \item Critical slowing down \cite{Wissel2004AUL}: As the system approaches such critical points, it becomes increasingly slow in recovering from small perturbations. In these regimes, the dynamical system will show an increase in lag-1 autocorrelation and increased variance in the pattern of fluctuations as the recovery rate from a small perturbation is reduced; 
    \item Skewness and flickering before transitions: Asymmetry in fluctuations increase near a critical transition. Flickering is indicated in the frequency distribution of states as increased variance and skewness as well as bimodality \cite{Carpenter2006RisingVA}; and
    \item  Increased spatial coherence: An analog to critical slowing down for spatial data, it has been shown that systems nearing critical transitions exhibit increased spatial correlation \cite{Dakos2010SpatialCA}. In systems consisting of numerous coupled units, slowing down near a critical transition will equalize differences between the units as each will tend to take the state of the units to which it is connected. 
\end{itemize}

\noindent \cite{Scheffer_2018} calls these generic properties ``dynamic indicators of resilience.''

\begin{figure}[h]
\centering
\includegraphics[width=\textwidth]{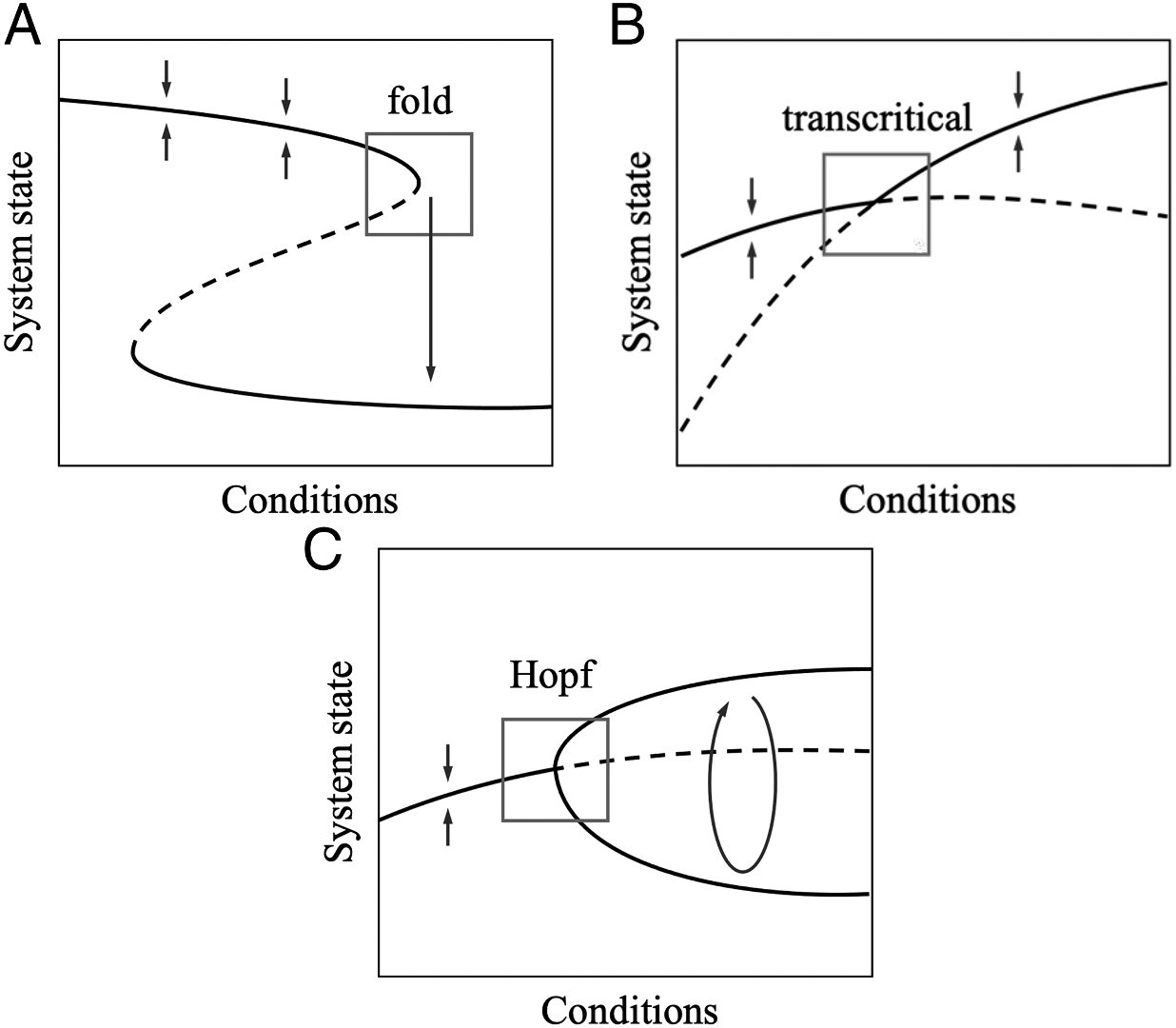}
\caption{Examples of types of bifurcations: (A) Fold; (B) (Supercritical) Hopf; and (C) transcritical bifurcation. Figure reproduced from \cite{Bury2021DeepLF}.} 
\label{bifurcation types figure}
\end{figure}

There are numerous types of bifurcation that each represent a critical transition. Figure \ref{bifurcation types figure} shows examples of fold, (supercritical) Hopf, and transcritical bifurcation. 

The type of bifurcation may be associated with different dynamics around the critical transition. For instance, system that undergoes a fold bifurcation exhibits an abrupt transition to a very different state whereas a transcritical bifurcation usually causes a smooth transition and a Hopf bifurcation can lead the system into a state of oscillatory behavior \cite{Bury2021DeepLF}. Despite the different dynamical behavior, as noted above the general properties preceding critical transitions can be associated with a wide variety of transition in complex systems \cite{Scheffer2009EarlywarningSF} and are exciting in their cross-disciplinary impact. 

Universality of these ideas is made apparent in the range of fields in which they have been studied and led to new understanding. In sociology, one of the most cited papers is Mark Granovetter's `Threshold Models of Collective Behavior' \cite{Granovetter1978ThresholdMO}, whose impact created a new domain of analyses (`threshold models') and that rely on critical transition theory to interpret model results. Similar developments have been observed in medicine \cite{Litt2001EpilepticSM}, finance \cite{Kambhu2007NewDF}, and climate \cite{Lenton2008TippingEI}), among numerous others. The similarity of the problems in those disciplines to the challenges in Heliophysics and various stellar-planetary systems and data (e.g., \cite{Schunk2021ChallengesIS, Srivastava2021EditorialSW, Palmerio2022CMEsAS}) suggest that the science of risk and critical transitions may have important utility to future leaps in our understanding.  Application of these theories in Heliophysics is nascent, but may represent an important frontier.




\subsection{Convergence research} \label{convergence research subsection}
Achieving these ambitious research directions will be a challenge.


The complexity paradigm, risk, and resilience are all bridging concepts \cite{baggio2015boundary, Burgess_Routledge_2016}. They create connections between domains, between areas of scholarship, between science and society that would otherwise not exist \cite{Granovetter1973TheSO, Allen2023JusticeBM}. To engage in them is to understand that the work needed is not only technical, but social and cultural, too. 

Pioneering domains like bioengineering \cite{nersessian2022} are scientific forebears in the creation of resilience research that treat the system as complex and integrate data-driven analyses. Their examples reveal that an emphasis on resilience and a risk science framework facilitate transdisciplinary approaches \cite{Promislow_2022} (here a distinction is made between transdisciplinary and interdisciplinary: interdisciplinarity brings different disciplines together, but maintains their identification; transdisciplinarity takes place in the context of a real world problem in which the disciplines are tightly integrated, their methods and epistemologies synthesized and blended into a novel approach, perhaps even a new discipline). One such approach, in many ways an instantiation of the transdisciplinary paradigm, is \textit{convergence research}. In 2016, the National Science Foundation (NSF) named ``Growing Convergence Research'' as one of its 10 Big Ideas for prioritizing future investments in science and engineering. At its most general, convergence has been defined ``an approach to problem solving that cuts across disciplinary boundaries. It integrates knowledge, tools, and ways of thinking from life and health sciences, physical, mathematical, and computational sciences, engineering disciplines, and beyond to form a comprehensive synthetic framework for tackling scientific and societal challenges that exist at the interfaces of multiple fields'' \cite[page 1]{NAP18722}. With convergence comes a new spectrum of challenges involving how we work across disciplinary lines, collaborate meaningfully in large groups, and develop healthy--meaning open, participatory, and resilient--connections among diverse stakeholders. Risk and hazard domains have been proving grounds for convergence research \cite{White1975AssessmentOR, Quarantelli1987DisasterSA, Prince2009CatastropheAS, Solnit2009APB, Peek2020AFF}. The full spectrum of risk science means spanning natural hazard (the physical phenomenon) to risk (likelihood times consequence) to resilience (understanding of the system's vulnerability and capacity) to societal impact. Crossing the spectrum requires \textit{convergence science}. Indeed, tackling the problems we face as a society, whether global pandemics or climate change or complex systems, requires new levels of cooperation, facilitation, and synthesis \cite{Mcgranaghan2022ConvergingTS}.  

In Heliophysics, these transdisciplinary convergent approaches offer the potential to integrate space physics, space weather, and society. Two developments are needed: 1) creating a \textit{knowledge commons}: a combination of intelligent information representation and the openness, governance, and trust required to create a participatory ecosystem whereby the whole community maintains and evolves this shared information space \cite{Hess_kcs, McGranaghan2021need}; and 2) acknowledging the need for new literacies on scientific teams for facilitation (either within Heliophysics researchers or through new roles on science teams)\footnote{\url{https://www.cscce.org/}; \url{https://openlifesci.org}}, which represent capacities on teams for facilitation and creating healthy communities. As our science questions grow in complexity, so, too, must the information and the knowledge we bring to them. As information grows, the costs and challenges of communication grow. Convergence research is a  and New roles to support convergence research 



A framework for risk science and a concomitant emphasis on resilience may be poised intellectually and institutionally to conduct complexity science \cite{Peek2020AFF}, bridge between predictive methods like AI/ML and fundamental science, and bring convergence research to Heliophysics. These inspire further studies on the nature of resilience from a systems level perspective of the Solar-Terrestrial connection, taking up the call of the 21st century ``to integrate the sciences of complexity with machine learning and artificial intelligence'' \cite{Krakauer_aeon}. 






\section{Conclusion}\label{conclusions section}

The 21st century will be marked by complexity, according to Stephen Hawking, and this is especially true for the field of Heliophysics. Heliophysics has traditionally been characterized by categorizing and separating domains, but with the advent of new sensing capabilities, data analysis, and computational tools, there is a growing need for a paradigm shift towards Complexity Heliophysics. Complexity science is the study of phenomena that emerge from a collection of interacting objects and requires a plurality of frameworks that move between levels of a system. This lived and living review details the network of complexity studies in Heliophysics and provides a definition of the Complexity Heliophysics paradigm. 

This review first outlined five dimensions of complexity science. Then, the analysis of the existing literature mapped into three parts: 1) a pivotal year for the paradigm: 1996; 2) transitional years that established dimensions of the paradigm between 1996-2010; and 3) emergent literature largely after 2010. For the final ternate, we drew on a much wider base of literature to situate Complexity Heliophysics in a broader context of the physical sciences, revealing trends and gaps. Several are proposed:

First, the ability to capture underlying structure and patterns in complex systems through coarse-graining is crucial in Heliohpysics. Two forms of coarse-graining, namely information theory and network science, are particularly important for the future of the field.

Second, reconciling the first principles with data-driven approaches, physics and complexity with artificial intelligence, is a grand challenge for the 21st century. We centered the discussion of Complexity Heliophysics in the tension between fundamental science vs. prediction-oriented science (e.g., basic science vs. applied science; physics-based modeling vs. artificial intelligence/machine learning) and suggest that this history of complexity science within Heliohpysics is instrumental in finding pathways between these extremes, therefore becoming inextricable from the future of Heliophysics research. The trend is clear: the future of Heliophysics and its applied counterpart, space weather, must explore the intersection between data-driven approaches with theory-driven science. Indeed, \cite{Klimas1996}, where the review begins, pointed to a key challenge for Complexity Heliophysics: converging the autonomous and the local linear prediction filter methods, merging the benefits of interpretability with the success of data-driven approaches. We provided a vision that could help respond to the challenge in a \textit{risk science} framework, which adopts probability and resilience as organizing concepts and identifies corresponding analysis methods. The technical challenges of complexity science are accompanied by socio-cultural challenges for which we conclude by relating the methods of convergence research. 




Ultimately, this review provides a foundation for how complexity science can help address outstanding questions in Heliophysics and space weather science. The artifacts from this work include this review article; a glossary of terms that define Complexity Heliophysics and can be useful to search and discovery of related resources, individuals, and groups; and a new corpus of Complexity Heliophysics that is likely full of further discovery and generative of new research questions. With the paradigm shift, we will gain new capacities to understand the Heliophysics system, and this will guide researchers towards directions that are better equipped to respond to the challenges of the 21st century.



\backmatter

\bmhead{Supplementary information}

Four appendices are included with this review, each included to facilitate more comprehensive understanding of this review and curated material to \textit{act} on it: 1) acronyms; 2) questions identified in the papers reviewed in this work to guide future research; 3) generation and analysis of the automated corpus of Complexity Heliophysics; and 4) key datasets explicitly mentioned in this review or that factor prominently in the results examined. 

Additionally, software to create the automated corpus along with the corpus itself are provided in a Github repository (\url{https://github.com/rmcgranaghan/Complexity_Heliophysics}). 

\bmhead{Acknowledgments}

Wide and deep thanks is due to a community of conversants, whose time, attention, and ideas were intellectual exhilaration and generativity for this manuscript. An incomplete list of those individuals includes (in no particular order):  Joseph Borovksy, Juan Valdivia, Simon Wing; Eric Donovan, Massimo Materassi, John Dorelli; Jeffrey Thayer, Vadim Uritsky, Sandra Chapman, Jay Johnson, Josh Semeter, Seebany Datta-Barua, Olga Verkhoglyadova, Giuseppe Consolini, Elizabeth Butler, Anthony Mannucci, Paul Wong, Jacob Bortnik, Enrico Camporeale, Barbara Thompson, Madhulika Guhathakurta, Jesper Gjerloev, Nick Watkins, Xing Meng, Brian Thomas, and numerous past guests on the Origins Podcast \url{https://www.originspodcast.co/}.

Several events were also formative for this work. To the organizers, conveners, and attendees I am deeply grateful: ``Exploring Systems-Science Techniques for the Earth’s Magnetosphere-Ionosphere-Thermosphere'' (July 2018 \cite{Mcgranaghan2018HowDW}), the Santa Fe Institute's Complexity Interactive held January 2022 \url{https://www.santafe.edu/engage/learn/programs/complexity-interactive}, the Lorentz Center event ``Space Weather: A Multidisciplinary Approach'' held in September 2017 \url{https://event.cwi.nl/spaceweather2017/}, the series of NASA Living With a Star Jack Eddy Symposia (especially the 3rd event held in June 2022 \url{https://cpaess.ucar.edu/meetings/eddy-symposium-2022}), and the National Science Foundation (NSF) Convergence Hub for the Exploration of Space Science (CHESS) event ``Simulating Space Weather Extremes: A Workshop to Identify Research Needs to Improve Power Grid Resilience to Geomagnetic Activity'' held April 2022 (\url{https://www.nsf.gov/awardsearch/showAward?AWD_ID=2131047}). 

The author gratefully acknowledges the support of the NASA Early Career Investigator Program (ECIP) Program (NASA Grant Number: 80NSSC21K0622) for the resources to research, pursue, and write articles on the philosophy of Heliophysics science and how to make breakthroughs in our epistemology such as this one. Additionally, the author is deeply appreciative of the NASA Center for HelioAnalytics, funded by NASA ISFM, for supporting this work and creating a community in which conversations like these occur.

Data and software supporting this review are available from a Github repository (\url{https://github.com/rmcgranaghan/Complexity_Heliophysics}) that provides information about generating the corpus for automated or natural language processing in support of this work as well as the glossary used to filter the articles and automated corpus itself. We acknowledge Omar Shalaby for the development of tools to programmatically query ADS and organize results. 

Thank you to the NASA Center for HelioAnalytics and Heliophysics Digital Resource Library for creating, maintaining, and making available HelioCloud, a cloud computing service for Heliophysicists. The corpus generation and natural language processing analysis for this publication were carried out on that platform.

Inordinate thanks is due to Semantic Scholar \url{https://www.semanticscholar.org/} built by the Allen Insitute for AI \url{https://allenai.org/} for its aid in finding literature and appropriately and conveniently citing it for inclusion in this review. 

Many thanks to the Lingo4G \url{https://carrotsearch.com/lingo4g/} team for support in getting their software up and running and for providing numerous trial licenses to complete the analysis detailed in this manuscript. 

The research was in-part carried out at the Jet Propulsion Laboratory, California Institute of Technology, under a contract with the National Aeronautics and Space Administration (80NM0018D0004). \copyright 2023. California Institute of Technology. Government sponsorship acknowledged.

\section*{Declarations}


Conflict of interest/Competing interests: Not applicable 



\begin{appendices}


\section{Appendix A: Acronyms}\label{acronyms appendix section}
\begin{center}
\begin{tabular}{ |l|c| } 
 \hline
 Term & Abbreviation \\ 
 \hline
 agent-based modeling & ABM \\
 \hline
 artificial intelligence and machine learning & AI/ML  \\ 
 \hline
 auroral electrojet & AE \\ 
 \hline 
 conditional mutual information & CMI \\
 \hline 
 Coupled Map Lattice & CML \\
 \hline
 disaster risk reduction & DRR \\
 \hline
 disturbance storm-time & Dst \\
 \hline
 European Space Agency & ESA \\ 
 \hline
 interplanetary magnetic field & IMF \\
 \hline
 Geocentric Solar Ecliptic & GSE \\
 \hline
 Geocentric Solar Magnetic & GSM \\
 \hline
 geomagnetically induced currents & GIC \\
 \hline
 large language model & LLM \\
 \hline
 local-linear predictor & LLP \\
 \hline 
 mutual information & MI \\ 
 \hline
 named entity recognition  & NER \\
 \hline 
 National Aeronautics and Space Administration & NASA \\
 \hline 
 natural language processing & NLP\\
 \hline
 nonlinear prediction filter & NPF \\
 \hline
 Polar Ultraviolet Imager & Polar UVI \\
 \hline
 probability distribution function & PDF \\
 \hline
 Santa Fe Institute & SFI \\
 \hline 
 self-organized criticality & SOC \\
 \hline
 state-space reconstruction & SSR \\
 \hline
 substorm current wedge & SCW \\
 \hline
 total electon content & TEC \\
 \hline
 Time History of Events and & THEMIS \\
 \hline
 transfer entropy & TE \\
 \hline
 
 \hline
\end{tabular}
\end{center}

\section{Appendix B: Questions identified in the papers reviewed in this work to guide future research}\label{open questions appendix section}
This appendix lists a few of the most potent questions either explicitly identified or implied by the studies reviewed for this paper. They are meant to be generative of future Complexity Heliophysics work. 

\begin{itemize}
    \item \cite{Chapman2000AvalanchingAS} What constitutes successful prediction of some observed activity?; 
    \item \textit{(From a collection of articles that each speak to this point)} What are appropriate measurables for magnetospheric dynamics (e.g., AE)? What can we determine given the limitations of these measurables (e.g., for the AE index, the limitations are well-documented \cite{Kamide1983NotesOT, Bargatze_1985, Klimas1996})?; 
    \item In distinguishing magnetospheric dynamics from those convolved with solar wind driving, are bursts in AU/AL causally related to those in $\epsilon$ and $vB_s$ (e.g., \cite{Freeman2000})? As we bring the complexity paradigm from the solar wind and magnetosphere into the upper atmosphere, what methods permit to connect or distinguish causal driving from internal dynamics?; 
    \item \textit{(Collection of questions)} ``Nine Outstanding Questions of Solar Wind Physics'' \cite{Viall_2020};
    \item \textit{(Collection of questions)} ``Outstanding questions in magnetospheric plasma physics'' \cite{Borovsky2020OutstandingQI}; 
    \item Have observational networks capable of finer resolution (e.g., local time-dependent auroral electrojet indices from the SuperMAG network \cite{Gjerloev_2009, Newell2011EvaluationOS, Newell2011SubstormAM, Newell2014LocalGI}) provided the necessary data to resolve open questions about the dynamic behavior of the magnetosphere that have been ambiguous in existing data (e.g., limitations of the AE index \cite{Bargatze_1985, Klimas1996})?
    \item \cite{Uritsky_2002} Given the observed power law statistics of the dynamic magnetosphere in auroral data and the corresponding inference of stationary critical dynamics in the magnetosphere, what can be learned about the role of cross-scale coupling in the development of geomagnetic disturbances?;
    \item \cite{Uritsky_2002} Do statistics of mesoscale magnetosphere simulations match observed statistics from the SOC paradigm?; 
    \item Which of our models of the magnetosphere and geospace exhibit complexity (e.g., emergent behavior)? How can those behaviors be observed; 
    \item What are the time scales for the phenomena that define Heliophysics in the Sun-Earth system (e.g., solar flares, coronal mass ejections, bursty bulk flows, substorms, auroral precipitation, ionospheric conductivity) and how are these consistent across a wide continuum of stellar-planetary systems? 
    \item \cite{McGranaghan_2017c, Orr_2019}Are topologies of Heliophysics systems (e.g., solar corona, ionosphere, magnetic environment) qualitatively different between low, moderate, and extreme periods of activity?  
    \item \cite{Mcgranaghan2022ConvergingTS, Peek2020AFF} To what extent can the study of space weather be transformed by adopting frameworks for natural hazards that enable convergence research and objectify resilience?  
    \item \cite{Mcgranaghan2022TheEO} What are the new literacies required of Heliophysicists and space scientists to embrace the Complexity paradigm? 
\end{itemize}

\section{Appendix C: Generation and Analysis of the Complexity Heliophysics Corpus}\label{corpus appendix section}

The references that constitute this review article were largely manually curated. The reference list includes well over 300 articles. However, current best estimates \cite{Bornmann2021GrowthRO} place annual growth rate of scientific literature at 4\% with a doubling time every 17 years. The impact is a significantly larger number of papers a researcher must read to be up to date on a field. More revealing is the growth in the number of papers a researcher must sift through to decide what to read. This `to read' pile can and does easily become thousands of papers long. 

The growth in publication and number of research artifacts is a trend that will only continue. The complexity paradigm exacerbates the problem, in which systems understanding demands information from more numerous and diverse sources to be integrated. It is likely that the effective growth rate when many disciplines must be considered is even greater. The outcome is that researchers must augment their traditional manual approach to curating and synthesizing literature with automated methods using databases of literature and research artifacts (e.g., Clarivate’s Web of Science\footnote{\url{https://www.webofknowledge.com}} and those tailored to more specific contexts like NASA’s Astrophysics Data System (ADS)\footnote{\url{https://ui.adsabs.harvard.edu/}}). Artificial Intelligence and Machine Learning (AI/ML) will play a role in augmentation and synthesis. 

Indeed, researchers are already squarely in the middle of this challenge to keep pace with the academic literature. To make this review an example of how to augment manual methods with automated approaches, we have used natural language processing (NLP) to examine the NASA ADS to construct a Complexity Heliophysics corpus (database of documents). We refer to this as the automated corpus, which is in addition to the manual corpus that is the reference list of this review. We describe our approach to create the automated corpus and provide the end product as an artifact with which future researchers can examine this new paradigm and also to experiment with AI/ML methods. 

\begin{figure}[h]
\centering
\includegraphics[width=\textwidth]{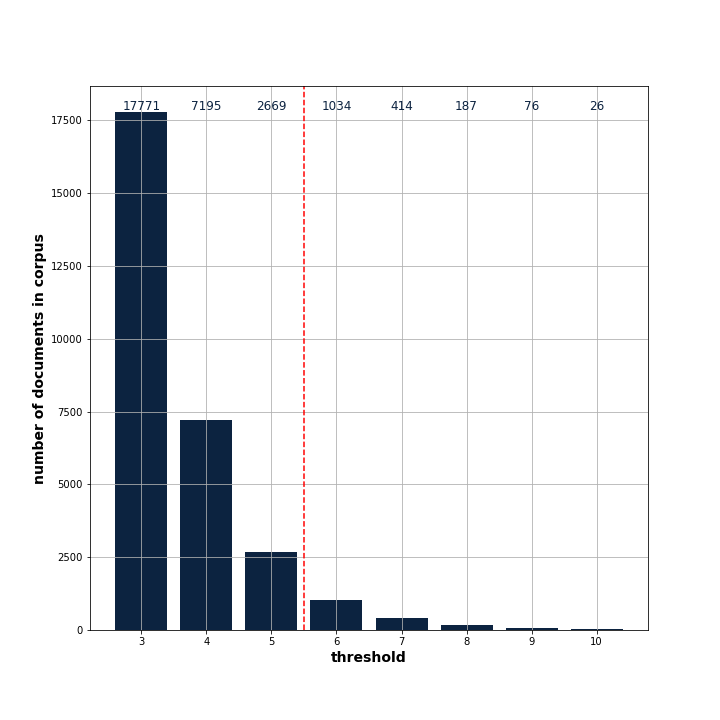}
\caption{Number of documents in the automated corpus vs. the threshold applied. The threshold indicates the number of distinct terms in the Complexity Heliophysics glossary that must be present in the abstract or title of a document for it to be included in the automated corpus. Numbers at the top of the bars indicate the size of the resultant corpus for that threshold. Dashed red line indicates the selected threshold.}
\label{corpus thresholds comparison}
\end{figure}

The process is as follows: 
\begin{enumerate}
    \item Construct a ‘Complexity Heliophysics’ terms list. These terms attempt to encompass relevant topics, phenomena, and concepts of the paradigm. The terms list contains 153 terms and is provided in the Github repository that accompanies this manuscript (\url{https://github.com/rmcgranaghan/Complexity_Heliophysics}); 
    \item Select a database of literature and research artifacts to explore. We chose the NASA ADS for its close context to Heliophysics and space physics research as opposed to something like Web of Science, which is much broader. We consider the years 1996 (when we picked up the thread of Complexity Heliophysics with \cite{Klimas1996}) to the present in this review; 
    \item Apply a `Heliophysics filter.’ From ADS we identify a subset of artifacts by selecting Heliophysics journals from the collection of journals that ADS indexes \footnote{\url{https://adsabs.harvard.edu/abs_doc/journals1.html}}. We manually and in cooperation with the Heliophysics community selected 33 journals. We intentionally broadly conceived of Heliophysics in selecting these journals, effectively casting a wide net for articles we would identify. The journals are listed in the Github repository. 
    \item Next we applied a ‘complexity science filter.’ For each article, we looked at the abstract and title and performed a simple match: if the abstract and title contained $>$N terms in the Complexity Heliophysics terms list, then it was considered a match and added to the corpus. N can be tuned based on desired inclusivity/exclusivity in the final corpus. Figure \ref{corpus thresholds comparison} shows the number of documents in the corpus for each threshold between three and ten, falling off roughly exponentially with the number of terms required to match. The graph indicates a slight knee between four and six matching terms. In an attempt to balance number of documents with relevance to Heliophysics, we chose five terms (i.e., only documents whose abstracts and titles include five or more distinct terms in the Heliophysics glossary were included). This give an automated database that is roughly one order of magnitude greater than the number considered manually ($\sim$3000 vs. 300). 
\end{enumerate}

In our analysis, $\sim$120k articles were obtained from ADS between 1996-present. Among those articles, Geophysical Research Letters (36k), Advances in Space Research (15k), Journal of Geophysical Research (14k), and Journal of Geophysical Research: Atmospheres (14k) contributed more than 13k articles. After matching based on the glossary, the automated corpus consisted of $/sim2600$ documents.

The corpus can then be examined manually or programmatically. We provide a few examples that give a glimpse into the corpus. Readers can freely explore the corpus via the Github repository \url{https://github.com/rmcgranaghan/ComplexityHelio_LivingReviews/tree/main/data}. 

First, Table \ref{table word occurrence} provides all words from the glossary that were found more than 100 times in the corpus (remember that this corpus only keeps papers with $\geq$5 words from the glossary, so this table represents the number of times that glossary word appears among such papers). 

\begin{center}
\begin{tabular}{ |l|c| } 
 \hline
 Glossary Word & Number of Appearances \\ 
 \hline
 system & 1424 \\
dynamics & 1015\\
evolution  & 944\\
distribution &  773\\
nonlinear &  772\\
systems &  694\\
dynamical  & 674\\
instability &  565\\
complex &  531\\
turbulence  & 511\\
interactions &  497\\
phenomena  & 449\\
instabilities &  406\\
turbulent  & 400\\
critical  & 379\\
scaling  & 375\\
equilibrium  & 292\\
environment &  276\\
component  & 255\\
bifurcation &  187\\
threshold  & 186\\
heterogeneity  & 171\\
complexity  & 161\\
phenomenon  & 151\\
network  & 149\\
stationary  & 128\\
chaos &  127\\
fluctuation  & 121\\
fractal  & 119\\
feedback &  113\\
nonlinearity &  111\\
community &  109\\
networks &  103\\
directed  & 102\\
multifractal  & 101\\
boundaries  & 100\\
exponential &  96\\
 \hline
 
\end{tabular}
\label{table word occurrence}
\end{center}

As an example, we used a text clustering tool Lingo4G\footnote{\url{https://carrotsearch.com/lingo4g/}} applied to the automated corpus to determine document clusters and topics. The mode used was Lingo4G's phrase extractor that extracts frequent words and sequences of words from the corpus (in this case the titles and abstacts from the corpus, not their full texts). Those frequent words and sequences are assigned as labels. 

\begin{figure}[h]
\centering
\includegraphics[width=\textwidth]{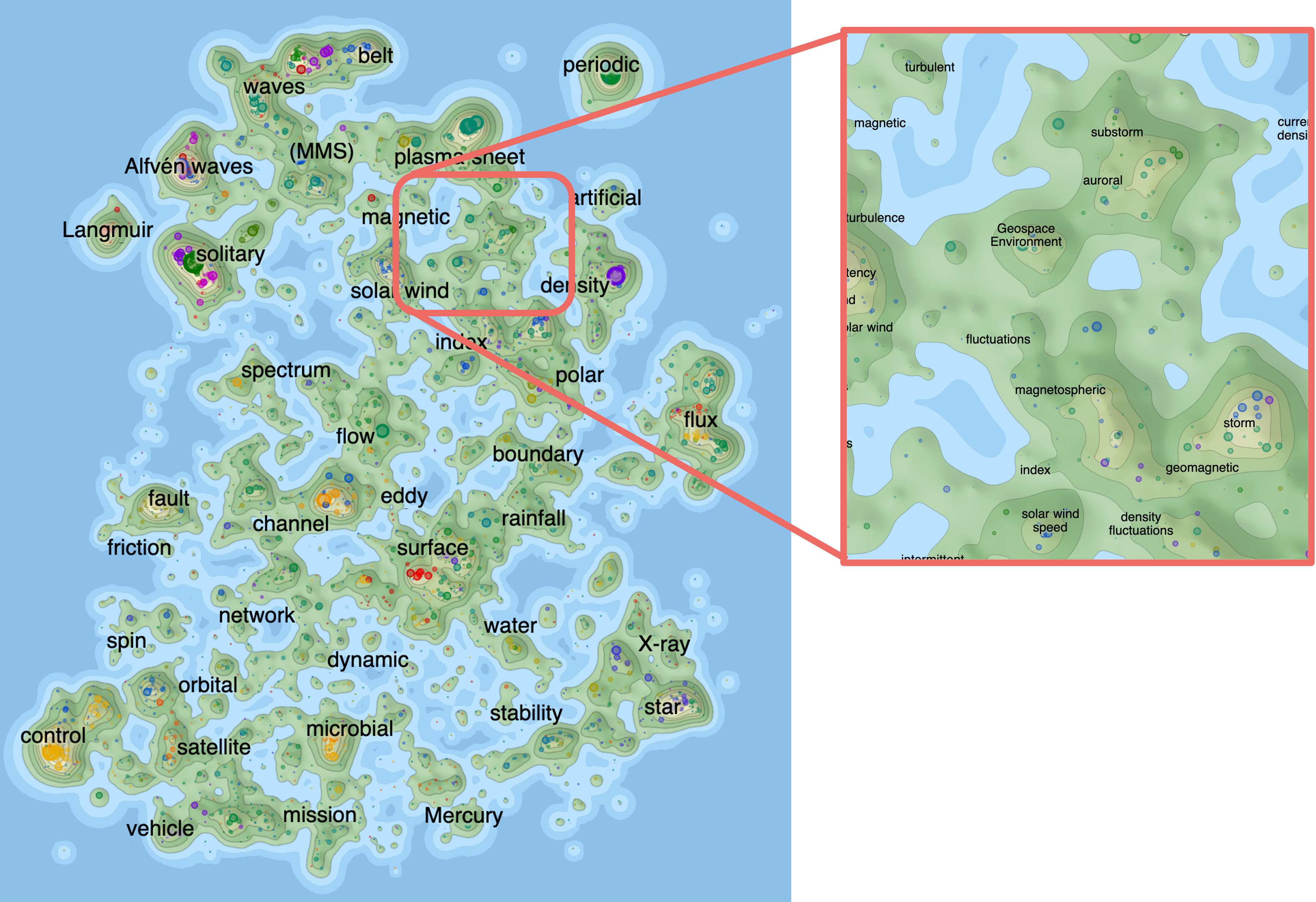}
\caption{Lingo4G (\url{https://carrotsearch.com/lingo4g/}) cluster map analysis of the automated corpus. Groups of thematically related labels are provided as high-level themes (overview plot on the left) and sub-themes (zoomed-in plot on the right). Labels are frequent words and sequences of words from the corpus.}
\label{Lingo4G clusters}
\end{figure}

\begin{figure}[h]
\centering
\includegraphics[width=\textwidth]{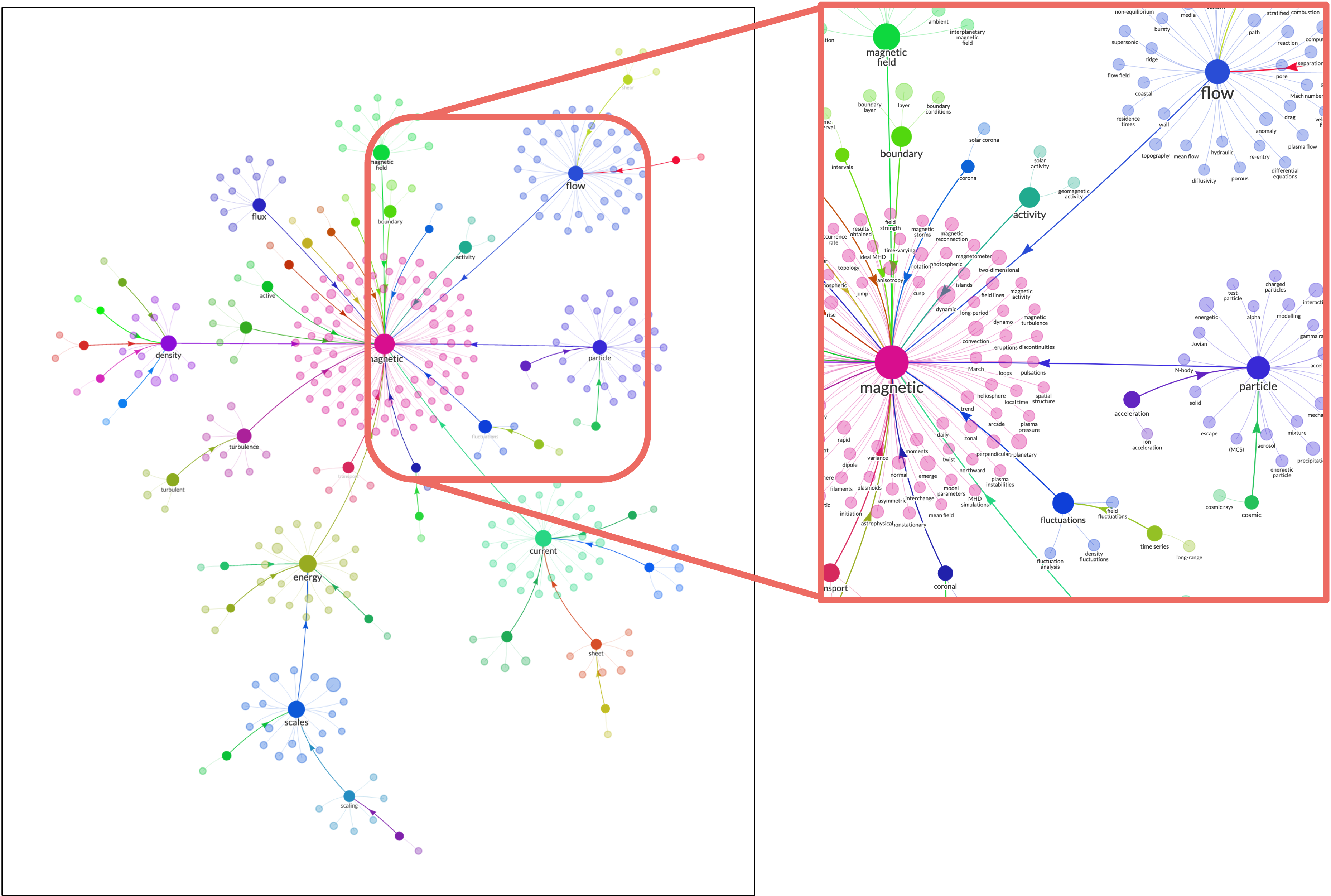}
\caption{Lingo4G (\url{https://carrotsearch.com/lingo4g/}) network map of the document clusters of the automated corpus. `Exemplar' labels for clusters are drawn on the map and connections to other exemplar labels are provided. The left shows the network for the exemplar `magnetic' and the right zooms in on a part of the network.}
\label{Lingo4G network}
\end{figure}

Figure \ref{Lingo4G clusters} shows a cluster map created from the automated corpus. Groups of thematically related labels are provided as high-level themes (overview plot on the left) and sub-themes (zoomed-in plot on the right). These clusters are helpful in myriad ways, including identifying content-wise similar documents (you can find each document within the cluster map), for identifying outliers as possible themes in need of more exploration, and `bridge' themes that link two or more prominent themes in the corpus. To further explore the cluster map, we created a network map of the clusters. Each cluster was identified by an \textit{exemplar} label, which serves as the description for the entire cluster. Then, related clusters and labels are connected to one another. The exemplar label of one cluster can be a member of another cluster. Figure \ref{Lingo4G network} illustrates the network map created for the `magnetic' label. Here, `flux,' `flow,' `particle,' etc. are all members of the magnetic cluster, but also are exemplars of their own clusters. One use of this network map would be to discover resources related to time series fluctuations that are also part of the broader class of magnetic phenomena. This permits rapid identification of documents in specific areas and gap analysis for what has not been extensively explored. 

Topic modeling methods like latent dirichlet allocation (LDA) \cite{Blei_2003} could easily be applied to the corpus as an engine for recommendation and insight and different semantic mapping of the information in the corpus (e.g., \cite{Papadimitriou1998LatentSI}). Additionally, more complex analyses of the automated corpus are possible. 

\begin{figure}[h]
\centering
\includegraphics[width=\textwidth]{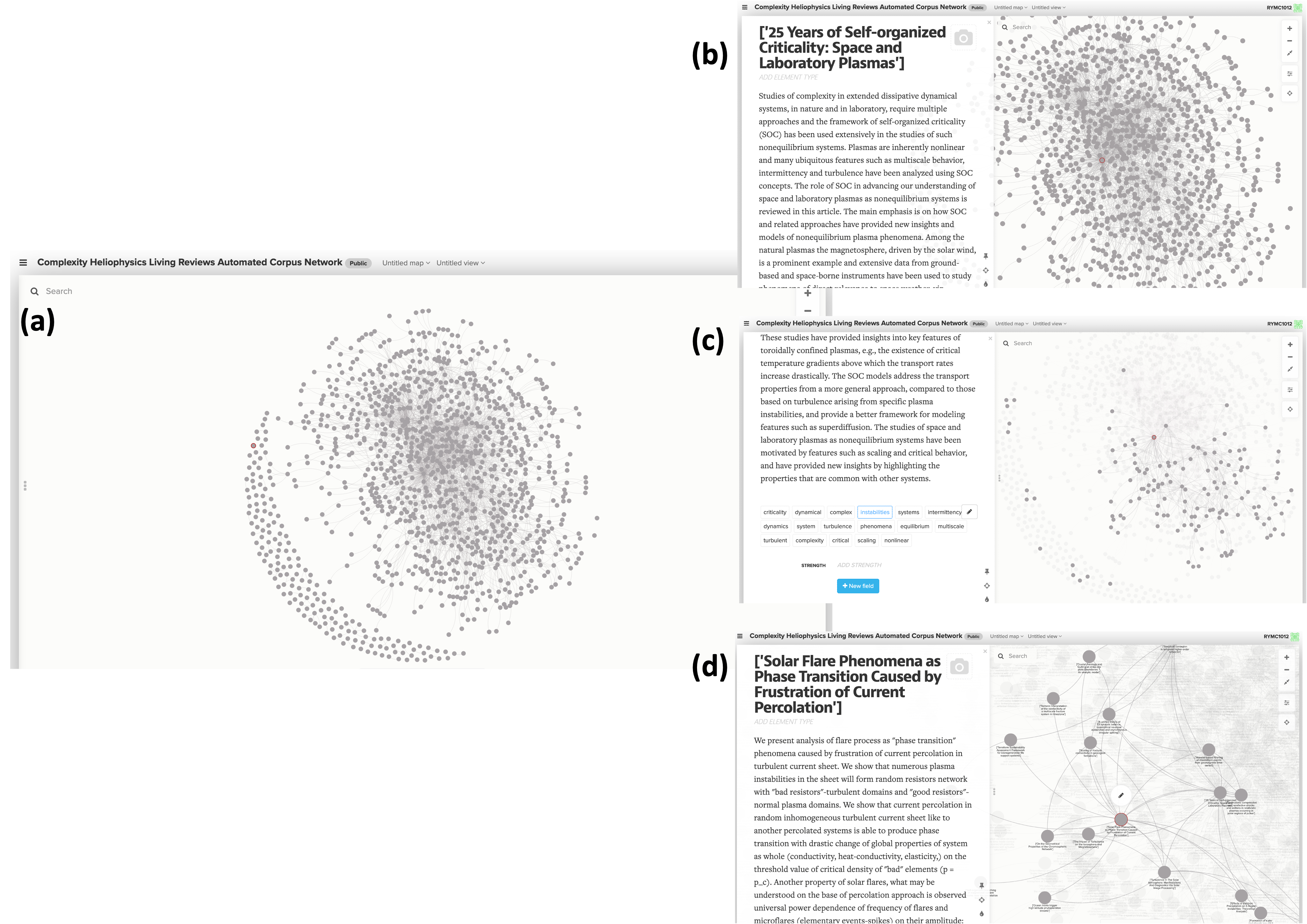}
\caption{A few screenshots of the \url{kumu.io} visualization. (b) Any node can be selected, revealing title, abstract, and words from the glossary found in them. (c) Any word can be selected and all nodes that contain it in the network will be highlighted. (d) Zooming in, one can explore local structures and clusters.}
\label{kumu figure}
\end{figure}

We next analyzed the automated corpus using network analysis. To construct the network, we considered each paper in the automated corpus using the threshold of five (i.e., only papers whose titles and abstracts contained five distinct terms from the Complexity Heliophysics glossary). For each paper, we looked at all other papers and a connection was defined if two papers shared more than four terms from the glossary. For instance, `paper1' and `paper2' would be connected if their respective set of terms from the glossary are: 1) [`\textbf{complexity}', '\textbf{systems}', 'fractal', '\textbf{multiscale}', '\textbf{network}', '\textbf{nonlinearity}', 'boundaries'] and 2) [exponential, bifurcation, `\textbf{complexity}', '\textbf{systems}', 'dynamical', `equilibrium', '\textbf{multiscale}', '\textbf{network}', '\textbf{nonlinearity}']. We constructed a directed network (meaning the edges have a preferred direction) that points from the earlier paper to the later paper, perhaps more capably capturing a flow of ideas. To analyze and interpret the network, we constructed a random network containing the same number of nodes and connections as the resultant network. Perhaps unsurprisingly, we find a much higher average clustering coefficient \cite{Schank2005ApproximatingCC} for the automated corpus than the random network, indicating the presence of much more densely connected clusters or communities and stronger local structure. This network contains rich potential for discovery. To enable the community to explore the possibilities, we have visualized and made fully interactive the network using \url{kumu.io}, which can be accessed at \url{https://embed.kumu.io/82d58b5453d1c4ba4f05d7240d142102}. Figure \ref{kumu figure} shows a few screenshots of the kumu visualization. Any node can be selected (Figure \ref{kumu figure}b), revealing title, abstract, and words from the glossary found in them. Any word can be selected and all nodes that contain it in the network will be highlighted (Figure \ref{kumu figure}c). Zooming in, one can explore local structures and clusters (Figure \ref{kumu figure}d). 

The automated corpus and topic analysis augmented the manual identification and review of articles and is an additional artifact of this review; perhaps even one that can become standard for future reviews. It should be considered a resource that complements the extensive references cited in the body of this review and contains high potential for discovering trends and knowledge about Complexity Heliophysics. It is important to note that the manual and automated corpora are not disjoint nor is the manual corpus strictly a subset of the automated corpus. Many references are shared across them, lending validation to the process of generating the automated set, but there are many references in the manual set that are not included in the automated one. This points to the flexibility of the scientist-driven discovery process, pulling in relevant references and material that might be more distant or irregularly connected to the research at hand than the necessarily more rigid automated process. This review, in particular, read widely in gathering material, many connections of which an automated approach would likely not have captured. The point is there must be an intersection of manual and automated gathering of resources, the manual approach benefiting from flexibility and capacity to range widely and be discerning and the automated approach benefiting from the volume of resources it can examine. 


Finally, given the corpus of articles from both manual and automated compilation, `mind maps' \cite{Buzan1994TheMM, Brner2015AtlasOK} were constructed from the main ideas in the articles. We do not present those mind maps, but a direct result from that mapping activity is the structure of this review such that the very sections and progression of this document reflects the achronological development of the ideas of Complexity Heliophysics.

\section{Appendix C: Key Datasets}\label{datasets appendix section}

This appendix compiles a list of important datasets, and their original appearance in the literature, that appear across this review to aid readers who wish to compile datasets and explore data-driven research across the datasets that have factored importantly in the Complexity Heliophysics paradigm. This list is merely an introduction, certainly not exhaustive. 

\begin{center}
\begin{sidewaystable}
\begin{tabular}{ |c|c| } 
 \hline
 Citation(s) & Description \\ 
 \hline
 \multirow{2}{*}{\cite{Bargatze_1985} } &  34 intervals of high time resolution Solar Wind (IMP8)--AL index dataset. \\
 & These data are the standard or benchmark dataset from most of the work addressed in \cite{Klimas1996} \\ 
 \hline
  \cite{Torr_1995} & Polar spacecraft ultraviolet imager (UVI) observations. \\
  \cite{BRITTNACHER19971037}  & High quality global images of auroral activity. \\
 \hline
 \multirow{3}{*}{\cite{Uritsky_2002} } & 15,500 POLAR UVI frames showing activity in the nighttime sector of the aurora (55 to 90$^{\circ}$ \\
 & MLAT, 2000 to 0400 MLT) in the Lyman-Birge-Hopfield-long filter mode. \\
 & These data permitted a spatio-temporal technique that proves vital in auroral data analyses \\
 \hline
  \multirow{5}{*}{\cite{Gjerloev_2009} } & A worldwide collaboration of organizations and national agencies\\
  & that operate more than 200 ground-based magnetometers. \\
  & Provides measurements of magnetic field perturbations\\
  & from all available stations in common coordinate frames,\\
  & identical time resolution, and a common baseline removal \\
   \hline
  \cite{Leger2015InflightPO}  & Fluctuations of the Earth's magnetic field as observed in-situ.\\
  \cite{deMichelis2015MagneticFF}  & European Space Agency (ESA) Swarm satellites vector field and absolute scalar magnetometers. \\
 \hline
 \hline
 
\end{tabular}
\end{sidewaystable}
\end{center}





\end{appendices}


%

\end{document}